\documentclass{article}[draft]

\usepackage{booktabs}       
\usepackage{arxiv}
\usepackage{float}
\usepackage[usenames,dvipsnames]{color}
\usepackage{placeins}
\usepackage{multirow}

\RequirePackage{uclthesis}

\usepackage[sorting=none,style=numeric-comp]{biblatex}
\renewbibmacro*{in:}{%
  \iffieldundef{journaltitle}
    {}
    {\printtext{\bibstring{in}\intitlepunct}}}

\newcommand{\printtitle}{Reconstruct everything everywhere all at once:\texorpdfstring{\\}{ }An end-to-end model for future colliders}

\title{\printtitle}

\author{
Max Hart$^{*}$,
Kwok Yiu Wong$^{*}$,
Samuel Van Stroud,
Philippa Duckett,
Tim Scanlon,
Gabriel Facini
\\
\textit{Centre for Data Intensive Science and Industry, University College London} \\
$^{*}$Equal contribution, primary authors.
}

\renewcommand{\shorttitle}{Reconstruct everything everywhere all at once: An end-to-end model for future colliders}
\hypersetup{
    colorlinks,
    pdftitle={\printtitle},
    pdfsubject={},
    pdfauthor={Max Hart, Kwok Yiu Wong, Samuel Van Stroud, Philippa Duckett, Tim Scanlon, Gabriel Facini},
    pdfkeywords={Particle Physics, ML Reconstruction, Object Detection},
    citecolor=blue,
    urlcolor=blue,
    linkcolor=blue,
}

\begin{document}

\maketitle

\vspace{8em}
\begin{abstract}
We present the Reconstruct Everything Model (\ours{}), an end-to-end multitask model for sparse, multimodal sensor data, demonstrated in high-energy physics. To our knowledge, \ours{} is the first demonstration of a single learned framework simultaneously reconstructing multiple particle types directly from hit-level measurements across the silicon tracker, calorimeters, and muon system in a full simulation. It performs hit assignment, particle identification, kinematic regression, and resonance reconstruction without externally reconstructed tracks or calorimeter clusters as inputs. \ours improves particle-reconstruction efficiency and purity and jet-energy resolution relative to both conventional and machine-learned particle-flow approaches. In diboson events, the learned resonance reconstruction improves $W/Z$ mass separation relative to conventional jet reconstruction. These results demonstrate a reconstruction chain from detector hits to particles to resonances, with potential applications to structured prediction in other large-scale, sparse, multimodal sensor systems.
\end{abstract}
\vfill


\clearpage
\setcounter{tocdepth}{2}
\begingroup
\renewcommand{\null}{\hbox{}}
\endgroup
\clearpage
\clearpage
\section{Introduction}
\label{sec:intro}

Collider experiments study the fundamental constituents of matter by measuring and analysing the particles produced in high-energy collisions.
Modern detectors~\cite{ATLAS,CMS,LHCb,ALICE} are building-sized instruments with multiple subsystems, each of which may contain hundreds of millions to billions of readout channels.
Their reconstruction pipelines combine specialised algorithms across these subsystems and comprise millions of lines of code maintained by large experimental collaborations~\cite{SOFT-2022-02,CMSSoftware}.

Raw detector signals are first grouped into local measurements, such as charge clusters in silicon sensors and energy deposits in calorimeters.
These measurements are used to identify charged-particle trajectories (tracks) and calorimeter cascades (showers), which are then combined into reconstructed particles. 
The stages are typically optimised separately.
This separation can limit the use of information across the detector and complicate adaptation to changing collider conditions, detector updates, or new signatures sought by downstream analyses.

Machine learning entered this workflow in the late 1980s, when Denby~\cite{Denby:1988} and Peterson~\cite{Peterson:1989} independently proposed neural networks for charged-particle track reconstruction in bubble and wire chambers~\cite{Denby:1999}. 
By the 2000s, boosted decision trees and shallow networks were widely used for specialised tasks and contributed to the discovery of the Higgs boson in 2012~\cite{HIGG-2012-27,CMS-HIG-12-028}. 
Deep learning subsequently extended these applications across pattern recognition, and by the early 2020s machine learning was used throughout detector simulation, reconstruction, and analysis~\cite{hepmllivingreview}. 
Most reconstruction methods nevertheless remain confined to individual subsystems or operate on previously reconstructed objects, improving individual components of the chain.

\begin{figure}[t]
    \centering
    \includegraphics[width=0.95\linewidth]{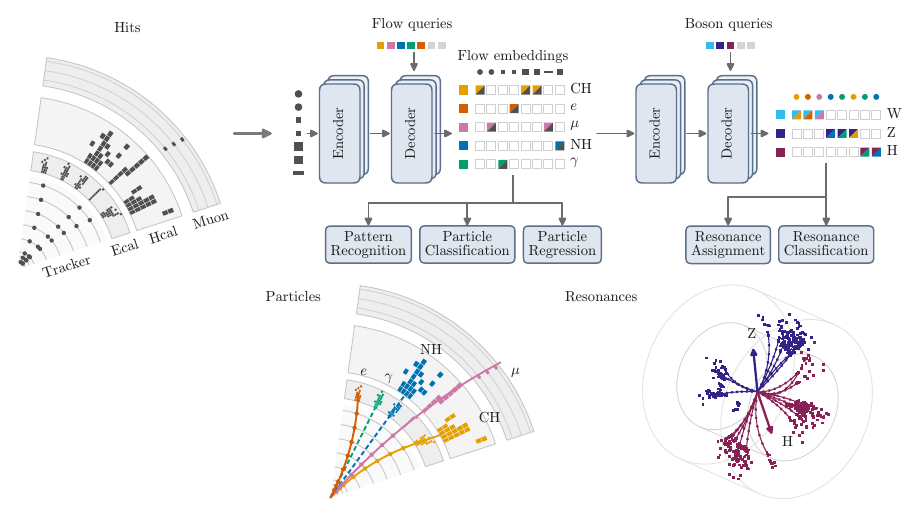}
    \caption{
    The \ours{} reconstruction pipeline, comprising two encoder-decoder MaskFormer blocks. 
    The first takes detector hits as input and jointly performs tracking and clustering to reconstruct particles. 
    The second uses the resulting particle-query embeddings to assign particles to resonance bosons.
    }
    \label{fig:pipeline}
\end{figure}

The Reconstruct Everything Model (\ours{}) combines these stages into a unified, end-to-end learned reconstruction framework, as illustrated in Figure~\ref{fig:pipeline}.
To our knowledge, it is the first demonstration of a single framework simultaneously reconstructing multiple particle types directly from hit-level inputs across the silicon tracker, calorimeters, and muon system in a full collider-detector simulation. 
\ours{} assigns hits to particles, identifies the particles, and estimates their kinematic properties without externally reconstructed tracks or calorimeter clusters as inputs. 
It retains intermediate outputs, including hit assignments, tracks, calorimeter clusters, particle identities, and kinematic estimates, for inspection, validation, and downstream analyses. 
A further learned stage assigns particles to resonance bosons and predicts the boson identities, providing a path from ``hits to Higgs''.

We demonstrate \ours{} on the CLD detector concept~\cite{Bacchetta:2019fmz}, proposed for the electron--positron stage of the Future Circular Collider (FCC-ee)~\cite{FCC:2025lpp,FCC:2025uan,FCC:2025jtd}. 
The detector contains more than $10^{10}$ readout channels.
For the 240~GeV benchmark it performs the full reconstruction pass in around \SI{30}{\ms} on a single GPU (Appendix~\ref{sec:results_timing}).

Training proceeds in stages, with earlier networks kept frozen (Section~\ref{sec:training}).

The motivation is to exploit complementary information across detector subsystems within a unified reconstruction framework. 
We demonstrate that reconstruction from hit-level measurements is feasible and improves performance across a range of observables while retaining intermediate outputs for validation. 
The approach allows rapid testing of detector designs and may also be used in other systems with sparse, multimodal sensor data.

\section{Related Work}
\label{sec:related}

\subsection{Detector Reconstruction at Colliders}
Detectors at the Large Hadron Collider~\cite{LHC} and proposed future colliders~\cite{FCC:2025lpp} typically have a layered geometry around the collision point.
A silicon-based inner tracking system, immersed in a magnetic field, records the interaction points of charged particles as they traverse successive layers of finely segmented sensors, with sensor dimensions, occupancies, and technologies varying across detector regions.
These silicon measurements are used to reconstruct the curved trajectories of charged particles, while neutral particles pass through the tracker without leaving a signal.
Surrounding calorimeters absorb both charged and neutral particles, converting their energies into cascades of secondary particles and light.
Electromagnetic calorimeters are optimised for the compact showers of electrons and photons, while hadronic calorimeters measure the broader showers produced by hadrons such as protons, neutrons, and pions.
Muons traverse the calorimeters with minimal energy loss and are typically reconstructed by matching inner-detector tracks to signals in outer muon detectors.
Neutrinos pass through all detector layers without leaving a direct signal, resulting instead in an imbalance of reconstructed momentum.

Reconstruction converts detector hits into a consistent set of particles: charged and neutral hadrons, muons, photons, and electrons.
These particles are used to reconstruct jets and missing momentum. Their multiplicities and properties provide information about the underlying collision process.
The reconstruction chain combines algorithms specialised for individual subsystems with algorithms that associate measurements across them. The following sections review conventional and machine-learning approaches to its main stages.

\paragraph{Inner detector hit assignment.}

The first tracking step groups nearby silicon signals to estimate where a charged particle crossed a sensor.
In pixel detectors this means clustering adjacent activated pixels, while in strip detectors it means combining coherent strip signals into a single measurement~\cite{ATLAS-ITk,CMS-TRK-11-001}. ATLAS has developed neural-network-based cluster-characterisation algorithms for pixel detectors~\cite{ATLAS:2014lvy}, while CMS has used template-based pixel-hit reconstruction~\cite{CMS-TRK-11-001}.
These approaches achieve excellent local position resolutions of $\mathcal{O}(10~\mu\mathrm{m})$ and high efficiency in resolving complex hit patterns, but they are detector-specific and make limited use of complementary information from neighbouring tracking layers or other detector subsystems.

\paragraph{Tracking.}

Charged-particle tracking has traditionally relied on combinatorial pattern recognition followed by Kalman filtering~\cite{Fruhwirth:1987fm,Rudolf21}, with variants such as conformal tracking~\cite{hansroul1988fast} and cellular automata-based approaches~\cite{Bocci:2020pmi,CMS-TDR-Tracker}.
These algorithms often require multiple passes to reconstruct different categories of tracks and in some cases are seeded by calorimeter information~\cite{IDTR-2021-03,CMS-TRK-11-001}.
Track reconstruction is also one of the most computationally demanding steps in event processing, and the combinatorial complexity of traditional algorithms becomes rapidly prohibitive at the particle densities expected at the upgraded LHC starting in 2030~\cite{atlas_hllhc_comp,cms_hllhc_comp}.

The TrackML challenge~\cite{8588707,2019EPJWC.21406037K,amrouche2020tracking,amrouche2023tracking} catalysed a wave of machine learning approaches, particularly graph neural networks~\cite{Ju:2021ayy,Biscarat:2021dlj,trackhgnn,ju2020graph,caillou2024physics,ATL-PHYS-PUB-2024-018}.
These models have demonstrated strong performance but still rely on pre-constructed inner detector clusters and complex graph structures.
More recently, detector-agnostic pipelines such as Geometric Graph Track Finding (GGTF)~\cite{GGTF} unify hits from heterogeneous tracking technologies into a single representation, though current implementations face significant memory and timing challenges.
The present work uses mask-transformer architectures inspired by object detection~\cite{maskformer,mask2former}. Applied directly to hits, these architectures have demonstrated favourable memory and inference-time scaling in tracking~\cite{MFTrk,PippaAttn,Miao:2026yvl}.
They have also improved tracking in the cores of high-$\pT$ jets, where multiple particles can contribute to the same detector measurement~\cite{ATL-PHYS-PUB-2025-045}.

\paragraph{Calorimeter clustering.}

Neutral and charged particles are measured in calorimeters that surround the inner tracker.
Energy deposits above noise thresholds in individual calorimeter cells are grouped into topological clusters~\cite{ATLAS:2017topoclusters,CMS-PRF-14-001}.
These clusters can correspond either to single particles or to spatially overlapping particle showers.
Recent work has explored neural-network-based calibrations of calorimeter clusters~\cite{JETM-2024-01} and particle reconstruction using deep-learning techniques~\cite{Valsecchi:2022rla,ATL-PHYS-PUB-2022-040,Qasim:2021hex,Canudas:2022lsf,Akchurin:2021afn}.
These include graph-based and point-cloud approaches to high-granularity calorimetry, as well as detector-specific applications such as electromagnetic calorimeter clustering in LHCb~\cite{Canudas:2022lsf}.
Such methods can achieve strong calorimeter performance, but they generally remain confined to calorimeter inputs or reconstructed clusters and do not fully exploit complementary tracking information.

\paragraph{Particle flow.}

Particle-flow reconstruction combines inner-detector tracks with calorimeter clusters to reconstruct charged particles, then uses the remaining calorimeter energy to reconstruct neutral particles. The result is a unified particle collection.
This approach was used in the ``energy-flow'' techniques at LEP~\cite{ALEPH,DELPHI}, was developed into full particle reconstruction at CMS~\cite{CMS-PRF-14-001} and ATLAS~\cite{PERF-2015-09}, and is the standard approach for future colliders through frameworks such as \pandora{}~\cite{Thomson:2009rp}.
More recently, particle flow has been enhanced by machine-learning methods such as MLPF~\cite{Pata:2021oez}, HGPFlow~\cite{Kakati:2024dun,DiBello:2022iwf}, and GLOW~\cite{glow}.
However, these methods still rely on reconstructed tracks and calorimeter clusters as inputs, rather than operating directly on raw detector signals.
\hitpf~\cite{Garcia:2026sqb} moves towards end-to-end reconstruction by working directly with calorimeter hits, removing the need for independent calorimeter clustering.
It still requires externally reconstructed tracks, uses heuristic clustering to form particle candidates, and needs subsequent reconstruction steps to obtain analysis-level objects.

\paragraph{Particle and jet identification.}

Electrons and photons are identified primarily from electromagnetic calorimeter clusters, with electrons additionally matched to inner-detector tracks~\cite{ATLAS:2019qmc,CMS:EGM-17-001}.
Muons are reconstructed by combining tracks in the inner tracker with segments in the outer muon spectrometer~\cite{ATLAS:2016iqc,CMS:MUO-16-001}.
Energetic quarks and gluons give rise to collimated sprays of particles, reconstructed as jets.
Jets are further classified through dedicated tagging algorithms that identify their origin, for example whether they were initiated by a specific type of quark, gluon, or tau-lepton~\cite{FTAG-2023-05,ATL-PHYS-PUB-2026-001,Qu:2022mxj,CMS:2017wtu,atlas_tau_reco,CMS-TAU-16-003}, or by a high-momentum $W$, $Z$, or Higgs boson decaying to quarks~\cite{JETM-2018-03,ATL-PHYS-PUB-2023-021,CMS-DP-2020-002}.
Physics analyses use the energies, multiplicities, and relative positions of these particles and jets to distinguish collision processes. Their performance therefore depends on the upstream reconstruction.

\begin{figure}
    \centering
    \includegraphics[width=0.95\linewidth]{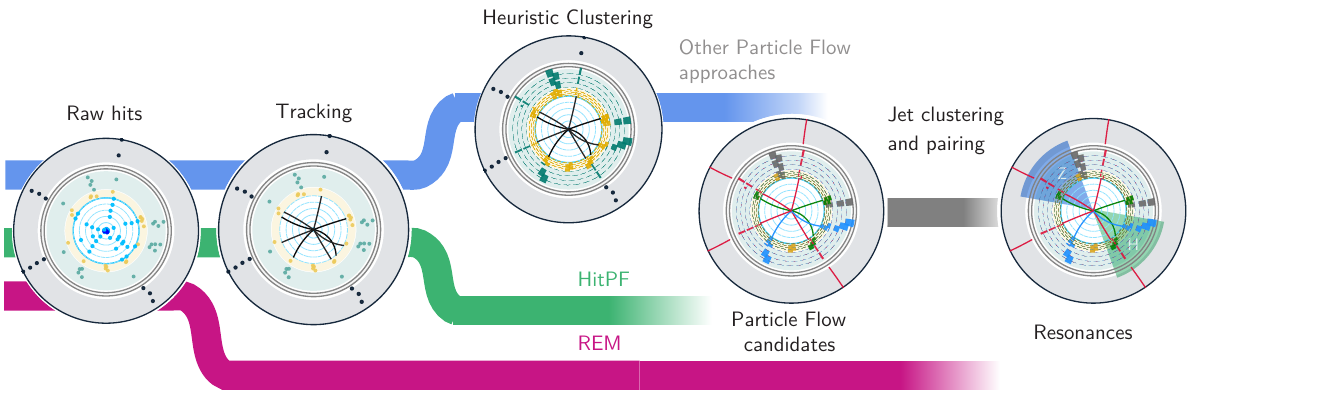}
    \caption{
    Input requirements and reconstruction stages of particle-flow methods.
    Traditional approaches (top) first reconstruct charged-particle tracks, muon tracks, and calorimeter clusters separately before combining them.
    \hitpf (middle) groups calorimeter and muon-system hits into particle candidates, but requires reconstructed charged-particle tracks as an input. 
    \ours{} (bottom) performs all major reconstruction steps, including tracking and resonance reconstruction, simultaneously.
    Figure adapted from Ref.~\cite{Garcia:2026sqb}.}
    \label{fig:reconstruction_approaches}
\end{figure}

Figure~\ref{fig:reconstruction_approaches} illustrates the differences in input requirements and reconstruction stages between conventional particle-flow approaches, \hitpf{} and \ours{}.

\subsection{Machine Learning Methods}

The design of \ours{} also draws on methods for structured prediction, set prediction, and multimodal learning.

\paragraph{Graph neural networks and sparse representation.}
Graph neural networks (GNNs)~\cite{kipf2016semi,battaglia2018relational} represent data as nodes connected by edges.
This representation is well suited to detector data, where hits, clusters, tracks, and particles form irregular geometric structures with sparse connectivity.
GNNs have consequently been widely adopted for tracking, calorimeter reconstruction, jet tagging, and particle-flow reconstruction.
Graph construction determines which entities can exchange information, while message passing introduces computational costs that must be controlled at high hit multiplicities.

\paragraph{Object detection, segmentation, and set prediction architectures.}

Transformer architectures formulate object detection and segmentation as direct set-prediction problems.
DETR~\cite{carion2020end} introduced an end-to-end object-detection paradigm in which a fixed set of learned object queries is matched to ground-truth objects using bipartite matching, avoiding hand-designed proposal generation and non-maximum suppression.
MaskFormer~\cite{maskformer} and Mask2Former~\cite{mask2former} extend this idea to segmentation by predicting a set of class-labelled masks, thereby unifying semantic, instance, and panoptic segmentation under a mask-classification formulation.
KMaX-DeepLab~\cite{kmax_deeplab_2022} further recasts the decoder interaction between object queries and image features as a $k$-means-like clustering step, with queries acting as cluster centres and pixels assigned to queries through cross-attention.
\ours{} builds on this object-detection and mask-prediction viewpoint, replacing image pixels with detector hits and replacing visual object masks with particle-to-hit assignments across tracker, calorimeter, and muon systems.

\paragraph{Multimodal learning and end-to-end pipelines.}

Multimodal learning provides methods for combining heterogeneous inputs in a shared representation.
Contrastive models such as CLIP~\cite{radford2021learning} and large-scale multimodal transformers~\cite{alayrac2022flamingo} demonstrate joint learning from images and language.
Point-cloud methods such as PointNet~\cite{qi2017pointnet} and Set Transformer~\cite{lee2019set} provide flexible processing of irregular geometric data.
Sutton's ``bitter lesson''~\cite{sutton2019bitter} motivates the use of general methods that can benefit from increasing data and computation.
\ours{} follows this direction by treating the detector as a sparse, heterogeneous sensor array and learning to reconstruct the particle content of the event directly.

\section{Dataset}
\label{sec:dataset}

Electron--positron collision events are generated with \textsc{PYTHIA}~\cite{bierlich2022comprehensive}, passed through full detector simulation with \textsc{Geant4}~\cite{Agostinelli:2002hh}, and digitised in \textsc{Key4hep}~\cite{Ganis:2021vgv} using the \texttt{2026-02-01} release.
The detector response is simulated using the CLD (CLIC-like Detector) concept~\cite{Bacchetta:2019fmz}, a proposed general-purpose experiment for the electron–positron stage of the Future Circular Collider~\cite{FCC:2025lpp,FCC:2025uan,FCC:2025jtd}, using the \texttt{CLD\_o2\_v05} geometry.

The main evaluation uses simulated $WW$, $ZZ$, and $ZH$ events at $\sqrts=240$~GeV, where $W$ and $Z$ are the weak gauge bosons and $H$ is the Higgs boson.
Complementary benchmarks use light-quark $\Zqq{}$ events at $\sqrts=91$~GeV and $\ttbar{}$ events at $\sqrts=365$~GeV.
The training and test samples are listed in \cref{tab:samples}.

\begin{table}[t]
    \centering
    \caption{
    Summary of the simulated event samples used to train the backbone hit-assignment and regression stage and the resonance-reconstruction stage of \ours{} and \flowf{}. 
    The columns $N_{\mathrm{train}}$ and $N_{\mathrm{test}}$ give the number of events used for training and evaluation respectively.
    For processes labelled as ``all'', particles decay according to their Standard Model branching fractions.
    The notation $\Hqq{}$ denotes $\Hbb{}$ or $c\bar c$, while
    $q=uds$ denotes only the light-flavour quarks $u$, $d$, and $s$.
    }
    \label{tab:samples}

\begin{tabular}{clccc}
        \toprule
        $\sqrts$ [GeV]
        & Process
        & $N_{\mathrm{train}}^{\mathrm{backbone}}$
        & $N_{\mathrm{train}}^{\mathrm{resonance}}$
        & $N_{\mathrm{test}}$ \\
        \midrule

        \multirow{7}{*}{365}
            & $\ttbar{} \to \mathrm{all}$
            & 112\,000
            & --
            & 5\,000 \\

            & $\ttbar{} \to \ell^{\pm}\ell^{\mp}b\bar{b}\nu\nu$
            & 77\,000
            & --
            & -- \\

            & $\WWqqqq{}$
            & 36\,000
            & 116\,000
            & -- \\

            & $\WWmunuqq{}$
            & 11\,000
            & 37\,000
            & -- \\

            & $\ZZqqqq{}$
            & 41\,000
            & 108\,000
            & -- \\

            & $\ZZnunuqq{}$
            & 21\,000
            & 54\,000
            & -- \\

            & $\ZHnunuqq{}$
            & 33\,000
            & 104\,000
            & -- \\

        \cmidrule(lr){1-5}

        \multirow{5}{*}{240}
            & $\WWqqqq{}$
            & 21\,000
            & 100\,000
            & 6\,000 \\

            & $\WWmunuqq{}$
            & --
            & 12\,000
            & 6\,000 \\

            & $\ZZqqqq{}$
            & 21\,000
            & 85\,000
            & 6\,000 \\

            & $\ZZnunuqq{}$
            & --
            & 12\,000
            & 6\,000 \\

            & $\ZHqqqq{}$
            & 57\,000
            & 177\,000
            & 6\,000 \\

        \cmidrule(lr){1-5}

        91
            & $\Zgammaqq{}$, $q=uds$
            & 70\,000
            & --
            & 5\,000 \\

        \midrule
            & Total
            & 500\,000
            & 805\,000
            & 40\,000 \\
        \bottomrule
    \end{tabular}
\end{table}

The 240~GeV evaluation samples contain $\WWqqqq{}$, $\WWmunuqq{}$, $\ZZnunuqq{}$, $\ZZqqqq{}$, and $\ZHqqqq{}$.
Here, $q$ denotes a quark, with charge-conjugate decays implied. The larger samples used for boson reconstruction are described in Section~\ref{sec:results_bosons}.

\subsection{CLD Detector}
\label{sec:detector}
The CLD detector combines an all-silicon pixel tracker, highly granular calorimeters, and a large muon system, with approximately $2.5\times 10^{10}$ readout channels in total, an order of magnitude more than the upgraded LHC detectors.
It follows the typical cylindrical design of collider experiments, with cylindrical central ``barrel'' layers capped by ``endcap'' discs. The overall dimensions are about 12~m in height and 11~m in length.

The silicon tracker comprises three subsystems: a vertex detector built from $25\times 25~\mu\mathrm{m}^2$ pixels; an inner tracker with fine elongated pixels; and an outer tracker with large-area elongated pixels.
Together, the all-silicon tracker provides at least eight measurement points (or hits) per charged particle, enabling momentum resolution of $\sigma(1/\pT)\sim 2\times 10^{-5}$~GeV$^{-1}$.
The electromagnetic calorimeter (ECAL) has 40 alternating silicon and tungsten layers. 
The surrounding hadronic calorimeter (HCAL) consists of 44 layers of steel absorber with interleaved scintillator tiles.
Together, they provide fine three-dimensional images of particle showers and support a particle-flow jet energy resolution of 3--4\%.
Outside the 2~T solenoid and iron yoke, a system of resistive plate chambers records the positions of penetrating muons. Table~\ref{tab:cld_summary} summarises the cell sizes and channel counts of each sub-detector.

\begin{table}[h!]
\centering
\small
\begin{tabular}{lcccc}
\toprule
Sub-detector & Barrel layers & Endcap discs & Cell size [mm$^2$] & Channels [$10^6$] \\
\midrule
Vertex detector (VTX) & 3 double layers & 3 double discs/side & $0.025\times0.025$ & 840 \\
Inner tracker (IT)    & 3 layers        & 7 discs/side        & $0.03\times0.3^{*}$    & 4609 \\
Outer tracker (OT)    & 3 layers        & 4 discs/side        & $0.03\times0.3$    & 19190 \\
ECAL                  & 40 layers  & Endcaps + plugs   & $5\times5$         & 158 \\
HCAL                  & 44 layers  & Endcaps + rings   & $30\times30$       & 9.2 \\
Muon system           & 6 layers   & Endcaps           & $30\times30$       & 3.6 \\
\midrule
Total &  &  &  & 24,810 \\
\bottomrule
\end{tabular}
\caption{
Summary of CLD sub-detector parameters assuming pixel geometry for the tracker. 
Numbers of channels are totals across barrel and endcap regions. $^{*}$ The innermost disks have a cell size of $0.025\times0.025$~mm$^2$
}
\label{tab:cld_summary}
\end{table}

\subsection{Reconstruction Baselines}

Our baseline is the standard CLD reconstruction pipeline: conformal tracking~\cite{hansroul1988fast} for charged-particle trajectories and \ppfa{} \texttt{v4.11.2}~\cite{Thomson:2009rp,Marshall:2015rfa} for particle-flow reconstruction.
Both are configured with their default CLD settings~\cite{CLICdp:2017vju}. 
The baseline and \ours{} are evaluated on the same events with the same object definitions and truth-matching criteria.
We also compare with the learned particle-flow model \hitpf~\cite{Garcia:2026sqb}, which takes conformal tracks and calorimeter hits as inputs.
We evaluate its published checkpoint, pre-trained on \Zqq{} events at $\sqrts=91$~GeV, on the 91~GeV light-quark evaluation sample used for the other reconstruction methods.

\subsection{Target Particles}
\label{sec:targets}

Truth particles include the stable final-state particles of the event generator, particles decayed by the generator, and the secondary particles created by the \textsc{Geant4} detector simulation. 
Beam remnant electrons and photons are excluded. 
Truth particles are kept only when they have a transverse momentum of at least 10~MeV and leave at least one hit in the detector.

Following the CLD design report in Ref.~\cite{Bacchetta:2019fmz}, we define \emph{reconstructable particles} as particles that do not decay in the tracker, have $\pT > 100~\mathrm{MeV}$ and $|\cos{\theta}|<0.99$, and leave at least 4 detector hits. For \emph{prompt-like tracks}, we additionally require the truth particle to have $10^\circ < \theta < 170^\circ$ and $R_{\rm production}<50\ \mathrm{mm}$.

Reconstruction targets are simulated particles at the calorimeter entrance, following the definition used by \hitpf~\cite{Garcia:2026sqb}. Hits from particles produced inside the calorimeter are grouped with their corresponding primary particle.
A \textit{tight} selection, following \hitpf~\cite{Garcia:2026sqb}, requires at least 10~MeV of visible energy deposited in the calorimeter region.
This excludes charged particles that leave sufficient tracker activity but do not arrive at the calorimeter. We use tight targets to evaluate all reconstruction methods, unless otherwise stated.\footnote{For the evaluations in this work, we additionally require that the simulated particle trajectory does not end inside the tracker volume. This excludes particles that decay inside the tracker and low-energy particles produced in the calorimeter that backscatter into it; the latter constitute about 1\% of tight targets at $\sqrts=240$~GeV.}
A \textit{loose} selection further includes particles that leave at least 4 silicon hits with at least 10~MeV transverse momentum and $|\eta| \leq 4$.

Algorithms may nevertheless reconstruct genuine particles outside the tight evaluation selection. To identify these objects, unmatched reconstructed particles are matched in a second step to particles selected only by the loose definition.
Both matchings use the double-majority (DM) criterion: shared hits must form at least half of the hits of each particle, counted over all sub-detectors together (\cref{sec:metrics}).
Objects matched in this second step are excluded from the purity denominator and from the event-level sums and jet inputs, for every algorithm alike.
These objects are excluded using simulation truth information to ensure a consistent evaluation against the tight target definition.
At $\sqrts=240$~GeV, the excluded particles account for an average reconstructed energy of 1.5, 0.2 and 0.4~GeV per event for \ours{}, \flowf{} and \ppfa{}, respectively. Their exclusion changes the relative visible-energy resolution by at most 4\%.
As a sensitivity check, extending the second truth matching to particles with three silicon hits reduces the \ours{} fake count by at most 3.2\%, with negligible impact on event energy.

Before the event-level sums and jet inputs are formed, \ours applies a selection that removes reconstructed particles with less than 5\% of their energy in the calorimeter, unless their predicted \IoU{} is at least 0.9. It aims at filtering out particles whose decay products are reconstructed separately and also removes fake energy.\footnote{In the 240~GeV evaluation set, this selection removes 4.5\% of the \ours objects, 91\% of which the second matching would also have identified, so that it reproduces most of the truth-based identification without truth information, as is required at inference. Most of the remainder are fakes. Applied to \flowf with its own \IoU{} score veto, the same cut would remove more real energy than fakes.} The selection uses no truth information and is part of the \ours output used for all event-level and jet results.

\subsection{Evaluation Metrics}
\label{sec:metrics}
We evaluate performance at the hit, particle, event, jet, and resonance levels.
Each reconstructed particle is first matched to the simulated particle that contributes the largest number of hits to the predicted mask using a one-to-one assignment. 

Hit-assignment efficiency and purity are computed separately for the tracker, ECAL, HCAL, and muon system.
Hit efficiency is the fraction of hits from a true particle that are recovered in its reconstructed counterpart, while hit purity is the fraction of hits assigned to a reconstructed particle that originate from the correct truth particle. 
The hit intersection-over-union is defined as
\[
\mathrm{IoU} = \frac{|H_{\mathrm{true}} \cap H_{\mathrm{pred}}|}{|H_{\mathrm{true}} \cup H_{\mathrm{pred}}|},
\]
where $H_{\mathrm{true}}$ is the set of detector hits associated with the truth particle and $H_{\mathrm{pred}}$ is the set of hits assigned to the matched reconstructed candidate. 
The \IoU{} can be calculated for a subset of systems or the entire detector.

A particle (or charged-particle track) is considered \emph{reconstructed} when both its hit-assignment efficiency and purity are at least $50\%$.
The corresponding object is labelled \textit{pure}.
\textit{Duplicate} objects are those matched to a truth particle that has already been matched by another reconstructed object with a higher \IoU{}.
\textit{Fake} objects do not match any truth particle under the double-majority criteria.
They can arise from combinations of unrelated hits, incomplete hit assignments, or insufficient detector activity to satisfy the matching criteria.

The average efficiency is the fraction of reconstructable truth particles that have been reconstructed. 
Purity is the fraction of reconstructed particles that are neither duplicates nor fakes.
In addition, the fake rate is defined as the fraction of objects that do not match to any truth particle. 
We assess particle identification both separately from and together with the pattern-recognition requirements.

For the visible particles selected for event-level observables (Section~\ref{sec:targets}), the total energy and invariant mass are
\begin{equation}
    E_{\mathrm{vis}} = \sum_{i\in\mathrm{visible}} E_i,
    \qquad M_{\mathrm{vis}}^2 = \left(\sum_{i\in\mathrm{visible}} p_i^\mu\right)^2,
    \label{eq:visible_mass}
\end{equation}
where $p_i^\mu$ is the particle four-momentum.

Energy performance is quantified through the response, defined as the relative difference between the reconstructed and true particle energy,
\begin{equation}
    \frac{E_{\mathrm{pred}} - E_{\mathrm{true}}}{E_{\mathrm{true}}}.
    \label{eq:fractional_energy_response}
\end{equation}
The median of the response distribution measures the bias. 
Different width estimators are used for different observables.
Tracking residuals and event-level responses are characterised by the interquartile range, $\mathrm{IQR}=Q_{75}-Q_{25}$, where $Q_x$ denotes the $x$th percentile. 
In the particle-resolution plots, the relative energy resolution is $(Q_{84}-Q_{16})/(2Q_{50})$ of $E_\mathrm{reco}/E_\mathrm{true}$, and the angular resolution is the 68th percentile of the opening angle between reconstructed and true directions. 
Jet-energy and boson-mass resolutions use iteratively fitted Gaussian widths, quoted as $\sigma/\mu$.
The separation of the reconstructed $W$ and $Z$ mass distributions is
\begin{equation}
    S = \frac{|\mu_Z-\mu_W|}{\bar\sigma},
    \qquad \bar\sigma = \frac{\sigma_W+\sigma_Z}{2},
    \label{eq:wz_separation}
\end{equation}
where $\mu_W$, $\mu_Z$ and $\sigma_W$, $\sigma_Z$ are the fitted peak positions and widths.
The fit procedures and additional measures of the response tails are specified in the corresponding Results sections.

Boson-label performance is evaluated independently of the particle assignments. In each fully hadronic event, the two predicted labels are paired with the two true bosons to maximise the number of agreeing labels. The efficiency of a class is the fraction of its true bosons paired with the correct label; its purity is the fraction of its predicted labels paired with a true boson of that class. Purity depends on the channel mixture. Since each evaluated event contributes two true bosons and two predicted labels, the overall efficiency and purity are equal.

\section{Model}
\label{sec:model}

\ours{} maps detector hits to particle masks, class labels and kinematic properties within a transformer framework.
It adapts MaskFormer~\cite{maskformer,mask2former} to sparse, heterogeneous, multimodal inputs, building on recent transformer-based approaches to charged-particle reconstruction~\cite{MFTrk,PippaAttn,Miao:2026yvl}.

\subsection{Attention}
Transformers~\cite{VaswaniAttention} allow information to be exchanged between input elements through attention.
Each input element (a detector hit or calorimeter cell) is projected into three learned representations: a \textit{key}, a \textit{value}, and a \textit{query}. 
The dot product between queries and keys produces similarity scores that determine how information is aggregated from the values. 
In the encoder, self-attention (SA) exchanges information between hits.
In the decoder, cross-attention (CA) connects latent particle queries to the encoded hits, allowing each query to assemble a particle candidate.

\subsection{Input Representation}
Inputs are raw hits from six detector subsystems: vertex detector, inner tracker, outer tracker, electromagnetic calorimeter (ECAL), hadronic calorimeter (HCAL), and muon system. 
Each hit is represented by Cartesian coordinates $(x,y,z)$ and radial and angular features $(r, s, \theta,\phi)$ relative to the beamline, where $r$ is the radial distance to the beam axis and $s$ is the spherical radius.
For silicon and muon hits, the input features consist of these coordinates. Calorimeter hits additionally include the logarithm of the measured energy deposit. The model takes digitised detector hits without external track finding or calorimeter clustering; coordinate transformations, energy transforms, sorting, and positional encodings form its internal input representation.
Each hit type is embedded separately into a $d=256$ dimensional space, before being augmented with sinusoidal positional encodings in $(x,y,z,r,\theta,\phi)$, with cyclic encodings for $\phi$~\cite{2022arXiv220209206L} to capture periodicity. Positional encoding is performed at full precision and cast down to \texttt{bfloat16} to avoid floating point resolution issues.

\subsection{Architecture Configuration}
Full self-attention scales quadratically with the number of input hits, making it costly at collider multiplicities.
We exploit approximate locality in the azimuthal angle $\phi$.
Hits are ordered by $\phi$, and attention is restricted to a sliding window of size $w$~\cite{2020arXiv200405150B,MFTrk}. 
This reduces complexity to $\mathcal{O}(M \times w)$ for $M$ hits while preserving sensitivity to nearby structures such as curved tracks or overlapping showers. 
Boundary effects at $\pm \pi$ are handled by sequence wrapping. 
We further employ efficient fused kernels~\cite{2023arXiv230708691D} and SwiGLU activations~\cite{2020arXiv200205202S} for scalability.

The encoder consists of eight layers of self-attention and feed-forward blocks applied to the embedded hits. 
The decoder maintains a fixed set of 512 learned object queries, each corresponding to a candidate particle. 
It contains four decoder layers, each comprising a $k$-means-inspired cross-attention step~\cite{kmax_deeplab_2022} between encoded hits and object queries, followed by self-attention among the queries to exchange information between particle candidates with overlapping detector signatures.
The final decoder outputs feed four types of task heads:
\begin{enumerate}
  \item hit masks across all sub-detectors,
  \item mask quality, expressed as a predicted \IoU{},
  \item particle class (electron, muon, photon, charged hadron, neutral hadron),
  \item kinematic properties (track parameters, when available, and calibrated energy).
\end{enumerate}
In total, the model has $\sim$10.8M parameters including the encoder, decoder, and task heads. 
An additional encoder-decoder transformer reconstructs resonance bosons by assigning reconstructed particles to candidate bosons and predicting the boson identities.
This block has 6.4M parameters, giving 17.2M parameters in total.

\subsection{Particle Parameter Regression}
\label{sec:regression}
The kinematic-regression head is trained with the encoder and decoder frozen. It learns corrections to estimates derived from the assigned tracker and calorimeter hits.
For particles with at least three assigned silicon hits, a simple helix fit provides track parameters at the perigee point.
This approach was inspired by Ref.~\cite{coccaro2026Fit} and recently further demonstrated in a full detector simulation in Ref.~\cite{renusch2026fast}. We denote the conventional five track parameters derived from the fit as the \textit{track-seed}.
For each reconstructed particle, the assigned calorimeter-hit energies are summed to obtain the uncalibrated energy estimate $E_{\text{raw}}$. Together with the energy-weighted centroid in $(\eta,\phi)$, this forms the \textit{calo-seed} $(E_{\text{raw}},\eta,\phi)$.

The regression head takes the final decoder query embeddings and 60 scalar features derived from the tracks, calorimeter hits, and corresponding seeds.
The scalar features are transformed using either $\ln(1+x)$ or z-score normalisation to bring them to comparable scales.
An MLP maps these features to a 128-dimensional representation, which is concatenated with the query embedding and passed to a second MLP.
Three output heads handle charged particles with tracks, charged particles without tracks, and neutral particles.
The heads predict multiplicative corrections to the corresponding track- and calo-seeds, targeting the inverse transverse momentum $1/\pT$ and the particle energy $E$. The inverse transverse momentum is unsigned; the charge sign is predicted separately to obtain $\qOverPt$.
Both track and calorimeter features enter the shared learned representation, allowing either source to inform the corrections predicted by the three heads.

To limit the effect of unreliable seeds, training is restricted to track-seeds with at least 5 assigned silicon hits and calo-seeds with at least 6 assigned calorimeter hits.
Training uses the Barron loss~\cite{BarronCVPR2019} for robustness to outliers, applied to the targets
\begin{equation}
    t_{1/p_{T}} = \frac{1}{\mathrm{MAD}}\ln\frac{(1/p_T)_\mathrm{truth}}{(1/p_T)_\mathrm{seed}}, \qquad t_{E} = \ln\frac{E_\mathrm{truth}}{E_\mathrm{seed}}, \qquad t_{x} = \frac{x_\mathrm{truth}-x_\mathrm{seed}}{\sigma_x}, \label{eq:regr_targets}
\end{equation}
where $x$ are the track parameters excluding $p_T$, $\mathrm{MAD}$ is the median absolute deviation of the inverse transverse momentum target in bins of the seed transverse momentum, and $\sigma_x$ is the seed resolution (IQR/1.349) as a function of momentum and polar angle. The network predicts these targets, and the relations are inverted at inference. Both $\mathrm{MAD}$ and $\sigma_x$ are measured once on simulated held-out $t\bar{t}$ events and serve as a scale factor.

\subsection{\flowf Comparison}
We introduce \flowf{} for comparison with \hitpf{} using the same input types: conformal tracks (SiTracks), calorimeter hits, and muon-system hits.
Its architecture closely follows \ours, with separate input networks for SiTracks, ECAL hits, HCAL hits, and muon hits, followed by a transformer encoder and a $k$-means-inspired decoder with the four types of task head described above.
\flowf uses 384 learned object queries, compared with 512 for \ours.

\subsection{Resonance Reconstruction}
The resonance stage uses an encoder-decoder transformer with queries representing candidate bosons.
Its decoder maintains 8 learned resonance queries, updated through cross-attention with the particle queries from the reconstruction backbone. A softmax output assigns probabilities to the $W$, $Z$, $H$, and null classes.
Dot products between the particle and resonance embeddings are used to predict particle-to-boson membership probabilities.
This task tests whether learned constituent assignments can improve resonance-mass reconstruction.

Each resonance query predicts a mask over reconstructed particles. 
The resonance four-momentum is obtained by summing the four-momenta of the particles assigned to that mask,
\begin{equation}
    P_r^\mu = \sum_{i\in\mathcal{M}_r} p_i^\mu,
    \qquad m_r^2 = P_r^\mu P_{r,\mu},
    \label{eq:resonance_four_momentum}
\end{equation}
where $\mathcal{M}_r$ is the particle mask of resonance $r$.
This reconstructs the parent boson directly from its assigned particles, without an intermediate jet-clustering and jet-pairing step.

The four-momentum sum is stable under soft additions and collinear splittings when the constituent assignment is preserved; infrared and collinear safety of the learned assignment has not been established. We treat the resulting mass as a detector-level reconstruction estimator and have not tested its generator dependence. 
The central aim is to demonstrate that the entire reconstruction chain, from detector hits to composite objects, can be learned. 
The target masks could alternatively be defined by applying the VLC jet algorithm to truth particles and using the resulting constituent assignments as supervision.

\subsection{Training Setup}
\label{sec:training}
The model has three training stages: backbone hit assignment, kinematic regression, and resonance reconstruction. Earlier networks are kept frozen while each subsequent stage is trained. \ours is trained to reconstruct loose particles (\cref{sec:targets}), while \hitpf and \flowf are trained to reconstruct tight particles.

Reconstructed particles and resonances are matched to ground-truth particles and bosons respectively, using the Hungarian assignment algorithm~\cite{kuhn1955hungarian,10.1145/3442348} to ensure permutation invariance. 
The training loss for \ours is a sum of three terms, one for each training stage,
\begin{align}
  L ={}& L_\text{PR} + L_\text{reg} + L_\text{res} \nonumber \\
  ={}&\underbrace{0.5\, L_\text{CE} + L_\text{Dice} + 0.5\, L_\text{BCE} + 0.5\, L_\text{Focal} + L_\text{\IoU{}}}_{L_\text{PR}} \nonumber \\
   &+ \underbrace{L_{1/\pT} + 5\, L^\text{neutral}_{E} + 5\, L^\text{trackless}_{E} + L_\text{track state} + L_\text{charge}}_{L_\text{reg}} \nonumber \\
    & + \underbrace{L_\text{CE} + L_\text{Dice} + 0.5\, L_\text{BCE}}_{L_\text{res}},
\end{align}
where $L_\text{PR}$ is the pattern-recognition term, $L_\text{reg}$ is the regression term, and $L_\text{res}$ is the resonance term.

The pattern-recognition term combines a particle-classification cross-entropy $L_\text{CE}$ loss with hit-assignment Dice $L_\text{Dice}$~\cite{2017arXiv170703237S}, binary cross-entropy $L_\text{BCE}$ and Focal $L_\text{Focal}$~\cite{2017arXiv170802002L} losses, and the object-quality head $L_\text{IoU}$~\cite{PippaAttn} loss, which regresses the intersection-over-union (\IoU{}) of each predicted hit mask.

The regression term contains the Barron loss computed on the targets of Eq.~\eqref{eq:regr_targets}, where $L_{1/\pT}$ is for the inverse transverse momentum of tracks, $L^\text{neutral}_{E}$ and $L^\text{trackless}_{E}$ for the energy of neutral and of charged particles without a reconstructed track, respectively. The track-state loss $L_\text{track state}$ regresses the track parameters excluding $\qOverPt$, and the charge loss $L_\text{charge}$ is a binary cross-entropy on the charge sign for particles with a track.

The resonance term combines a boson-classification cross-entropy $L_\text{CE}$ loss with reconstructed particle-boson assignment Dice $L_\text{Dice}$ and binary cross-entropy losses $L_\text{BCE}$.

All terms except $L_\text{IoU}$ and $L_\text{reg}$ are applied after each decoder layer, each with its own Hungarian assignment~\cite{mask2former}. The Hungarian assignment cost function combines the predicted object validity and a Dice cost for the hit assignment.

The \flowf{} objective adds a Lovász loss~\cite{berman2018lovasz} with weight 1.0 to $L_\text{PR}$ for the ECAL and HCAL masks, and a supervised contrastive loss~\cite{khosla2020supervised} with weight 0.3 for the encoder's ECAL hit embeddings. It omits $L_\text{track state}$ and $L_\text{charge}$ because the input conformal tracks are Kalman-fitted and provide the track parameters directly.

Both \ours and \flowf were trained on 500\,000 events given in~\cref{tab:samples} during the backbone hit-assignment and regression stage\footnote{\hitpf was trained on $\Zqq{}$ events ($q=u,d,s$) at $\sqrts=91$~GeV, and its energy correction and particle identification on single-particle samples~\cite{Garcia:2026sqb}.}. The Lion~\cite{chen2023symbolic} optimiser is used with a weight decay of $10^{-5}$, and a one-cycle learning rate schedule~\cite{smith2018superconvergence} with initial and peak values of $ 10^{-5} $ and $5\times10^{-5}$, respectively. The final value matches the initial value, and the peak is reached within the first $0.1\%$ of the training steps. Training runs for five epochs in \texttt{bfloat16} mixed precision with 10 events per optimiser step.

The resonance stage is trained on a larger event corpus composed of 800\,000 diboson and $ZH$ events at $\sqrts=240, 365$~GeV (see ~\cref{tab:samples}).
The AdamW optimiser~\cite{loshchilov2019decoupled} was used with weight decay of $10^{-5}$ and a one-cycle learning-rate schedule with initial, final and peak values of $2\times10^{-5}$, $10^{-6}$ and $10^{-4}$, respectively. The peak is reached within the first 5\% of the training steps. Training runs for five epochs in \texttt{bfloat16} mixed precision with 16 events per optimiser step. All training and evaluation are carried out on NVIDIA L40S GPUs.

\subsection{Code Availability}
The full implementation is publicly available~\cite{hepattn}.

\section{Results}
\label{sec:results}

The main benchmark is unified reconstruction in diboson and $ZH$ events at $\sqrts=240~\mathrm{GeV}$, following the chain from detector hits to particles to resonance bosons.
We then compare with \hitpf~\cite{Garcia:2026sqb} in light-quark $\Zqq{}$ events at $\sqrts=91~\mathrm{GeV}$. Appendix~\ref{sec:results_ttbar} presents a complementary high-occupancy $\ttbar{}$ benchmark at $\sqrts=365~\mathrm{GeV}$.

\subsection{Unified Reconstruction at \texorpdfstring{$\sqrts=240$~GeV}{sqrt(s) = 240 GeV}}
\label{sec:results_240}

We evaluate particle reconstruction and its propagation to event, jet, and resonance observables.
The results follow the reconstruction chain, beginning with tracking (Section~\ref{sec:results_240_tracking}), calorimeter hit assignment (Section~\ref{sec:results_240_calo}), and object quality (Section~\ref{sec:results_240_quality}).
We then examine particle classification (Section~\ref{sec:results_240_particles_class}), particle reconstruction (Section~\ref{sec:results_240_particles}), and particle kinematic regression (Section~\ref{sec:results_240_particles_regr}).
Finally, we assess event-level observables (Section~\ref{sec:results_240_event}), jet-level observables (Section~\ref{sec:results_240_jet}), and boson reconstruction and particle-to-resonance assignment (Section~\ref{sec:results_bosons}).

\subsubsection{Tracking}
\label{sec:results_240_tracking}

Tracking performance is assessed through charged-particle reconstruction efficiency, track purity, and track-parameter resolution using silicon-detector measurements.
The baseline is conformal tracking before any selection by \ppfa{}.

\begin{figure}[!htbp]
    \centering
    \includegraphics[width=\textwidth]{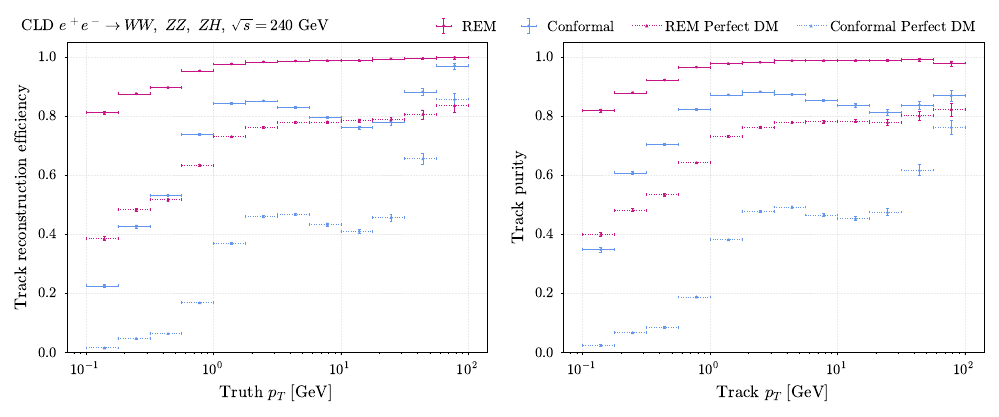}
    \caption{
    Charged-particle tracking efficiency (left), purity (right) in simulated $e^+e^-\to WW,ZZ,ZH$ events at $\sqrts=240$~GeV, for \ours tracks and the conformal tracking baseline.
    The efficiency and purity are shown for the double-majority matching (solid) and for perfect matches (dotted), in which the silicon hits of the track are exactly those of the particle.}
    \label{fig:240_tracking_summary}
\end{figure}

\begin{figure}[!htbp]
    \centering
    \includegraphics[width=\textwidth]{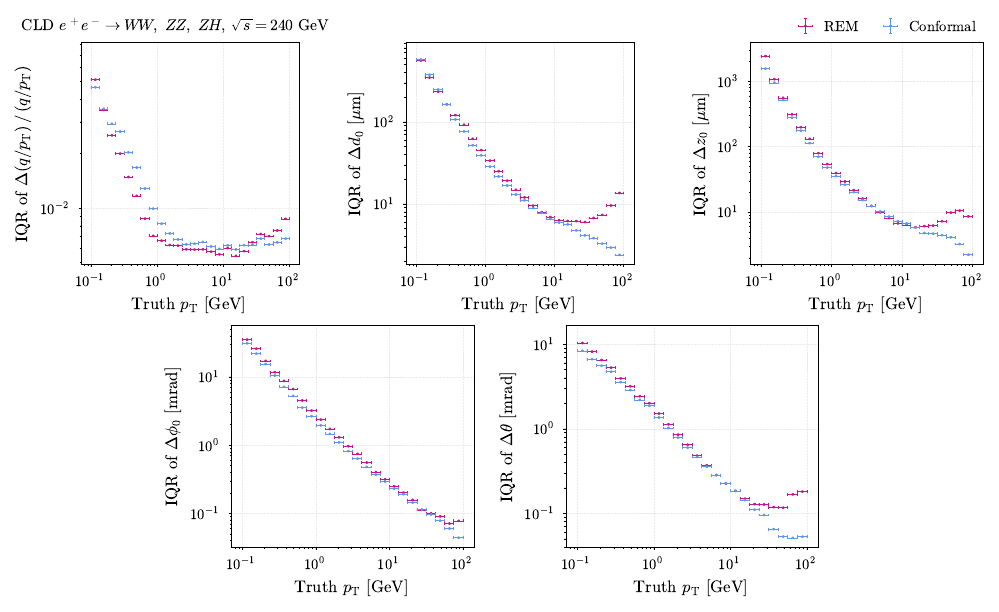}
    \caption{
    Charged-particle track-parameter resolution in simulated $e^+e^-\to WW,ZZ,ZH$ events at $\sqrts=240$~GeV.
    The panels show the interquartile range of the residuals of the track parameters at the IP, $(\qOverPt, d_0, z_0,\phi,\theta)$, in bins of truth $\pT$ for \ours and the baseline conformal-tracking reconstruction, on the reconstructable tracks that both reconstruct. 
    The $\qOverPt$ residual is relative to the truth value.
    }
    \label{fig:240_tracking_params}
\end{figure}

Figure~\ref{fig:240_tracking_summary} shows the tracking efficiency and purity, and Figure~\ref{fig:240_tracking_params} shows all five track-parameter resolutions for the 240~GeV benchmark. Table~\ref{tab:240_tracking_residuals} summarises the integrated residuals for the tracks reconstructed by both methods. 
Overall, \ours{} achieves parameter resolutions comparable to conformal tracking, with a narrower integrated $\qOverPt$ response and slightly broader impact-parameter and angular responses. 
The $\qOverPt$ resolution is better over intermediate transverse momenta, but worse at the highest $\pT$. At high transverse momentum, the impact-parameter and polar-angle resolutions show larger differences. As multiple scattering decreases, these observables become more sensitive to the precision of the track fit and parameter regression.
Additional phase-space and species dependence is detailed for a $\ttbar{}$ sample in Appendix~\ref{sec:ttbar_tracking}.
\ours recovers a larger fraction of each particle's silicon hits.
The average hit-on-track \IoU{} over the tracks is 0.95 for \ours and 0.87 for conformal tracking, which is manifest in the higher rate of perfect-match efficiency and purity in Figure~\ref{fig:240_tracking_summary}.
About 71\% and 41\% of \ours and conformal tracks, respectively, exactly match the silicon hits of the corresponding simulated particle. 
Counted over the reconstructable particles, 67.5\% have a track of \ours with exactly their silicon hits, compared with 29.6\% for conformal tracking (Figure~\ref{fig:240_tracking_summary}, left).

\begin{table}[!htbp]
    \centering
    \small
    \caption{
    Median and interquartile range (IQR) of track-parameter residuals for the 108\,000 tracks reconstructed by both \ours{} and conformal tracking in the 240~GeV benchmark. 
    Residuals are reconstructed minus true values. 
    The $\qOverPt$ residual is divided by its true value and expressed as a percentage.
    }
    \label{tab:240_tracking_residuals}
    \renewcommand{\null}{\hbox{}}
    \begin{tabular}{lrrrr}
        \toprule
        & \multicolumn{2}{c}{\ours{}} & \multicolumn{2}{c}{Conformal tracking} \\
        Parameter & Median & IQR & Median & IQR \\
        \midrule
        $\qOverPt$ (relative, \%) & 0.0050 & 0.71 & 0.011 & 0.86 \\
        $d_0$ [$\mu$m] & $-0.31$ & 23 & 0.038 & 21 \\
        $z_0$ [$\mu$m] & $-0.038$ & 28 & $-0.036$ & 26 \\
        $\phi$ [mrad] & 0.011 & 1.5 & $-0.0024$ & 1.3 \\
        $\theta$ [mrad] & $-0.0019$ & 0.95 & $-0.0018$ & 0.88 \\
        \bottomrule
    \end{tabular}
\end{table}

\FloatBarrier

\subsubsection{Calorimeter Hit Assignment}
\label{sec:results_240_calo}

\begin{figure}[!htbp]
    \centering
    \includegraphics[width=\textwidth]{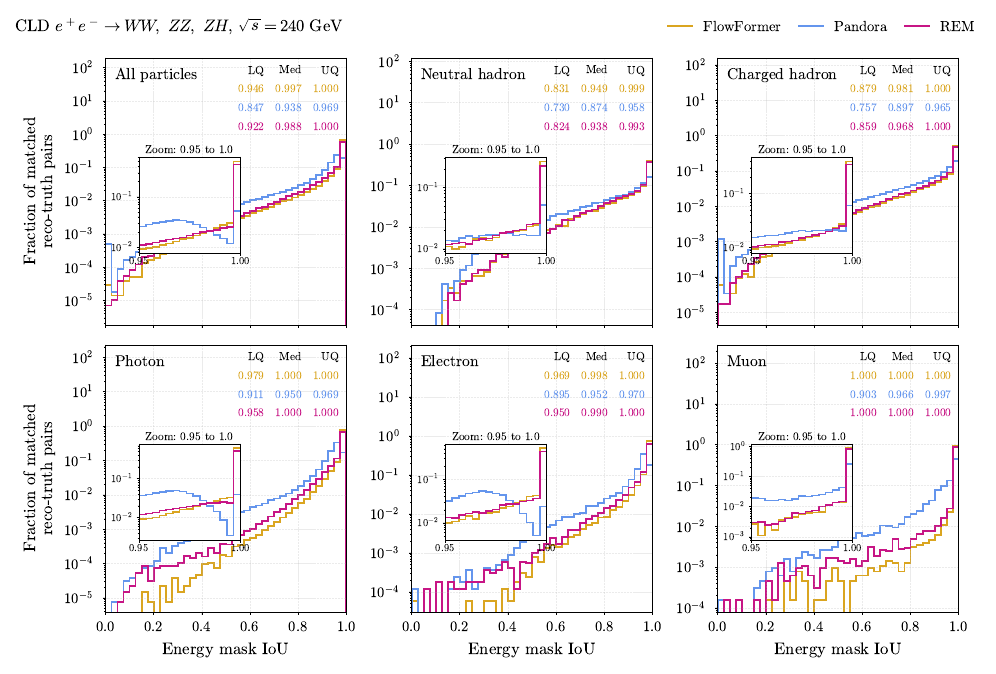}
    \caption{
    Energy-weighted calorimeter mask \IoU{}, $\mathrm{IoU}_E$, between reconstructed particles and their matched truth particles in simulated $WW,ZZ,ZH$ events at $\sqrts=240$~GeV, comparing \ours, \flowf and \ppfa on the truth particles that all three reconstructions match, without requiring the correct particle identification.
    From left to right, all particles, neutral hadrons and charged hadrons (top) and photons, electrons and muons (bottom), by truth class; the insets show the region from 0.95 to 1 and the tables the lower quartile, median and upper quartile.
    }
    \label{fig:240_energy_mask_iou}
\end{figure}

The calorimeter hits assigned to a particle determine how much of its deposited energy enters the energy estimate (see \cref{sec:regression}). 
The quality of the calorimeter pattern recognition is assessed in isolation using energy-weighted mask \IoU{} without requiring correct particle identification.
This \IoU{} variable is defined as $\mathrm{IoU}_E = I/(I+M+C)$, where $I$ is the calorimeter energy of the claimed hits that belong to the matched truth particle, $M$ the energy of the truth particle's hits that are not claimed and $C$ the energy of the claimed hits that do not belong to it.
Figure~\ref{fig:240_energy_mask_iou} compares $\mathrm{IoU}_E$ on the same set of truth particles for all three reconstructions. 
Both learned methods capture the deposited energy more completely and with less contamination than \ppfa{}, as summarised by the mean values in Table~\ref{tab:240_energy_mask_iou}.
The better energy capture is an origin of the narrower energy responses of the learned methods relative to \ppfa discussed below. 
It does not account for the differences between \ours and \flowf, whose calorimeter masks are similar, with \flowf slightly ahead.

\begin{table}[!htbp]
    \centering
    \small
    \caption{
    Mean energy-weighted calorimeter mask $\mathrm{IoU}_E$ in the 240~GeV benchmark, using the same particle selection as Figure~\ref{fig:240_energy_mask_iou}. 
    Bold marks the highest displayed values, including ties.
    }
    \label{tab:240_energy_mask_iou}
    \begin{tabular}{lrrr}
        \toprule
        Truth class & \ours{} & \flowf{} & \ppfa{} \\
        \midrule
        All particles & 0.94 & \textbf{0.95} & 0.88 \\
        Charged hadrons & 0.90 & \textbf{0.91} & 0.84 \\
        Neutral hadrons & \textbf{0.89} & \textbf{0.89} & 0.82 \\
        Photons & 0.97 & \textbf{0.98} & 0.92 \\
        \bottomrule
    \end{tabular}
\end{table}

\FloatBarrier

\subsubsection{Object Quality}
\label{sec:results_240_quality}

\begin{figure}[!htbp]
    \centering
    \includegraphics[width=\textwidth]{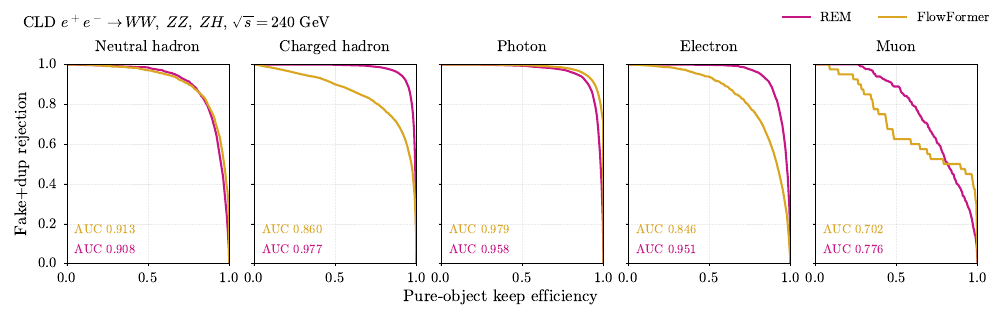}
    \caption{
    Fake and duplicate rejection against pure-object keep efficiency for a threshold on the predicted \IoU{} of \ours and \flowf in simulated $e^+e^-\to WW,ZZ,ZH$ events at $\sqrts=240$~GeV, per predicted class (left to right: neutral hadrons, charged hadrons, photons, electrons, muons), with the area under each curve.
    Object classes follow Section~\ref{sec:metrics}; objects matched in the second step to particles outside the tight evaluation selection belong to neither population (Section~\ref{sec:targets}).
    }
    \label{fig:240_quality}
\end{figure}

Both learned methods predict a hit-mask \IoU{} for each object. \ours uses a single output; \flowf combines one output per detector, weighted by the number of hits.
Pure objects carry most of their energy at high predicted \IoU{}, while fake and duplicate objects carry most at low values (Appendix~\ref{sec:results_240_app}).
The predicted \IoU{} therefore provides a quality score for selecting particles for downstream tasks.
Figure~\ref{fig:240_quality} shows the fraction of fake and duplicate objects rejected against the fraction of pure objects retained (pure-object keep efficiency) as the \IoU{} threshold is varied, separately for each predicted class.
Table~\ref{tab:240_quality_auc} summarises the areas under the curves. \ours gives better discrimination for charged hadrons, electrons and muons, while \flowf performs better for photons; the two methods are comparable for neutral hadrons.
\ppfa provides no analogous per-object score.

\begin{table}[!htbp]
    \centering
    \caption{Area under the object-quality curves in Figure~\ref{fig:240_quality}. Higher values indicate better discrimination between pure objects and fakes or duplicates. The best displayed value in each row is shown in bold, including ties.}
    \label{tab:240_quality_auc}
    \begin{tabular}{lcc}
        \toprule
        Particle class & \ours & \flowf \\
        \midrule
        Charged hadrons & \textbf{0.98} & 0.86 \\
        Neutral hadrons & \textbf{0.91} & \textbf{0.91} \\
        Photons & 0.96 & \textbf{0.98} \\
        Electrons & \textbf{0.95} & 0.85 \\
        Muons & \textbf{0.78} & 0.70 \\
        \bottomrule
    \end{tabular}
\end{table}

For the event-level sums and jet inputs, \ours uses these predictions to purify the reconstructed particles by removing them when their predicted \IoU{} is below 0.9 and their calorimeter energy fraction is below 5\%.
This selection uses no truth information. 

\FloatBarrier

\subsubsection{Particle Classification}
\label{sec:results_240_particles_class}

\begin{figure}[!htbp]
    \centering
    \includegraphics[width=\textwidth]{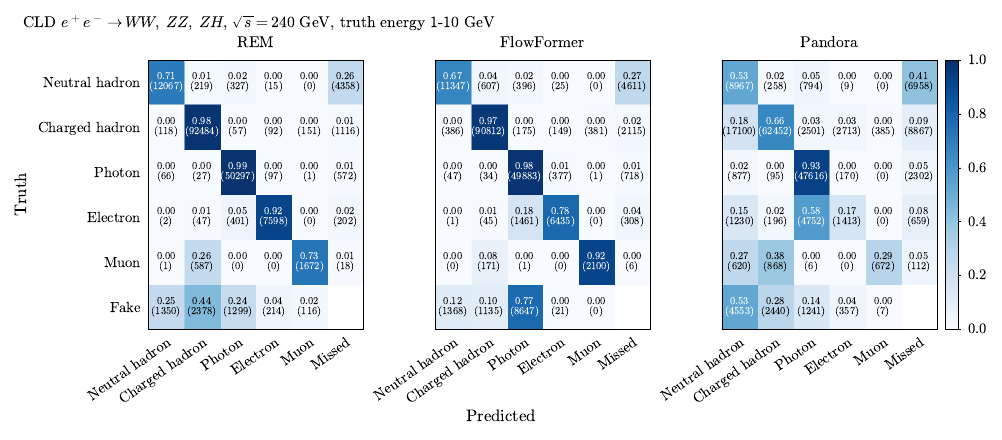} \\
    \includegraphics[width=\textwidth]{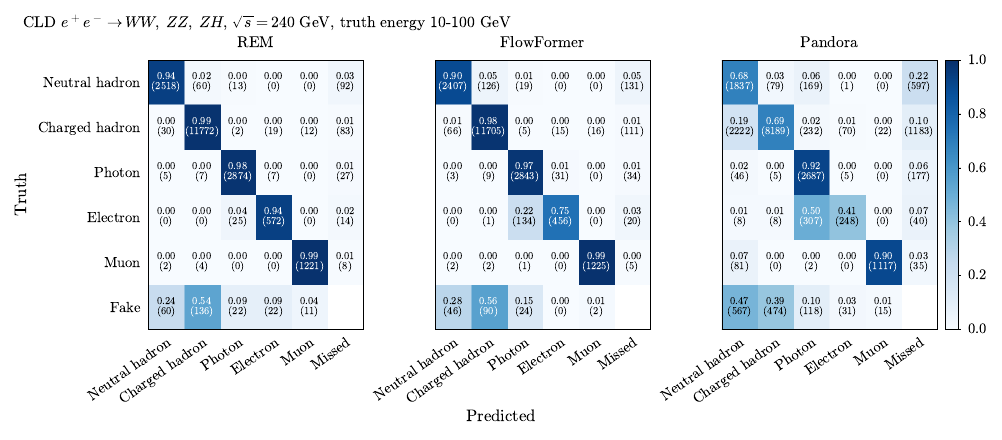}
    \caption{
    Particle-identification confusion matrices in simulated $e^+e^-\to WW,ZZ,ZH$ events at $\sqrts=240$~GeV for \ours, \flowf and \ppfa (left to right), for truth particles of 1--10~GeV (top) and 10--100~GeV (bottom).
    Rows show the true class of the truth particles.
    The fake row includes duplicates, i.e. all reconstructed objects that are not pure. 
    Columns show the predicted class of reconstructed objects. ``Missed'' counts the truth particles that are inefficient under the loose double-majority requirement. 
    Rows are normalised to unity, and the numbers of particles are given in brackets.
    }
    \label{fig:240_pid_confusion}
\end{figure}

Figure~\ref{fig:240_pid_confusion} shows the particle-identification response for truth energies of 1--10~GeV (top) and 10--100~GeV (bottom).
Both learned methods identify 97--99\% of charged hadrons and photons correctly in both ranges, compared with 66--69\% and 92--93\% for \ppfa{}, respectively; \ppfa{} misidentifies 18--19\% of charged hadrons as neutral hadrons. 
Since \flowf{} and \ppfa{} use the same reconstructed tracks, the charged-hadron improvement comes from the subsequent calorimeter hit assignment, track--calorimeter association, and particle formation.
The largest difference between the learned methods occurs for electrons: \ours{} identifies 92--94\%, compared with 75--78\% for \flowf{} and 17--41\% for \ppfa{}. The gain may arise from improved track reconstruction, track--shower association, or both.\footnote{The comparison does not isolate the benefit of joint learning from improved track quality: better standalone tracking could recover some of the gain.}
For low-energy muons, \flowf{} performs best at 92\%, compared with 73\% for \ours{} and 29\% for \ppfa{}, while above 10~GeV all three exceed 90\%.
Neutral-hadron identification reaches 71--94\% for \ours{} and 67--90\% for \flowf{}, compared with 53--68\% for \ppfa{}, consistent with the gains in calorimeter reconstruction.

\subsubsection{Particle Reconstruction}
\label{sec:results_240_particles}

\begin{figure}[!htbp]
    \centering
    \includegraphics[width=\textwidth]{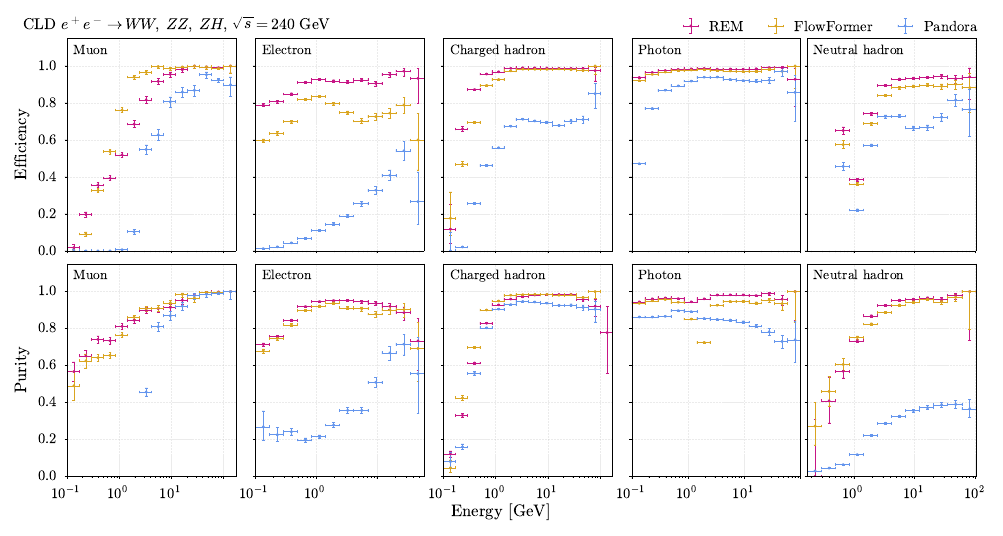}
    \caption{
    Particle reconstruction efficiency and purity after particle identification in simulated $e^+e^-\to WW,ZZ,ZH$ events at $\sqrts=240$~GeV, comparing \ours, \flowf and \ppfa.
    The top row shows the efficiency against the truth particle energy and the bottom row the purity against the reconstructed particle energy, for muons, electrons, charged hadrons, photons and neutral hadrons (left to right). Definitions follow Section~\ref{sec:metrics}.}
    \label{fig:240_particle_eff_purity}
\end{figure}

Figure~\ref{fig:240_particle_eff_purity} compares particle reconstruction efficiency and purity after particle identification, and so looks at hit pattern recognition and particle classification together. 
Both \ours and \flowf improve substantially on \ppfa across much of the energy range, with particularly clear gains for electrons and neutral-hadron purity. 
The two learned methods have similar charged-hadron and photon efficiencies. 
\ours generally achieves higher electron efficiency and photon purity; \flowf has higher muon efficiency over much of the intermediate-energy range. Table~\ref{tab:240_particle_eff_pur_by_class} gives the integrated efficiency and purity for all target particles and for each particle class, together with the number of fakes per event.

The increased charged particle performance in \ours is mainly due to the better track reconstruction.
It also is in part due to the fact that \ours does not require a reconstructed track to decide if an object is charged.
Evidence of charge can remain in tracker hits that are too few to form a track or fail the track fit, or in an ionising trajectory in the high-granularity calorimeter before the hadronic shower starts.

\begin{table}[htb]
    \centering
    \small
    \setlength{\tabcolsep}{3pt}
    \newcommand{\metricpair}[2]{%
        \makebox[3em][c]{#1}\makebox[3em][c]{#2}}
    \newcommand{\metrictriple}[3]{%
        \makebox[3.5em][c]{#1}%
        \makebox[3.5em][c]{#2}%
        \makebox[4em][c]{#3}}

    \caption{Integrated reconstruction efficiency and purity after particle identification for all target particles and for each particle class in 5{,}000 simulated $e^+e^-\to WW,ZZ,ZH$ events at $\sqrts=240$~GeV. Matching and metric definitions follow Section~\ref{sec:metrics}. The best displayed value in each column is shown in bold, including ties after rounding.}
    \label{tab:240_particle_eff_pur_by_class}

    \begin{tabular}{lc|c|c|c|c|c}
        \toprule
        \multirow{2}{*}{Algorithm}
        & All & CH & NH & $\gamma$ & $e$ & $\mu$ \\
        & \metrictriple{eff}{pur}{fakes/ev}
        & \metricpair{eff}{pur}
        & \metricpair{eff}{pur}
        & \metricpair{eff}{pur}
        & \metricpair{eff}{pur}
        & \metricpair{eff}{pur} \\
        \midrule
        \ours{}
        & \metrictriple{\textbf{0.92}}{\textbf{0.92}}{\textbf{4.2}}
        & \metricpair{\textbf{0.97}}{0.91}
        & \metricpair{\textbf{0.73}}{\textbf{0.88}}
        & \metricpair{\textbf{0.93}}{\textbf{0.95}}
        & \metricpair{\textbf{0.84}}{\textbf{0.85}}
        & \metricpair{0.55}{\textbf{0.84}} \\
        \flowf{}
        & \metrictriple{0.89}{0.91}{\textbf{4.2}}
        & \metricpair{0.94}{\textbf{0.95}}
        & \metricpair{0.69}{0.86}
        & \metricpair{0.92}{0.89}
        & \metricpair{0.69}{0.83}
        & \metricpair{\textbf{0.63}}{0.81} \\
        \ppfa{}
        & \metrictriple{0.61}{0.73}{4.5}
        & \metricpair{0.61}{0.90}
        & \metricpair{0.54}{0.19}
        & \metricpair{0.71}{0.87}
        & \metricpair{0.09}{0.29}
        & \metricpair{0.23}{0.81} \\
        \bottomrule
    \end{tabular}
\end{table}

\begin{figure}[!htbp]
    \centering
    \includegraphics[width=0.79\textwidth]{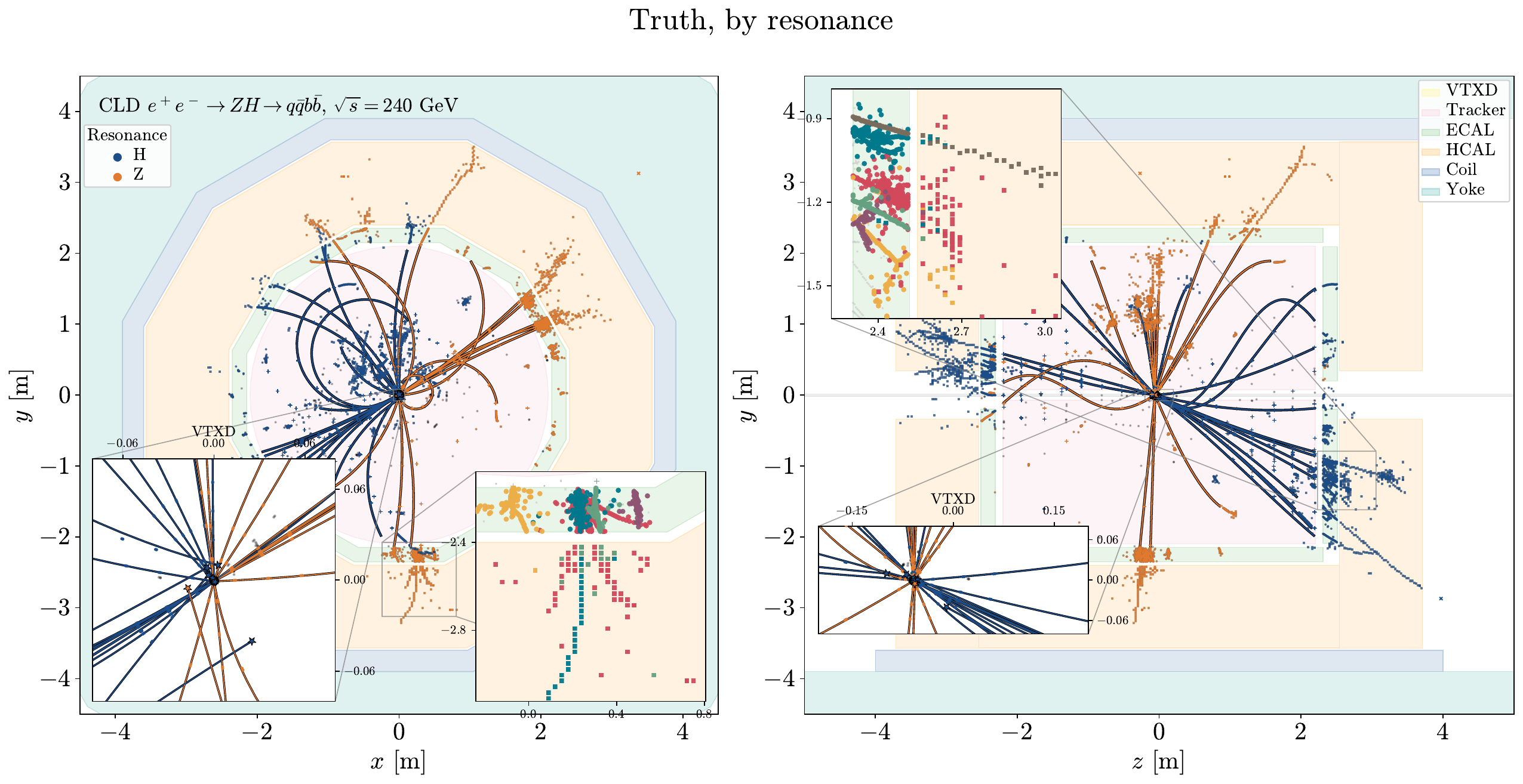}\\[0.1cm]
    \includegraphics[width=0.79\textwidth]{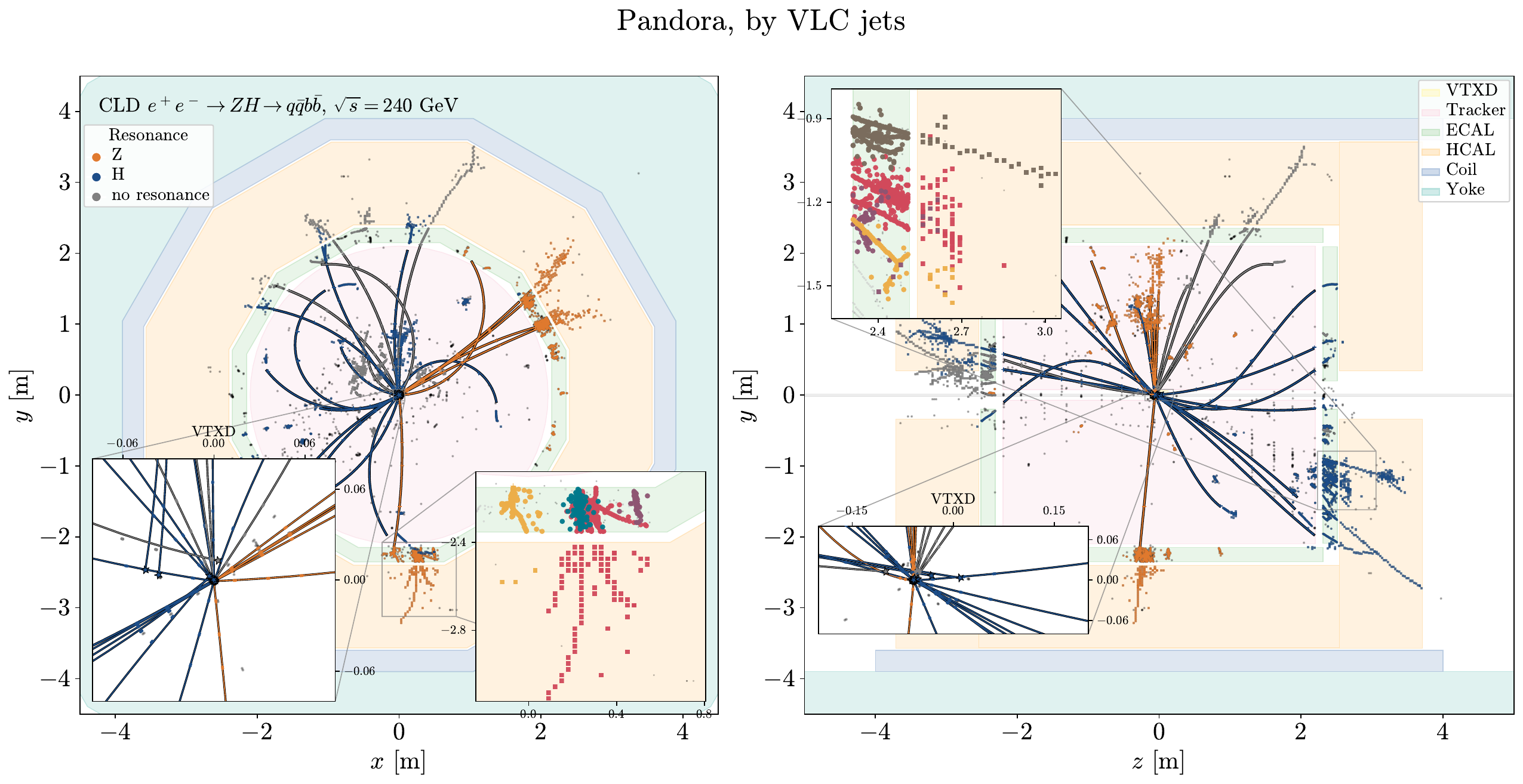}\\[0.1cm]
    \includegraphics[width=0.79\textwidth]{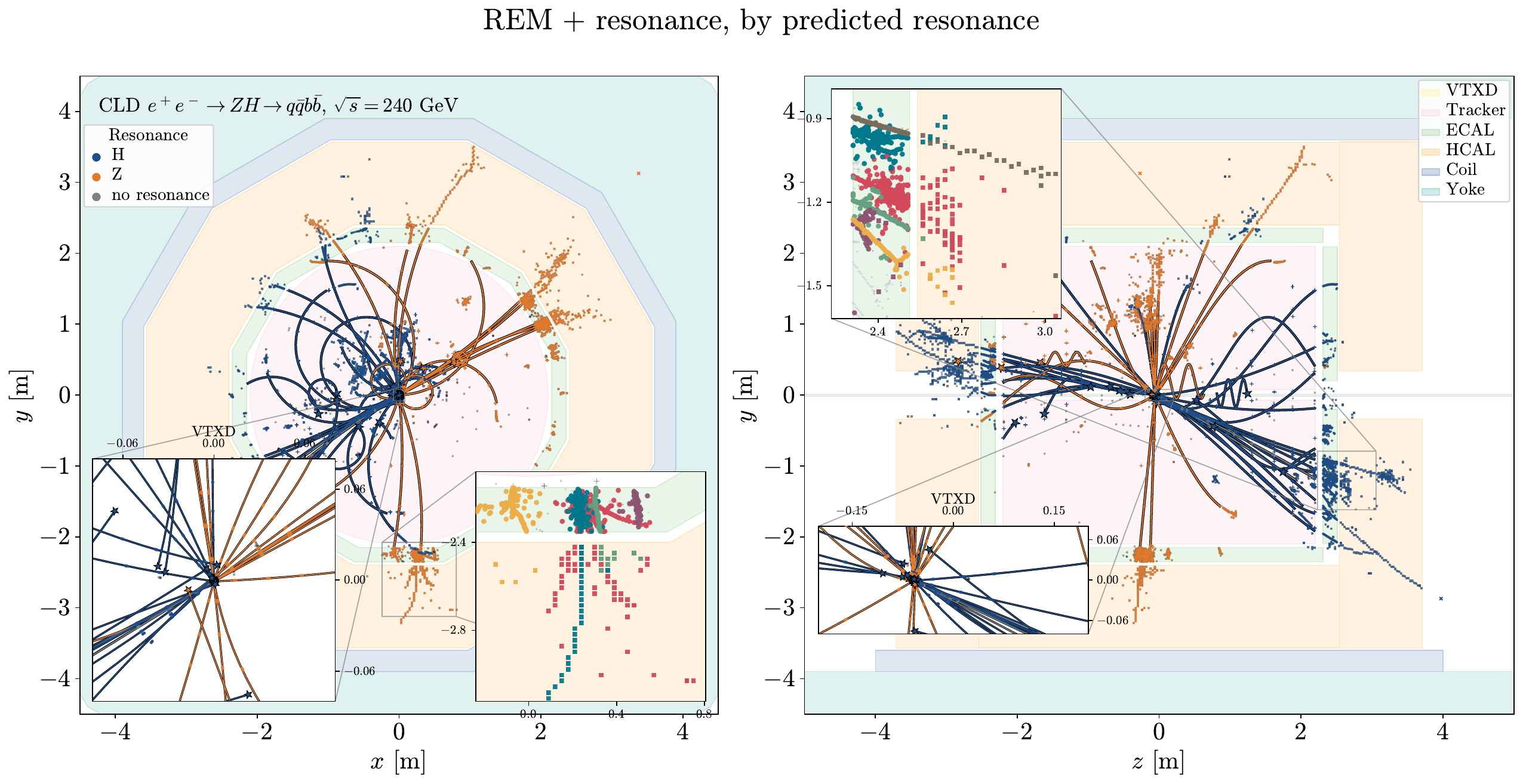}
    \caption{
    Event display of a simulated $e^+e^-\to ZH\to q\bar{q}b\bar{b}$ event at $\sqrts=240$~GeV in CLD, for truth particles (top), \ppfa (middle) and \ours (bottom), in the transverse $x$--$y$ (left) and $z$--$y$ (right) views.
    Tracks and calorimeter hits are coloured by their assigned boson. 
    For truth particles the true parent is used, for \ppfa the boson is assigned by clustering its particles into four exclusive VLC jets and pairing them into the $Z$ and $H$ candidates that minimise $(m_{Z}-91.2~\mathrm{GeV})^2+(m_{H}-125~\mathrm{GeV})^2$, and for \ours it is the boson predicted by the resonance task.
    The lower-left insets zoom on the vertex detector, while the other insets zoom on overlapping calorimeter showers, in which each hit is coloured by the particle or reconstructed object it belongs to.
    }
    \label{fig:240_event_display}
\end{figure}

Figure~\ref{fig:240_event_display} shows an example $ZH\to q\bar{q}b\bar{b}$ event, reconstructed by \ppfa and \ours, with the truth particles for reference.

\FloatBarrier

\subsubsection{Particle Kinematic Regression}
\label{sec:results_240_particles_regr}

With improved track reconstruction and/or improved calorimeter hit clustering, the learned models provide more accurate particle-energy estimates.
\ours{} gives a largely unbiased response for electrons and charged hadrons down to 300~MeV. \flowf{} shows a bias for electrons below 1~GeV, reaching about 10\% at 100~MeV, but this remains substantially smaller than for \ppfa{}. 
All three methods show a low-energy muon bias below about 3~GeV, with the learned models performing best. 
\ours{} underestimates photon energies above 30~GeV, while both learned methods retain an energy-dependent bias for neutral hadrons. 
Further details are given in Appendix~\ref{app:240_response}.

Figure~\ref{fig:240_particle_resolution} compares energy resolutions on the same truth particles, requiring every method to reconstruct and correctly identify them.
The learned models generally give better energy resolution, by more than a factor of two in some regions.
The charged-hadron structure around 1--2~GeV depends on particle species and the pion-mass assumption used to calculate their energies (Appendix~\ref{app:zqq_ch_momentum}).
The learned models also give substantially better neutral-hadron angular resolution above 5~GeV, where \ppfa{} reaches a plateau, although they are slightly worse below 2~GeV (Appendix~\ref{app:240_angular}).

\begin{figure}[!htbp]
    \centering
    \includegraphics[width=\textwidth]{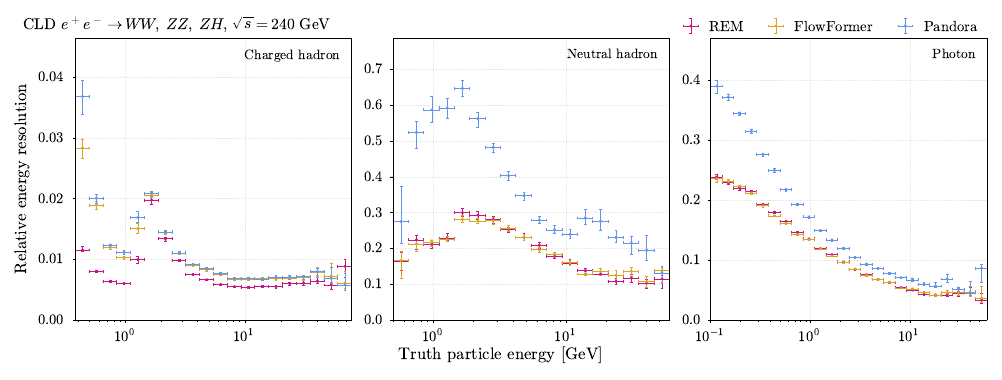}
    \caption{
    Energy resolutions for charged hadrons (left), neutral hadrons (middle) and photons (right) in simulated $WW,ZZ,ZH$ events at $\sqrts=240$~GeV, comparing \ours{}, \flowf{} and \ppfa{}.
    Only truth particles reconstructed under the DM requirement and correctly identified by all three methods are considered.
    Resolution definitions are given in Section~\ref{sec:metrics}. 
    Error bars span the 16th--84th percentiles from event bootstrap resampling; bins with fewer than 20 particles are omitted.
    }
    \label{fig:240_particle_resolution}
\end{figure}

\FloatBarrier

\subsubsection{Event-Level Observables}
\label{sec:results_240_event}

\begin{figure}[!htbp]
    \centering
    \includegraphics[width=0.49\textwidth]{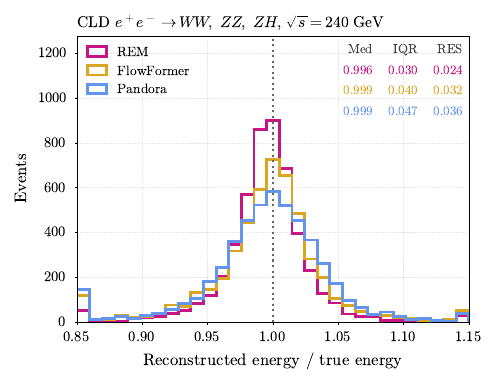}
    \includegraphics[width=0.49\textwidth]{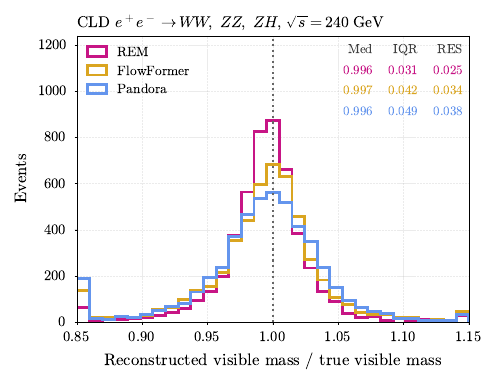}
    \includegraphics[width=0.49\textwidth]{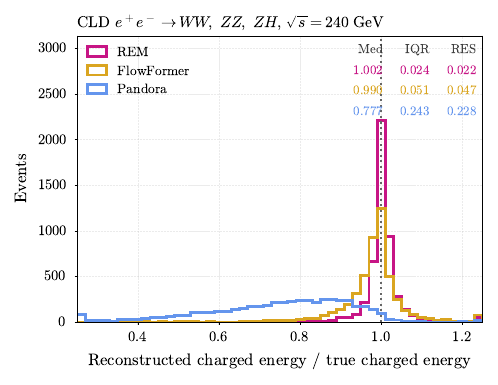}
    \includegraphics[width=0.49\textwidth]{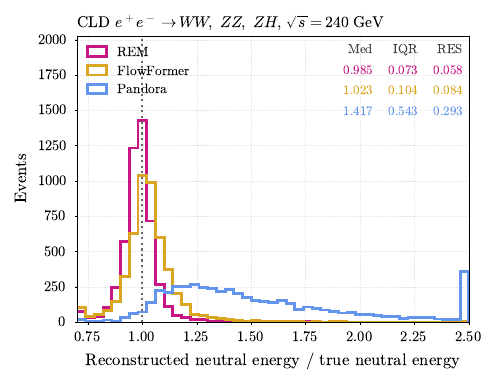}
    \caption{
    Event-level reconstructed-to-truth ratios in simulated $WW,ZZ,ZH$ events at $\sqrts=240$~GeV, comparing \ours{}, \flowf{} and \ppfa{}.
    Top: total visible energy (left) and visible invariant mass (right). 
    Bottom: charged (left) and neutral (right) energy. 
    Object selections follow Section~\ref{sec:targets}.}
    \label{fig:240_event_level}
\end{figure}

The visible event energy and invariant mass are calculated from the reconstructed particles as defined in Section~\ref{sec:metrics}. 
Figure~\ref{fig:240_event_level} compares the reconstructed-to-truth ratios for the visible energy, visible mass, and charged and neutral energy components.
All three methods give nearly unbiased total visible-energy and mass responses. \ours{} gives the narrowest distributions, reducing their IQRs by approximately 36\% and 37\%, respectively, relative to \ppfa{}. Table~\ref{tab:240_event_summary} summarises the total, charged and neutral responses, together with the energy carried by pure, fake and duplicate objects.

\begin{table}[!htbp]
    \centering
    \small
    \caption{
    Event-level responses and energy composition in the 240~GeV benchmark. 
    Responses are reconstructed-to-truth ratios for the corresponding energy or mass. 
    Object energy fractions are relative to the total true visible event energy. 
    Metric and object-class definitions follow Section~\ref{sec:metrics}. 
    Bold marks the narrowest responses.
    }
    \label{tab:240_event_summary}
    \begin{tabular}{lrrr}
        \toprule
        Observable & \ours{} & \flowf{} & \ppfa{} \\
        \midrule
        Visible-energy response IQR & \textbf{0.030} & 0.040 & 0.047 \\
        Visible-mass response IQR & \textbf{0.031} & 0.042 & 0.049 \\
        \midrule
        Charged-energy response median & 1.00 & 0.99 & 0.78 \\
        Charged-energy response IQR & \textbf{0.024} & 0.051 & 0.24 \\
        Neutral-energy response median & 0.99 & 1.02 & 1.42 \\
        Neutral-energy response IQR & \textbf{0.073} & 0.10 & 0.54 \\
        \midrule
        Pure-object energy [\%] & 97.7 & 96.7 & 93.1 \\
        Fake and duplicate energy [\%] & 1.8 & 2.7 & 6.1 \\
        Duplicate energy [\%] & 0.5 & 0.3 & 0.2 \\
        \bottomrule
    \end{tabular}
\end{table}

The charged and neutral components reveal differences that are not seen in the total energy. 
\ppfa{} underestimates charged energy and overestimates neutral energy, which cancels leaving the total energy unbiased. 
Both learned methods recover a more accurate partition, with \ours{} giving the narrowest responses. 
Since \flowf{} uses the same reconstructed tracks as \ppfa{}, its improvement points to better association of tracks with calorimeter activity, while \ours{} also learns the tracking stage.

\subsubsection{Jet-Level Observables}
\label{sec:results_240_jet}

Jets are clustered with the exclusive VLC algorithm~\cite{Boronat:2016tgd} ($R=1.1$, $\beta=\gamma=1$), requesting as many jets as there are quarks at matrix-element level, two or four. 
In semileptonic diboson events, the leptonic boson decay products and their matched reconstructed objects are excluded before jet clustering.
\textit{Reference jets} are constructed by clustering the tight targets defined in Section~\ref{sec:targets}. 
Reconstructed particles matched in the second step to real particles outside this selection are excluded from the jet inputs.
Jets created from reconstructed particles are paired one-to-one with the reference jets by minimising the summed $\Delta R=\sqrt{\Delta\eta^2+\Delta\phi^2}$ over all pairs and each pair must have $\Delta R<0.3$. 
Only events in which every reference jet is matched are considered. 
The jet energy scale (JES) and jet energy resolution (JER) are estimated from the mean $\mu$ and the relative width $\sigma/\mu$ of an iterative Gaussian fit to $E_\mathrm{reco}/E_\mathrm{ref}$ over $[\mu-1.25\sigma,\mu+1.25\sigma]$, which describes the core of a non-Gaussian response. 
Uncertainties are obtained by bootstrap resampling of the events. 
To characterise the response tails, $\sigma_{68}$ is the half-width of the narrowest interval containing 68\% of the jets, divided by the median response; $\mathrm{RMS}_{90}$ is the RMS within the narrowest interval containing 90\%, divided by its mean.

\begin{figure}[!htbp]
    \centering
    \includegraphics[width=\textwidth]{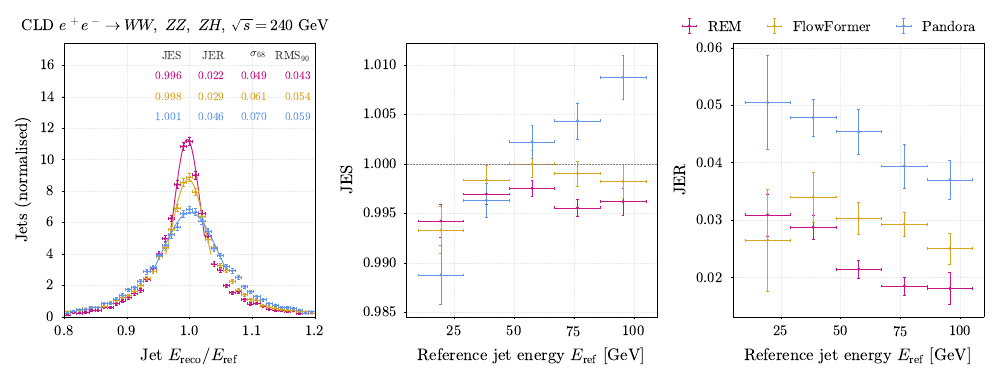}
    \caption{
    Jet performance in simulated $WW,ZZ,ZH$ events at $\sqrts=240$~GeV, comparing \ours{}, \flowf{} and \ppfa{}. 
    Left: normalised jet-energy response with Gaussian core fits and summary metrics. 
    Middle and right: JES and JER against reference jet energy. 
    Definitions and selections follow Section~\ref{sec:results_240_jet}.
    }
    \label{fig:boson_sample_jet_performance}
\end{figure}

Figure~\ref{fig:boson_sample_jet_performance} shows the jet response and its dependence on reference jet energy. 
All three methods have an average JES close to unity. 
\ours{} gives the narrowest response, with the same ordering for the fitted core and the measures that include more of the tails (Table~\ref{tab:240_jet_summary}).
Across reference jet energies, the learned models provide a more stable response and up to a 50\% improvement in resolution.

\begin{table}[!htbp]
    \centering
    \small
    \caption{
    Jet response in the 240~GeV benchmark. 
    Definitions and selections follow Section~\ref{sec:results_240_jet}. 
    JER uncertainties are obtained by event bootstrap resampling. 
    Bold marks the narrowest responses.
    }
    \label{tab:240_jet_summary}
    \begin{tabular}{lrrr}
        \toprule
        Metric & \ours{} & \flowf{} & \ppfa{} \\
        \midrule
        JES & 0.996 & 0.998 & 1.001 \\
        JER [\%] & $\boldsymbol{2.2\pm0.1}$ & $2.9\pm0.1$ & $4.6\pm0.1$ \\
        $\sigma_{68}$ [\%] & \textbf{4.9} & 6.1 & 7.0 \\
        $\mathrm{RMS}_{90}$ [\%] & \textbf{4.3} & 5.4 & 5.9 \\
        \bottomrule
    \end{tabular}
\end{table}

\FloatBarrier

\subsubsection{Boson Reconstruction and Particle-to-Resonance Assignment}
\label{sec:results_bosons}

The resonance task extends the MaskFormer approach from hit-to-particle assignment to particle-to-boson assignment. 
Hadronic boson masses and label predictions are evaluated using 5{,}000 events per channel for the 240~GeV processes listed in Table~\ref{tab:samples}.
A central performance target is the separation of the reconstructed mass distributions of hadronically decaying $W$ and $Z$ bosons, whose masses differ by only about \SI{11}{GeV}.
The separation $S$ is defined in Section~\ref{sec:metrics}. 
The peak position $\mu$ and width $\sigma$ of each mass distribution are obtained from an iterative Gaussian fit over $[\mu-\sigma,\mu+2\sigma]$, following the CLD design report~\cite{Bacchetta:2019fmz}. 
The uncertainty on $S$ is the standard deviation of $S$ over 300 bootstrap resamples of the events, each refitted.

Two methods of calculating the hadronic boson mass are compared. 
The jet-based method clusters particles with the VLC algorithm and calculates the mass from the combined jet four-momentum. 
We apply this procedure to truth particles, \ppfa outputs, and \ours outputs. 
The resonance-mask method instead sums the four-momenta of particles assigned to each resonance mask. 
At truth level, the mask follows the parent-boson assignment; for \ours, it is predicted by the resonance task\footnote{An alternative target mask could be defined by applying the VLC algorithm to truth particles.}.

\begin{figure}[!htbp]
    \centering
    \includegraphics[width=0.49\textwidth]{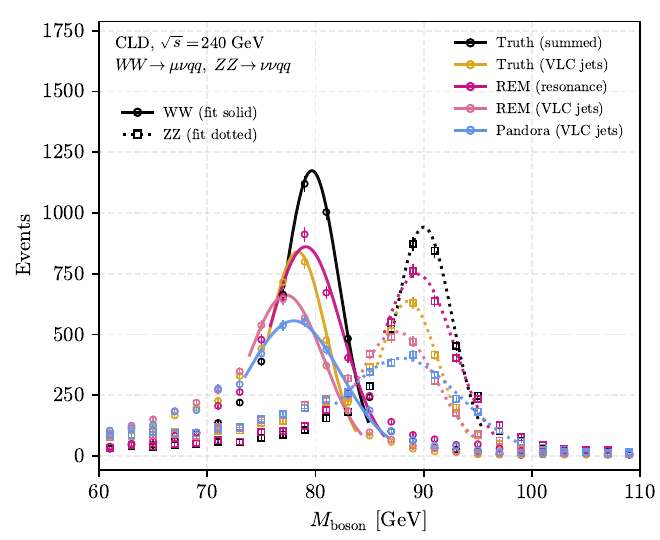}
    \includegraphics[width=0.49\textwidth]{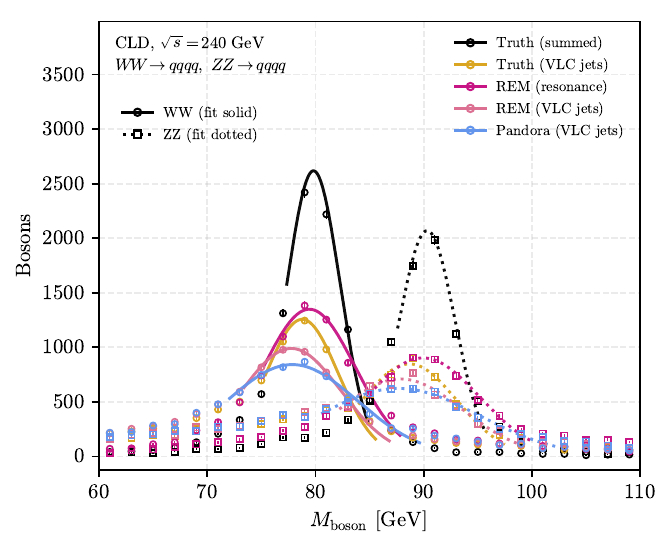}
    \caption{
    Reconstructed hadronic $W$ and $Z$ masses at $\sqrts=240$~GeV in $\WWmunuqq{}$ and $\ZZnunuqq{}$ events (left), and $\WWqqqq{}$ and $\ZZqqqq{}$ events (right), with one entry per hadronic boson.
    VLC jets and resonance masks are compared as described in Section~\ref{sec:results_bosons}. 
    Gaussian fits are solid for $W$ and dotted for $Z$; fit results are listed in Table~\ref{tab:boson_performance_summary}. 
    Error bars show Poisson uncertainties.
    }
    \label{fig:wz_mass_distributions}
\end{figure}

Figure~\ref{fig:wz_mass_distributions} compares the reconstructed masses, with fit results summarised in Table~\ref{tab:boson_performance_summary}. 
In the semileptonic samples, VLC jets of \ours{} particles give narrower mass peaks and better $W/Z$ separation than those of \ppfa{}\footnote{The CLD design report quotes $S=2.1$--$2.6$ with \ppfa in the same two channels for bosons with 125~GeV energy in an earlier detector model \texttt{FCCee\_o1\_v04}, software release (iLCSoft 2019), and VLC radii from 0.7 to 1.1~\cite{Bacchetta:2019fmz}; the comparisons here are made within a different, but consistent simulation and reconstruction setup.}. 
The resonance masks provide a further, smaller improvement over VLC jets of the same \ours{} particles.

In fully hadronic $\WWqqqq{}$ and $\ZZqqqq{}$ events, particles must also be assigned between two hadronic bosons. 
The jet-based method pairs four exclusive VLC jets by minimising the difference between the two dijet masses, whereas the resonance task learns the particle assignments directly. 
The predicted masks retain more $W/Z$ separation than jet clustering and pairing applied to the same reconstructed particles (Table~\ref{tab:boson_performance_summary}).

\begin{figure}[!htbp]
    \centering
    \includegraphics[width=1.0\textwidth]{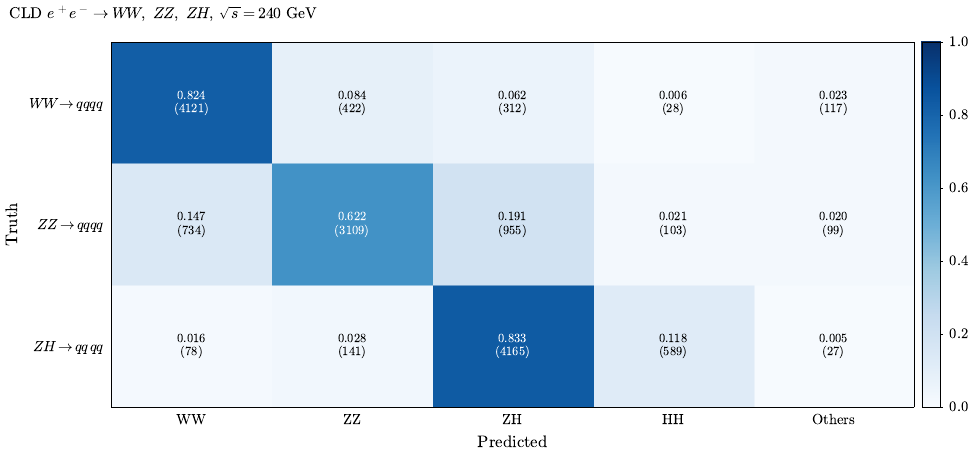}
    \caption{
    Pair of boson labels predicted by the resonance task in each event, against the true decay channel, for fully hadronic $\WWqqqq{}$, $\ZZqqqq{}$ and $\ZHqqqq{}$ events at $\sqrts=240$~GeV; rows are normalised to unity.
    The column ``Others'' collects the label pairs $WZ$ and $WH$ and the events with fewer than two predicted bosons.
    \label{fig:boson_labels}}
\end{figure}

The predicted boson labels (Figure~\ref{fig:boson_labels}) identify 82.4\% of the $WW$ events as containing two $W$ bosons, whereas 62.2\% of the $ZZ$ events are labelled $ZZ$, with most of the remainder labelled $ZH$ (19.1\%) or $WW$ (14.7\%). 
In fully hadronic $\ZHqqqq{}$ events, 83.3\% are labelled $ZH$ and 11.8\% $HH$. 
Table~\ref{tab:boson_label_eff_pur} gives the corresponding label efficiency and purity for each boson class.

\FloatBarrier

\begin{table}[!htbp]
    \centering
    \caption{
    Hadronic boson mass resolution and $W/Z$ separation at $\sqrts=240$~GeV. Semileptonic: $\WWmunuqq{}$ and $\ZZnunuqq{}$ events, one hadronic boson per event (Figure~\ref{fig:wz_mass_distributions}, left). 
    Fully hadronic: $\WWqqqq{}$ and $\ZZqqqq{}$ events, one entry per boson; the four VLC jets of an event are paired by the smallest difference between the two dijet masses (Figure~\ref{fig:wz_mass_distributions}, right). 
    The fit procedure follows Section~\ref{sec:results_bosons}; resolution and separation definitions are given in Section~\ref{sec:metrics}. 
    In the event bootstrap, the two bosons of each fully hadronic event enter together. 
    The truth rows apply the same procedure to the truth particles, clustered into VLC jets or summed over the particles of each true parent boson (true assignment). 
    Bold marks the best relative mass resolutions and separation among reconstructed methods in each event category.
    }
    \label{tab:boson_performance_summary}
    \renewcommand{\null}{\hbox{}}
    \small
    \setlength{\tabcolsep}{4pt}
    \begin{tabular}{llccccccc}
        \toprule
        & & \multicolumn{2}{c}{$W$ [GeV]} & \multicolumn{2}{c}{$Z$ [GeV]} & \multicolumn{2}{c}{$\sigma/\mu$ [\%]} & \\
        \cmidrule(lr){3-4}\cmidrule(lr){5-6}\cmidrule(lr){7-8}
        Events & Reconstruction & $\mu_W$ & $\sigma_W$ & $\mu_Z$ & $\sigma_Z$ & $W$ & $Z$ & $S$ \\
        \midrule
        \multirow{5}{*}{Semileptonic} & Truth particles, true assignment & 79.67 & 2.53 & 90.00 & 2.64 & 3.18 & 2.93 & $4.00\pm0.15$ \\
        & Truth particles, VLC jets & 78.35 & 2.62 & 88.58 & 2.79 & 3.34 & 3.15 & $3.78\pm0.20$ \\
        \addlinespace[3pt]
        & \ppfa, VLC jets & 78.00 & 4.37 & 87.93 & 5.44 & 5.60 & 6.19 & $2.02\pm0.09$ \\
        & \ours, VLC jets & 77.27 & 3.48 & 87.52 & 3.67 & 4.51 & 4.19 & $2.87\pm0.16$ \\
        & \ours, resonance masks & 79.10 & 3.35 & 89.39 & 3.44 & \textbf{4.24} & \textbf{3.85} & $\boldsymbol{3.03\pm0.12}$ \\
        \midrule
        \multirow{5}{*}{Fully hadronic} & Truth particles, true assignment & 79.82 & 2.45 & 90.31 & 2.56 & 3.07 & 2.84 & $4.19\pm0.13$ \\
        & Truth particles, VLC jets & 78.77 & 3.33 & 88.67 & 4.38 & 4.23 & 4.94 & $2.57\pm0.10$ \\
        \addlinespace[3pt]
        & \ppfa, VLC jets & 77.85 & 6.03 & 87.67 & 7.41 & 7.74 & 8.45 & $1.46\pm0.08$ \\
        & \ours, VLC jets & 77.76 & 4.60 & 87.86 & 5.70 & 5.91 & 6.49 & $1.96\pm0.10$ \\
        & \ours, resonance masks & 79.43 & 4.25 & 89.94 & 5.20 & \textbf{5.35} & \textbf{5.78} & $\boldsymbol{2.22\pm0.08}$ \\
        \bottomrule
    \end{tabular}
\end{table}
\begin{table}[!htbp]
    \centering
    \caption{Boson-label efficiency and purity of \ours{} in fully hadronic $\WWqqqq{}$, $\ZZqqqq{}$ and $\ZHqqqq{}$ events at $\sqrts=240$~GeV. Label matching and metric definitions follow Section~\ref{sec:metrics}; the corresponding event-label predictions are shown in Figure~\ref{fig:boson_labels}.}
    \label{tab:boson_label_eff_pur}
    \begin{tabular}{lcccc}
        \toprule
        Boson & Efficiency [\%] & Predicted labels & Purity [\%] \\
        \midrule
        $W$ & 83.6 & 10{,}109 & 82.7 \\
        $Z$& 77.2 & 12{,}947 & 89.4 \\
        $H$ & 95.5 & 6{,}944 & 68.7 \\
        \midrule
        All & 82.4 & 30{,}000 & 82.4 \\
        \bottomrule
    \end{tabular}
\end{table}

The particle-level comparisons help explain the gains in event and resonance observables, since the improved particle recovery, identification, and energy estimates provide more accurate inputs. 
Retaining the reconstructed particles and their hit assignments makes it possible to investigate the sources of these gains and the remaining limitations.

\FloatBarrier

\subsection{Comparison with \hitpf in Light-Quark Events}
\label{sec:results_hitpf}

The $\Zqq{}$ benchmark with $q=u,d,s$ at $\sqrts=91~\mathrm{GeV}$ compares unified reconstruction with specialised learned particle flow. Like \flowf, \hitpf uses externally reconstructed conformal tracks. \ours reconstructs tracks and particles jointly from hits.

\begin{figure}[!htbp]
    \centering
    \includegraphics[width=\textwidth]{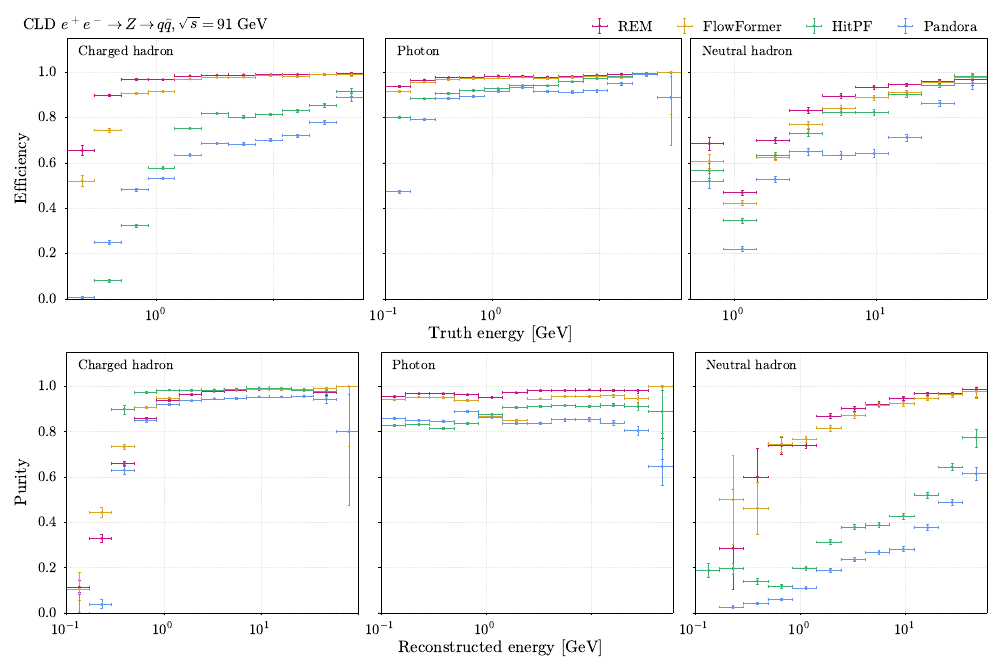}
    \caption{Reconstruction efficiency (top) and purity (bottom) after PID for charged hadrons (left), photons (middle) and neutral hadrons (right) in light-quark $\Zqq{}$ events at 91~GeV, comparing \ours, \flowf, \hitpf and \ppfa.}
    \label{fig:hitpf_summary}
\end{figure}

\begin{figure}[!htbp]
    \centering
    \includegraphics[width=\textwidth]{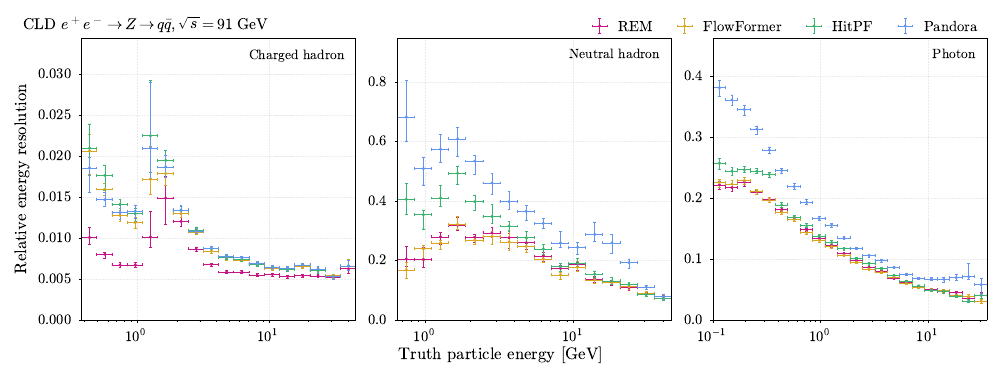}
    \caption{
    Energy resolution of charged hadrons (left), neutral hadrons (middle) and photons (right) as a function of truth energy in simulated $e^+e^- \to \Zqq{}$ events with $q=u,d,s$ at $\sqrts=91~\mathrm{GeV}$, comparing \ours, \flowf, \hitpf and \ppfa. 
    Only truth particles matched by every reconstruction with an object of their own particle class are used, classified by their truth class. 
    Resolution definitions follow Section~\ref{sec:metrics}. Error bars are 16th to 84th percentile intervals from bootstrap resampling of events, and bins with fewer than 20 particles are not shown. 
    The charged-hadron energy resolution is shown above 0.38~GeV. 
    For \ours, \flowf and \ppfa, the energy of a charged hadron with a track is computed from the track momentum assuming the charged-pion mass (Appendix~\ref{app:zqq_ch_momentum}).}
    \label{fig:zqq_particle_resolution}
\end{figure}

Figure~\ref{fig:hitpf_summary} compares charged-hadron, photon, and neutral-hadron efficiency and purity after particle identification.
Both \ours and \flowf generally achieve higher efficiency and purity than \hitpf. 
For low-energy charged hadrons, \hitpf has higher purity but roughly half the efficiency.
Figure~\ref{fig:zqq_particle_resolution} shows the energy resolution of charged hadrons, neutral hadrons and photons for the truth particles that every reconstruction matches with an object of their own particle class.
\ours and \flowf have better energy resolution at low energies, while the learned models tend to converge at higher energies.
The angular-resolution comparison is given in Appendix~\ref{app:zqq_angular}. 
The rise of the charged-hadron energy resolution between about 1 and 2~GeV is discussed in Appendix~\ref{app:zqq_ch_momentum}.
Figure~\ref{fig:zqq_event_level} shows the visible event energy, dijet mass, and charged and neutral energy components. 
For visible energy and dijet mass, \hitpf performs comparably to \flowf, while \ours gives narrower distributions. Like \ppfa, however, \hitpf substantially misestimates the partition of energy between charged and neutral particles.
Additional comparisons are given in Appendix~\ref{sec:results_zqq_breakdown}.

\begin{figure}[!htbp]
    \centering
    \includegraphics[width=0.49\textwidth]{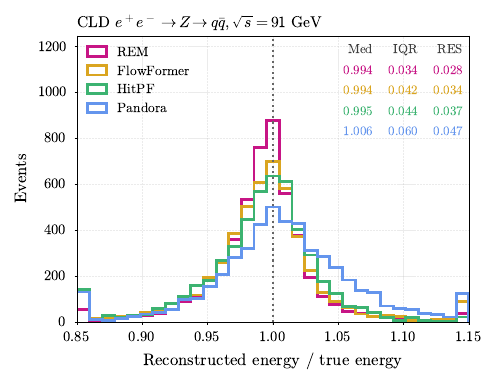}
     \includegraphics[width=0.49\textwidth]{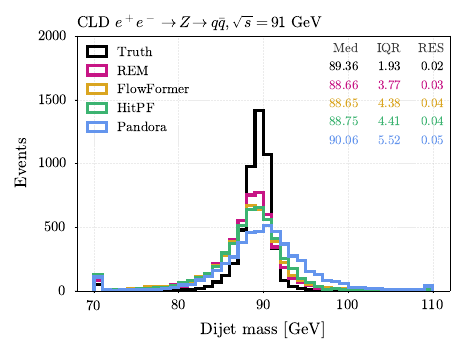}   
    \\
    \includegraphics[width=0.49\textwidth]{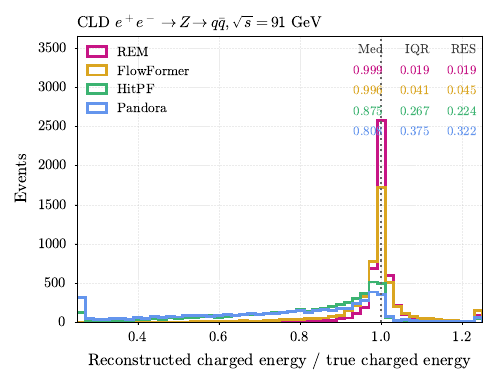}
    \includegraphics[width=0.49\textwidth]{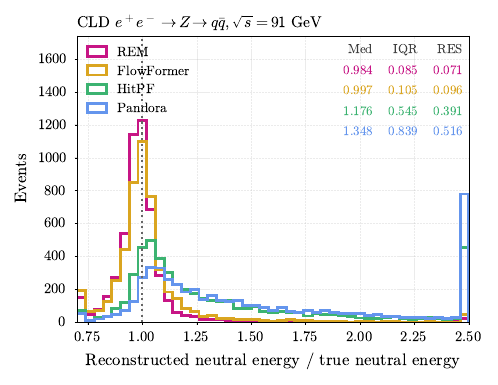}
    \caption{
    Event-level observables in simulated $e^+e^- \to \Zqq{}$ events with $q=u,d,s$ at $\sqrts=91~\mathrm{GeV}$.
    The top row shows the reconstructed-to-truth visible-energy ratio (left) and the reconstructed dijet mass (right). 
    The bottom row shows the reconstructed-to-truth energy ratios for charged (left) and neutral (right) particles. 
    The dijet mass is formed from two exclusive VLC jets per event. Particle selections follow Section~\ref{sec:targets}.}
    \label{fig:zqq_event_level}
\end{figure}

\FloatBarrier

\subsection{Complementary High-Occupancy Benchmark}
\label{sec:results_summary}

The $\ttbar{}$ benchmark at $\sqrts=365$~GeV tests reconstruction in events with dense hadronic activity, heavy-flavour decays, displaced charged particles and overlapping showers. 
Table~\ref{tab:ttbar_headline} summarises the particle and event-level performance. 
The learned methods retain high particle efficiency and purity in this environment, with \ours{} giving the narrowest visible-energy response. 
Fake counts remain similar across the three methods. 
Detailed species and tracking comparisons are given in Appendix~\ref{sec:results_ttbar}.

\begin{table}[!htbp]
    \centering
    \small
    \caption{
    Main results in simulated $\ttbar{}$ events at $\sqrts=365$~GeV. 
    Particle metrics use the tight evaluation targets; response widths use the visible-energy ratio. Definitions follow Section~\ref{sec:metrics}. Bold marks the highest efficiencies and purity and the narrowest response.}
    \label{tab:ttbar_headline}
    \begin{tabular}{lrrr}
        \toprule
        Metric & \ours{} & \flowf{} & \ppfa{} \\
        \midrule
        Particle efficiency before PID [\%] & \textbf{93.3} & 92.2 & 77.8 \\
        Particle efficiency after PID [\%] & \textbf{91.2} & 88.8 & 61.7 \\
        Particle purity after PID [\%] & \textbf{91.7} & 90.5 & 73.5 \\
        Fakes per event & 7.0 & \textbf{6.7} & 7.0 \\
        \midrule
        Visible-energy response IQR & \textbf{0.028} & 0.036 & 0.042 \\
        \bottomrule
    \end{tabular}
\end{table}

\FloatBarrier

\section{Discussion}\label{sec:discussion}

The results demonstrate an end-to-end learned reconstruction chain, from hit assignment across detector subsystems through particle identification and kinematic estimation to resonance reconstruction.
This integration offers a route to simpler reconstruction workflows while retaining the intermediate objects needed to diagnose performance. 
In this study, training is staged with earlier networks frozen, although the framework also allows joint optimisation of the full chain using the resonance objective.

Table~\ref{tab:results_summary} collects the principal 240~GeV results. 
The gains in particle reconstruction propagate to narrower event and jet responses, while the resonance task extends the learned chain to boson reconstruction. 
The fully hadronic comparison includes both particle reconstruction and the assignment of particles to bosons; its conventional baseline uses VLC jets paired by the smallest difference between the two dijet masses.

\begin{table}[!htbp]
    \centering
    \small
    \setlength{\tabcolsep}{4pt}
    \caption{
    Summary of the main 240~GeV results. 
    Efficiency and purity changes are absolute percentage-point increases; other changes are fractional increases in separation or reductions in width, calculated from the displayed values. 
    Metric definitions follow Section~\ref{sec:metrics}. 
    For $W/Z$ separation, \ours{} uses resonance masks and \ppfa{} uses VLC jets. 
    Boson-label efficiency concerns labels only, rather than the correctness of the particle assignment, and has no \ppfa{} counterpart in this study.
    }
    \label{tab:results_summary}
    \begin{tabular}{lccc}
        \toprule
        Metric & \ppfa{} & \ours{} & Change \\
        \midrule
        Particle efficiency after PID & 61.11\% & 91.74\% & $+30.63$ pp \\
        Particle purity after PID & 73.04\% & 92.17\% & $+19.13$ pp \\
        Visible-energy response IQR & 0.047 & 0.030 & $-36\%$ \\
        Jet energy resolution & $(4.61\pm0.14)\%$ & $(2.19\pm0.11)\%$ & $-52\%$ \\
        $W/Z$ separation, semileptonic & $2.02\pm0.09$ & $3.03\pm0.12$ & $+50\%$ \\
        $W/Z$ separation, fully hadronic & $1.46\pm0.08$ & $2.22\pm0.08$ & $+52\%$ \\
        Boson-label efficiency & -- & 82.4\% & -- \\
        \bottomrule
    \end{tabular}
\end{table}

Building on the \flowf{} comparison, further ablations should isolate which aspects of joint tracking and calorimeter reconstruction drive the remaining gains.

Understanding the reconstruction remains essential for diagnosing systematic uncertainties. 
Hit assignments and intermediate objects expose where performance changes, but do not by themselves explain how the learned model reaches its decisions. 
Reliable deployment will require tests of sensitivity to detector conditions and modelling assumptions, alongside tools for inspecting the information used by the model.

Robustness to beam-induced backgrounds, detector defects, and misalignment remains to be established. 
Validation with alternative event generators and shower/hadronisation models is also needed, particularly for the learned resonance assignments. 
Tests across a broader range of detector geometries and operating conditions are needed before deployment in an experiment. 
The event and jet comparisons use truth-defined selections, whose role must be revisited in downstream analyses.

The predicted boson-label pairs already provide event-level classification of the studied fully hadronic $WW$, $ZZ$, and $ZH$ channels. 
Further extensions could include secondary vertexing, jet classification by flavour, and classification of a broader range of event types. 
The intermediate objects allow these additions to be evaluated separately and their sensitivity to reconstruction errors to be studied. 
Coupling learned reconstruction to detector-design studies or simulation-based inference is another possible direction.

The combination of measurement assignment, object identification, and property regression may also be useful for structured prediction in other sparse, multimodal sensor systems.

\FloatBarrier

\section{Conclusion}\label{sec:conclusion}

This work demonstrates end-to-end learned reconstruction from hit-level measurements to resonance bosons in a full simulation of the CLD detector concept.
The Reconstruct Everything Model (\ours{}) takes silicon-tracker, calorimeter, and muon-system hits as inputs and reconstructs tracks and calorimeter showers, identifies particles, and estimates their kinematic properties. 
It requires no externally reconstructed tracks or calorimeter clusters, and retains intermediate outputs that can be inspected and validated. 
The framework also reconstructs composite objects via an additional learned stage that assigns particles to resonance bosons and predicts their identities, extending the chain from detector hits to particles to bosons.

In $WW$, $ZZ$, and $ZH$ events at $\sqrts=240$~GeV, the integrated particle-reconstruction efficiency after identification is 91.7\% for \ours{}, compared with 89.0\% for \flowf{} and 61.1\% for \ppfa{}.
The largest additional gain over \flowf{} is in electron identification: \ours{} identifies 92--94\% of electrons in the 1--10 and 10--100~GeV ranges, compared with 75--78\% for \flowf{}, which uses externally reconstructed tracks.
The improvements at particle level carry through to event and jet observables. 
Relative to \ppfa{}, \ours{} reduces the interquartile width of the visible-energy response by approximately 36\%, while the fitted core jet energy resolution is $(2.19\pm0.11)\%$, compared with $(2.88\pm0.12)\%$ for \flowf{} and $(4.61\pm0.14)\%$ for \ppfa{}. 
These gains accompany a more accurate charged--neutral energy partition and improved calorimeter hit assignment.

The resonance task demonstrates that the learned chain can extend beyond particle reconstruction. 
Summing the four-momenta of particles assigned to the predicted resonance masks gives a $W/Z$ separation of $3.03\pm0.12$ in the semileptonic diboson samples, compared with $2.87\pm0.16$ for VLC jets of \ours{} particles and $2.02\pm0.09$ for VLC jets of \ppfa{} particles. 
Fully hadronic events demonstrate the potential to learn the more complex assignment of particles between two bosons. 
Here the resonance masks outperform the jet-clustering and equal-mass pairing procedure used in this study. 
This technique could be particularly useful for analyses which require jet pairing, such as $HH\rightarrow4b$ analyses in hadron colliders.
Although theoretical guarantees remain to be established and generator dependence is untested, the result demonstrates how successive learned stages can extend reconstruction from particles to composite objects.

The light-quark $\Zqq{}$ benchmark at $\sqrts=91$~GeV also shows an advantage over specialised learned particle flow.
The fitted core jet energy resolution is $(1.17\pm0.07)\%$ for \ours{}, compared with $(1.61\pm0.10)\%$ for \flowf{} and $(2.25\pm0.11)\%$ for \hitpf{}. 
The gains are not uniform across all observables or kinematic regions, with high-momentum tracking and low-energy muon identification among the areas for further improvement.

The $\ttbar{}$ benchmark at $\sqrts=365$~GeV extends the evaluation to denser events with heavy-flavour decays and displaced particles.
Here, \ours{} achieves superior efficiency and purity compared \ppfa{}.

Together, the 91, 240 and 365~GeV benchmarks demonstrate strong reconstruction performance across three centre-of-mass energy regimes and different event topologies. 
\ours{} demonstrates a complete learned end-to-end reconstruction chain, from detector hits to particles to resonance bosons, while retaining the intermediate outputs to validate each stage and understand the origins of performance gains.

\FloatBarrier

\section*{Acknowledgments}
We extend our gratitude to Dolores Garcia for insightful discussion, support with CLD reconstruction, and evaluation of \hitpf. We thank Nikita Pond, Wei Sheng Lai, Edoardo Critelli, Mungo Mangat, Luke Johnson, Jackson Barr, and S\'ebastien Rettie for fruitful discussions. We also extend our thanks to UCL for the use of their high-performance computing facilities, with special thanks to Edward Edmondson for his expert management and technical support. 

We gratefully acknowledge the support of the UK's Science and Technology Facilities Council (STFC). S.V.S. is supported by ST/X005992/1. M.H., K.Y.W., and P.D. are supported by the STFC UCL Centre for Doctoral Training in Data Intensive Science (ST/W00674X/1) and by departmental and industry contributions. G.F. and T.S. receive support from the STFC (ST/W00058X/1) and T.S. is supported by the Royal Society (URF/R/180008).

\clearpage

\appendix
\section{Additional results at \texorpdfstring{$\sqrts=240$~GeV}{sqrt(s) = 240 GeV}}
\label{sec:results_240_app}

\subsection{Particle Energy Response}
\label{app:240_response}
\begin{figure}[!htbp]
    \centering
    \includegraphics[width=\textwidth]{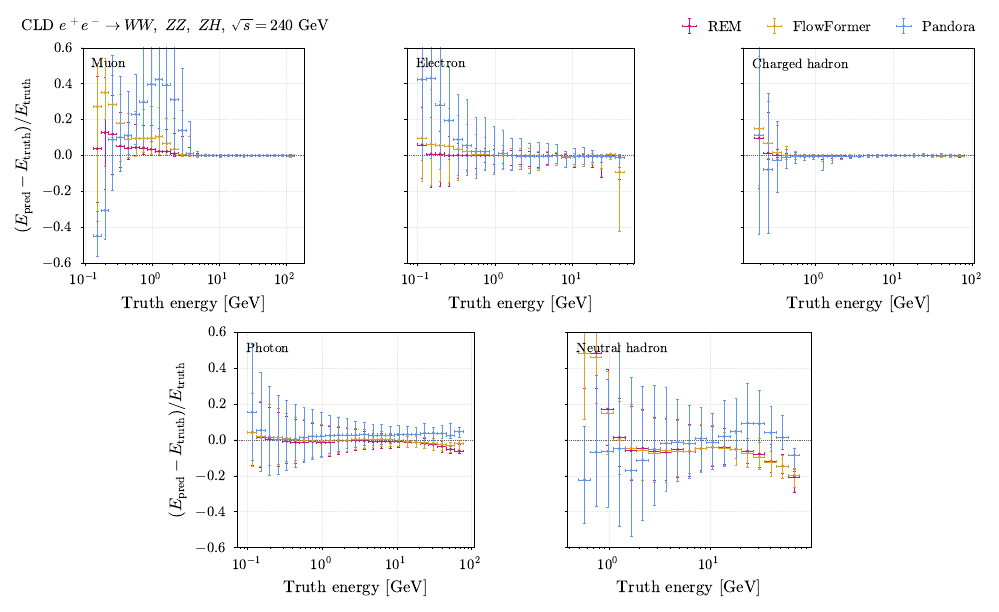}
    \caption{Fractional particle energy response in simulated $e^+e^-\to WW,ZZ,ZH$ events at $\sqrts=240$~GeV, comparing \ours, \flowf and \ppfa. The top row shows muons, electrons and charged hadrons (left to right), and the bottom row shows photons and neutral hadrons. Markers indicate the median response in each truth-energy bin; vertical bars span the 25th to 75th percentiles. The dashed horizontal line denotes an unbiased response.}
    \label{fig:240_particle_energy_response}
\end{figure}

Figure~\ref{fig:240_particle_energy_response} shows the median fractional energy response as a measure of bias. Vertical bars span the central half of each response distribution, with a full width equal to the interquartile range, $\mathrm{IQR}=Q_{75}-Q_{25}$, where $Q_{25}$ and $Q_{75}$ are the 25th and 75th percentiles. We use this width as a robust measure of resolution.
\ours gives a largely unbiased response for electrons and charged hadrons over most of the energy range, and reduces the low-energy muon bias relative to the other methods. Both learned methods generally produce narrower photon and neutral-hadron responses than \ppfa. Some calibration limitations remain: \ours underestimates photon energies in the high-energy tail, while both learned methods retain an energy-dependent bias for neutral hadrons. 

\FloatBarrier

\subsection{Particle Angular Resolution}
\label{app:240_angular}
\begin{figure}[!htbp]
    \centering
    \includegraphics[width=\textwidth]{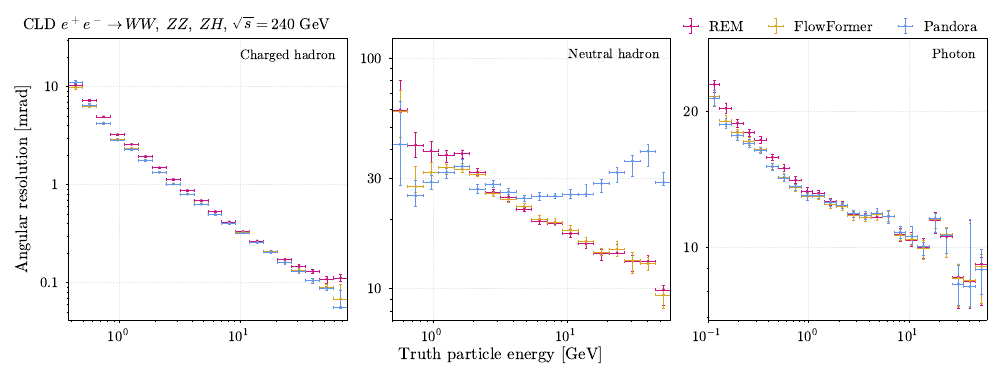}
    \caption{Angular resolutions of charged hadrons (left), neutral hadrons (middle) and photons (right) in simulated $WW,ZZ,ZH$ events at $\sqrts=240$~GeV, comparing \ours{}, \flowf{} and \ppfa{}. Particle selection and bootstrap uncertainties follow Figure~\ref{fig:240_particle_resolution}; resolution definitions follow Section~\ref{sec:metrics}.}
    \label{fig:240_particle_angular_resolution}
\end{figure}

Charged-hadron angular resolutions improve with energy, with \flowf{} and \ppfa{} closely aligned and \ours{} somewhat broader (Figure~\ref{fig:240_particle_angular_resolution}). For neutral hadrons, the learned models are slightly worse below 2~GeV but substantially better above 5~GeV.

\FloatBarrier

\subsection{Composition of the Reconstructed Objects}
\label{app:240_composition}

\begin{figure}[!htbp]
    \centering
    \includegraphics[width=\textwidth]{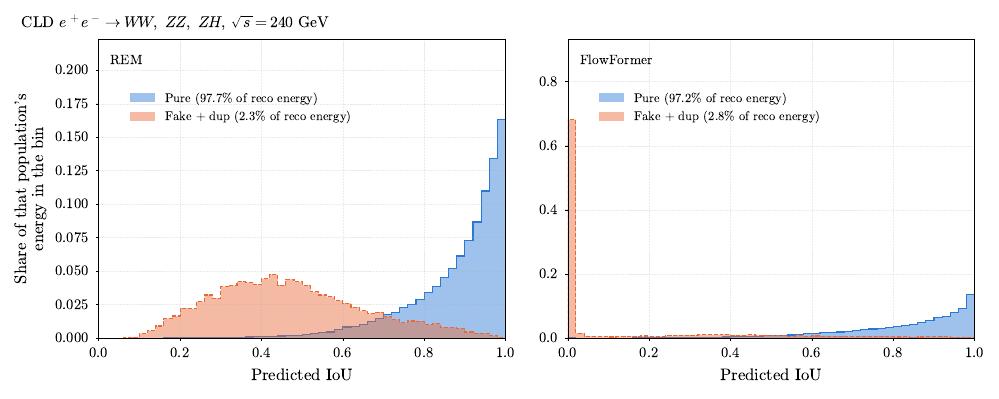}
    \caption{Energy of pure and of fake or duplicate objects against the predicted \IoU{} of \ours (left, single output) and \flowf (right, per-detector outputs combined weighted by the numbers of hits) in simulated $e^+e^-\to WW,ZZ,ZH$ events at $\sqrts=240$~GeV, as the share of each population's energy per bin, averaged over events, with the loose double-majority matching.}
    \label{fig:240_iou_energy}
\end{figure}

Figure~\ref{fig:240_iou_energy} shows how the energy of pure and of fake or duplicate objects is distributed in the predicted \IoU{} of \ours (left) and \flowf (right), the scores used for Figure~\ref{fig:240_quality}. All objects enter except those matched to particles outside the tight definition; the calorimeter requirement discussed in \cref{sec:results_240_quality} is not applied. Fake and duplicate objects carry 2.3\% of the reconstructed energy of \ours and 2.8\% of that of \flowf. For \ours their energy peaks near a predicted \IoU{} of 0.4, for \flowf about two thirds of it lies below 0.02; in both, the energy of pure objects rises towards a predicted \IoU{} of one.

\FloatBarrier
\section{Detailed particle-flow comparison in \texorpdfstring{$\Zqq{}$}{Z->qq} events}
\label{sec:results_zqq_breakdown}

This appendix supplements Section~\ref{sec:results_hitpf} with lepton reconstruction, energy response, hit assignment, charged-hadron species dependence, object formation before PID, and jet performance.

\subsection{Particle-Level Reconstruction after PID}
\label{sec:results_zqq_particle_level}

\begin{figure}[htb]
    \centering
    \includegraphics[width=0.68\textwidth]{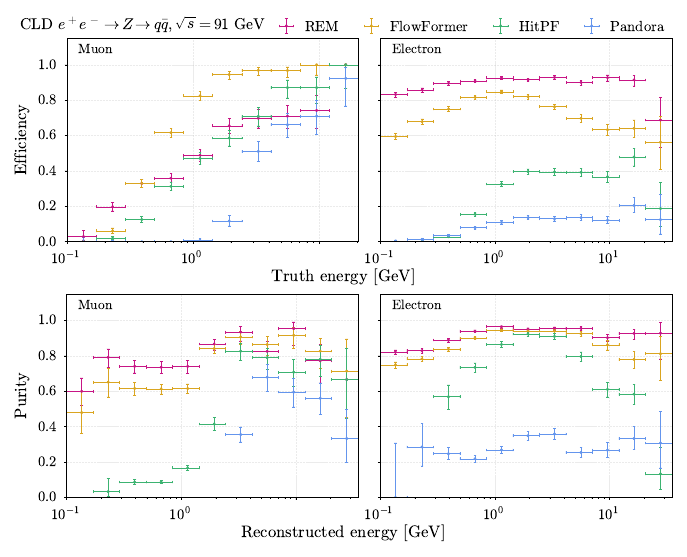}
    \caption{
    Particle-level reconstruction performance in simulated
    $e^+e^- \to \Zqq{}$ events with $q=u,d,s$ at $\sqrts=91~\mathrm{GeV}$.
    Top: reconstruction efficiency for muons (left) and electrons (right), as a function of truth energy.
    Bottom: reconstruction purity for muons and electrons, as a function of reconstructed energy.
    The full particle-identification requirement is applied. \ours is compared with \ppfa{}, \hitpf, and \flowf using the same truth-matching and object definitions.
    }
    \label{fig:zqq_pid_eff_purity_leptons}
\end{figure}

\begin{figure}[htb]
    \centering
    \includegraphics[width=\textwidth]{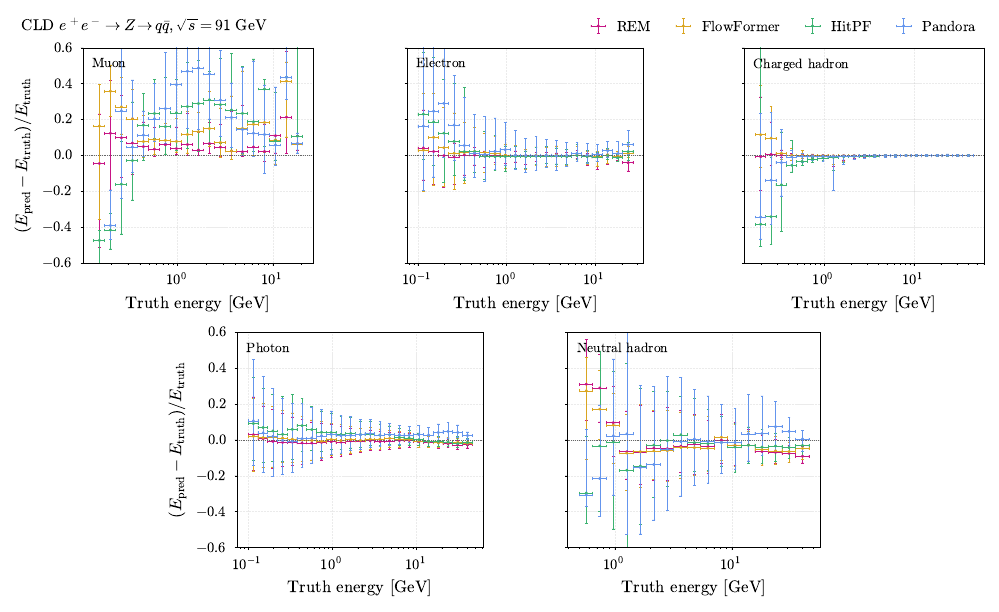}
    \caption{
    Fractional energy response (Section~\ref{sec:metrics}) in simulated
    $e^+e^- \to \Zqq{}$ events with $q=u,d,s$ at $\sqrts=91~\mathrm{GeV}$.
    The response is shown as a function of truth energy from left to right for muons, electrons, and charged hadrons on the top, and for photons and neutral hadrons on the bottom.}
    \label{fig:zqq_energy_response}
\end{figure}

Figure~\ref{fig:zqq_pid_eff_purity_leptons} extends the particle-identification comparison to muons and electrons; Figure~\ref{fig:zqq_energy_response} shows the energy response for all five particle classes.

Table~\ref{tab:hit_iou_ecal_hcal} compares calorimeter hit assignment using \IoU{}-based Hungarian matching. \flowf{} gives the highest mean \IoU{} for charged particles in ECAL and \ours{} in HCAL. For neutral particles, \hitpf{} and \flowf{} perform best in ECAL and \ours{} in HCAL.

\begin{table}[t]
    \centering
    \caption{
    Mean hit-\IoU{} between Hungarian-matched truth particles and predicted queries, restricted to calorimeter hits. 
    Results are shown separately for charged truth particles, CH+$e+\mu$, and neutral truth particles, NH+$\gamma$.
    }
    \label{tab:hit_iou_ecal_hcal}
    \renewcommand{\null}{\hbox{}}
    \begin{tabular}{lcc|cc}
        \toprule
        \multirow{2}{*}{Algorithm}
        & \multicolumn{2}{c|}{Charged}
        & \multicolumn{2}{c}{Neutral} \\
        \cmidrule(lr){2-3}
        \cmidrule(lr){4-5}
        & ECAL & HCAL & ECAL & HCAL \\
        \midrule
        \ppfa{}    & {0.757} & {0.638} & {0.740} & {0.416} \\
        \hitpf      & {0.933} & {0.802} & \textbf{{0.932}} & {0.543} \\
        \flowf      & \textbf{{0.948}} & {0.806} & \textbf{{0.932}} & {0.547} \\
        \ours       & {0.943} & \textbf{{0.823}} & {0.917} & \textbf{{0.651}} \\
        \bottomrule
    \end{tabular}
\end{table}

\subsection{Particle Angular Resolution}
\label{app:zqq_angular}
\begin{figure}[!htbp]
    \centering
    \includegraphics[width=\textwidth]{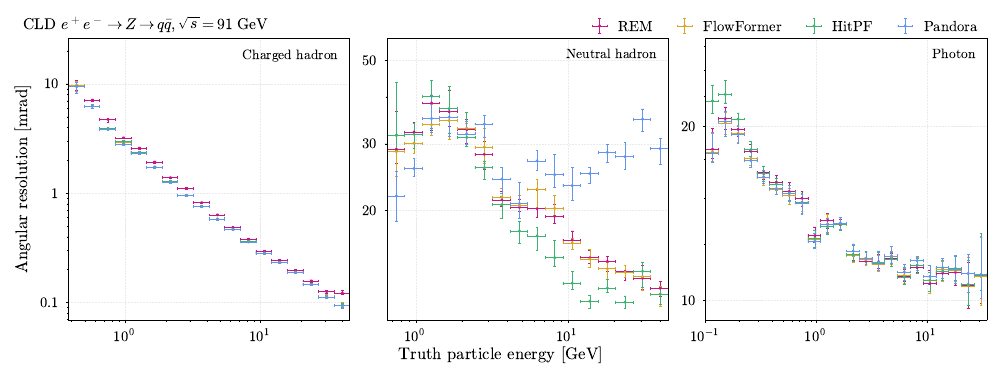}
    \caption{Angular resolutions of charged hadrons (left), neutral hadrons (middle) and photons (right) in light-quark $\Zqq{}$ events at $\sqrts=91$~GeV, comparing \ours{}, \flowf{}, \hitpf{} and \ppfa{}. Particle selection and bootstrap uncertainties follow Figure~\ref{fig:zqq_particle_resolution}; resolution definitions follow Section~\ref{sec:metrics}.}
    \label{fig:zqq_particle_angular_resolution}
\end{figure}

Figure~\ref{fig:zqq_particle_angular_resolution} compares the angular resolutions. The learned methods perform comparably, with \hitpf{} having a slight advantage for neutral hadrons.

\FloatBarrier

\subsection{\texorpdfstring{Charged-hadron momentum resolution}{Charged-hadron momentum resolution}}
\label{app:zqq_ch_momentum}

\begin{figure}[htb]
    \centering
    \includegraphics[width=\textwidth]{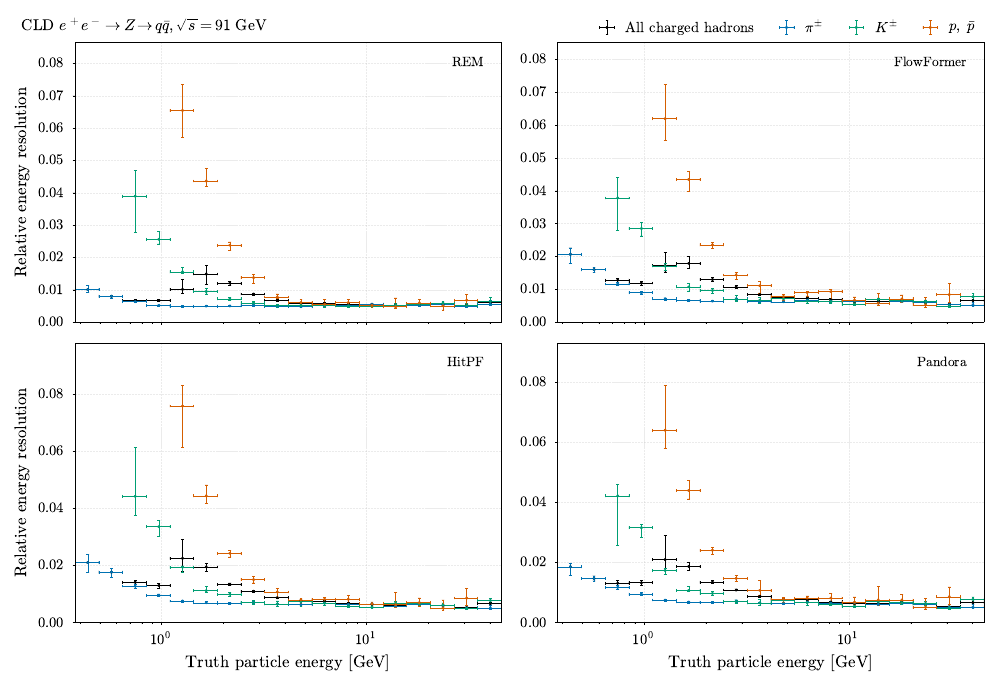}
    \caption{Relative energy resolution of charged hadrons as a function of truth energy, split by the species of the matched truth particle (pions, kaons and protons, with all charged hadrons for reference), for \ours (top left), \flowf (top right), \hitpf (bottom left) and \ppfa (bottom right) in simulated $e^+e^- \to \Zqq{}$ events with $q=u,d,s$ at $\sqrts=91~\mathrm{GeV}$. The charged hadrons and the resolution definition are those of Figure~\ref{fig:zqq_particle_resolution}.}
    \label{fig:zqq_ch_energy_resolution_by_species}
\end{figure}

\begin{figure}[htb]
    \centering
    \includegraphics[width=0.48\textwidth]{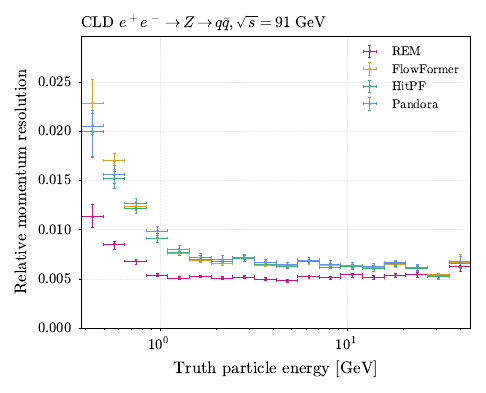}
    \caption{Momentum resolution of charged hadrons as a function of truth energy in simulated $e^+e^- \to \Zqq{}$ events with $q=u,d,s$ at $\sqrts=91~\mathrm{GeV}$, comparing \ours, \flowf, \hitpf and \ppfa, for the charged hadrons of Figure~\ref{fig:zqq_particle_resolution}. The momentum resolution is the half-width of the 16th to 84th percentile interval of $p_\mathrm{reco}/p_\mathrm{true}$ divided by its median, with $E_\mathrm{reco}/E_\mathrm{true}$ used for charged hadrons reconstructed without a track. Error bars are 16th to 84th percentile intervals from bootstrap resampling of events, and bins with fewer than 20 particles are not shown.}
    \label{fig:zqq_ch_momentum_resolution}
\end{figure}

The rise in charged-hadron energy resolution around 1--2~GeV depends on particle species (Figure~\ref{fig:zqq_ch_energy_resolution_by_species}). 
It appears for kaons and protons near their rest masses, while \ours{}, \flowf{} and \ppfa{} calculate charged-hadron energies from track momenta assuming the pion mass. The corresponding momentum resolution has no such rise (Figure~\ref{fig:zqq_ch_momentum_resolution}), also at 240 and 365~GeV.

\FloatBarrier

\subsection{Particle-Flow Object Formation before PID}
\label{sec:results_zqq_pre_pid}

\begin{figure}[!htbp]
    \centering
    \includegraphics[width=\textwidth]{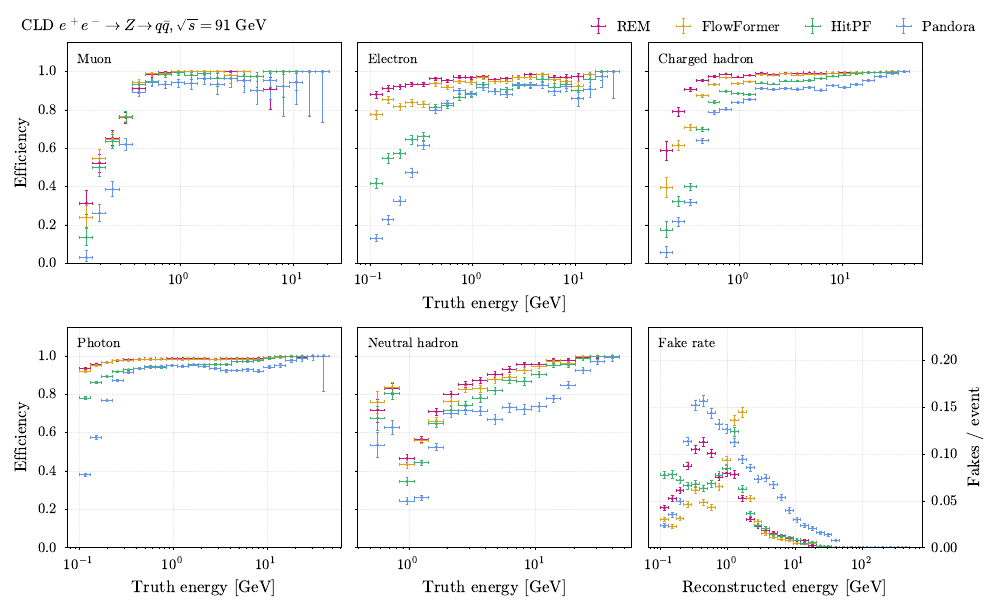}
    \caption{   
    Particle-flow object reconstruction efficiency before the final PID requirement in simulated
    $e^+e^- \to \Zqq{}$ events with $q=u,d,s$ at $\sqrts=91~\mathrm{GeV}$.
    The efficiency is shown as a function of truth energy for muons, electrons and charged hadrons at the top row; photons and neutral hadrons at the bottom row. 
    A class-agnostic fake rate per event is shown at the last column at the bottom row, which counts both fake and duplicate objects.
    This metric tests whether the object is reconstructed, independent of whether it is assigned the final correct particle class.
    }
    \label{fig:zqq_prepid_pfo_eff}
\end{figure}

The pre-PID comparison tests particle finding independently of the assigned class (Figure~\ref{fig:zqq_prepid_pfo_eff}).

\subsection{Jet-Level Observables}
\label{app:zqq_jets}

\begin{figure}[!htbp]
    \centering
    \includegraphics[width=\textwidth]{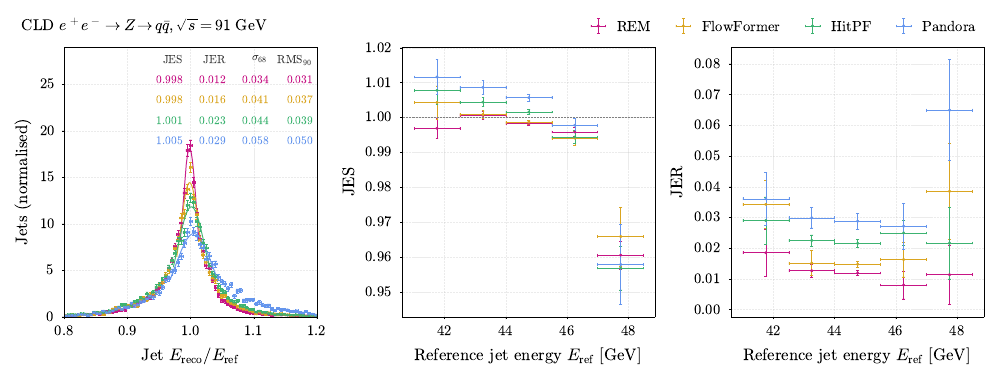}
    \caption{
    Jet performance in simulated $e^+e^- \to \Zqq{}$ events with $q=u,d,s$ at $\sqrts=91~\mathrm{GeV}$, comparing \ours, \flowf, \hitpf and \ppfa, with two exclusive VLC jets per event. 
    The panels mirror those in  Figure~\ref{fig:boson_sample_jet_performance}, showing the distribution of $E_\mathrm{reco}/E_\mathrm{ref}$ with its Gaussian core fit (left), and the JES (middle) and JER (right) against the reference jet energy.
    }
    \label{fig:zqq_jet_performance}
\end{figure}

Figure~\ref{fig:zqq_jet_performance} compares two exclusive VLC jets per event, using the reference jets, matching and fits of Section~\ref{sec:results_240_jet}. 
All four methods give a JES close to unity. \ours{} gives the narrowest response, with the same ordering for the core and tail measures (Table~\ref{tab:zqq_jet_summary}).

\begin{table}[!htbp]
    \centering
    \small
    \caption{Jet response in light-quark $\Zqq{}$ events at $\sqrts=91$~GeV. Definitions follow Section~\ref{sec:results_240_jet}; JER uncertainties use event bootstrap resampling. Bold marks the narrowest responses.}
    \label{tab:zqq_jet_summary}
    \begin{tabular}{lrrrr}
        \toprule
        Metric & \ours{} & \flowf{} & \hitpf{} & \ppfa{} \\
        \midrule
        JES & 0.998 & 0.998 & 1.001 & 1.005 \\
        JER [\%] & $\boldsymbol{1.2\pm0.1}$ & $1.6\pm0.1$ & $2.3\pm0.1$ & $2.9\pm0.2$ \\
        $\sigma_{68}$ [\%] & \textbf{3.4} & 4.1 & 4.4 & 5.8 \\
        $\mathrm{RMS}_{90}$ [\%] & \textbf{3.1} & 3.7 & 3.9 & 5.0 \\
        \bottomrule
    \end{tabular}
\end{table}

\subsection{Timing and Reconstruction Scope}
\label{sec:results_timing}

\begin{figure}[!htbp]
    \centering
    \includegraphics[width=\textwidth]{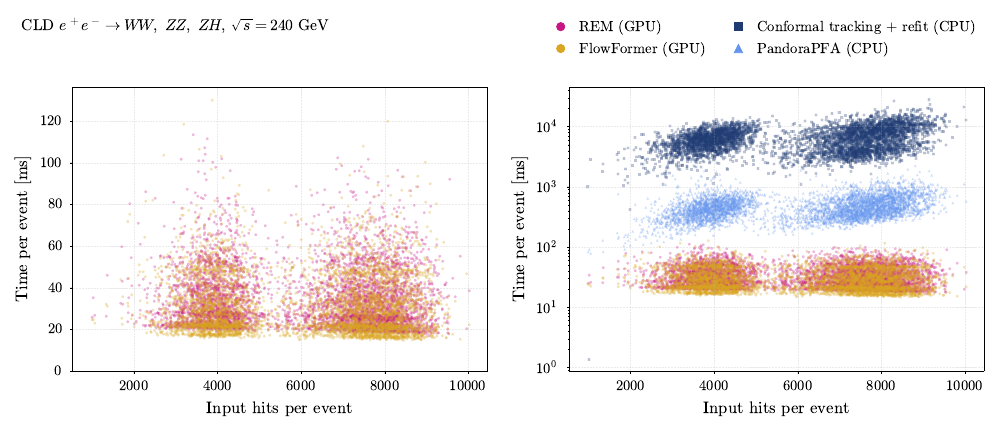}
    \caption{
    Per-event reconstruction time against the number of input hits in simulated $e^+e^-\to WW,ZZ,ZH$ events at $\sqrts=240$~GeV. Left: the GPU forward passes of \ours{} and of \flowf{}, the latter without the conformal tracking it takes as input, at batch size one. 
    Right: the same together with the CPU times of conformal tracking and \ppfa{}, on a logarithmic scale. 
    Model times above five times the median, from the compilation warm-up (at most five events per model), are not shown.
    }
    \label{fig:timing_240}
\end{figure}

Table~\ref{tab:timing_hitpf_comparison} records the scope of each method and its timing measured on the 5{,}000 events of the 240~GeV benchmark, and Figure~\ref{fig:timing_240} shows these times against the number of input hits. 
Comparisons must account for the external track reconstruction required by \flowf{}, \hitpf{} and \ppfa{}. 
At batch size one, the GPU times of \ours{} and \flowf{} depend little on the number of input hits.

\begin{table}[t]
    \centering
    \caption{
    Per-event inference time, hardware and reconstruction scope. Times are medians over the 5{,}000 simulated $e^+e^-\to WW,ZZ,ZH$ events at $\sqrts=240$~GeV (Figure~\ref{fig:timing_240}). 
    \ours{} and \flowf{} run on one NVIDIA L40S GPU with batch size one and \texttt{bfloat16} mixed precision. 
    The CPU times of conformal tracking and \ppfa{} are those recorded per event by the Key4hep reconstruction during the sample production. 
    \flowf{}, \hitpf{} and \ppfa{} take the conformal tracks as input. 
    $^\dagger$\hitpf{} was run only on the $\Zqq{}$ sample. 
    Its inference uses the GPU for encoding and CPU for its density peak clustering step. 
    }
    \label{tab:timing_hitpf_comparison}
    \begin{tabular}{lcccc}
        \toprule
        Algorithm & Input level & Tracks included & Hardware & Time / event \\
        \midrule
        \ours{} & detector hits & yes & GPU & 31~ms \\
        \flowf{} & tracks + calo/muon hits & no & GPU & 28~ms \\
        \hitpf & tracks + calo/muon hits & no & CPU + GPU & 0.64~s$^\dagger$ \\
        \ppfa{} & tracks + hits & no & CPU & 0.46~s \\
        Conformal tracking + refit & silicon hits & yes & CPU & 6.3~s \\
        \bottomrule
    \end{tabular}
\end{table}

\FloatBarrier
\section{End-to-end reconstruction in \texorpdfstring{\ttbar{}}{ttbar} events}
\label{sec:results_ttbar}

Simulated $e^+e^- \to \ttbar{}$ production at $\sqrts=365~\mathrm{GeV}$ provides a complementary test of the full reconstruction chain. Dense hadronic activity, heavy-flavour decays, displaced charged particles, and overlapping calorimeter showers make this sample a challenging benchmark. We compare \ours with \flowf and \ppfa using the tight truth definition and the second matching described in \cref{sec:targets}. Objects matched in the second step to genuine particles outside the tight evaluation selection are excluded from the fake and duplicate populations.

\subsection{Tracking}
\label{sec:ttbar_tracking}

The charged-particle tracking performance builds on the approach described in Ref.~\cite{MFTrk}, with additional improvements to the track-parameter regression. We compare \ours and the conformal tracking baseline on track reconstruction quality.

\begin{figure}[htb]
    \centering
    \includegraphics[width=\textwidth]{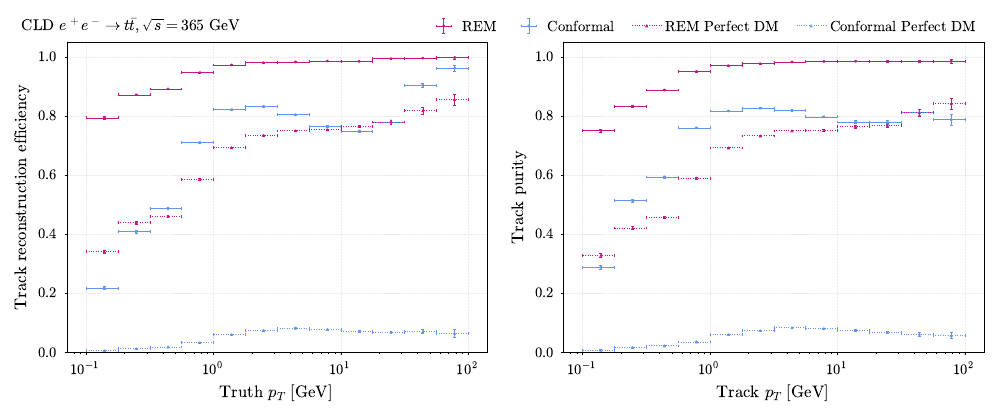}
    \caption{
    Charged-particle tracking efficiency and purity in simulated
    $e^+e^- \to \ttbar{}$ events at $\sqrts=365~\mathrm{GeV}$, comparing \ours tracks and baseline conformal tracks. Perfect matches require the track and truth particle to have exactly the same silicon hits.
    }
    \label{fig:ttbar_tracking_eff}
\end{figure}

\begin{figure}[htb]
    \centering
    \includegraphics[width=\textwidth]{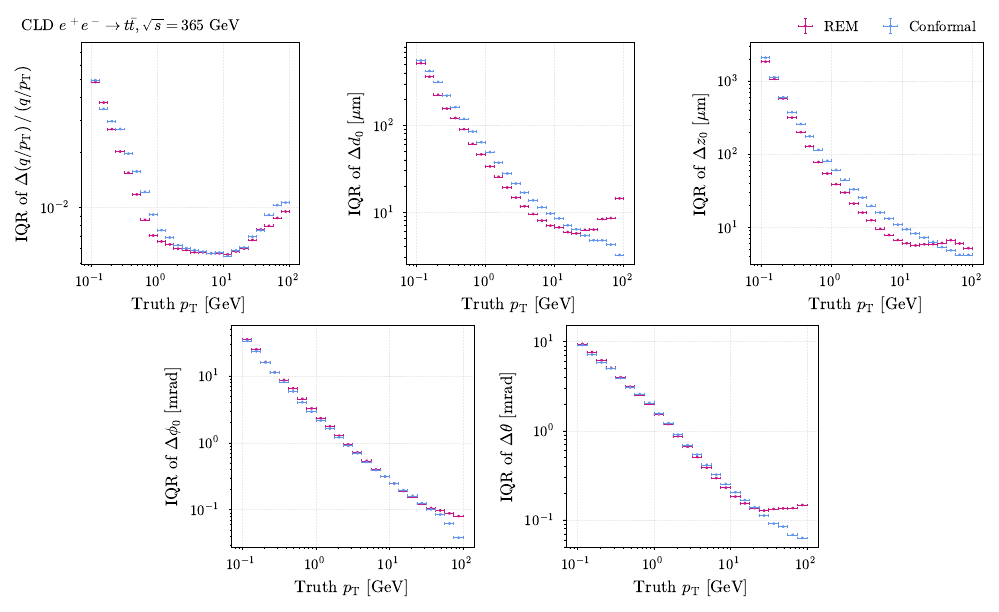}
    \caption{
    Charged-particle track-parameter resolution in simulated
    $e^+e^- \to \ttbar{}$ events at $\sqrts=365~\mathrm{GeV}$.
    The panels show the interquartile range of the residuals of the track parameters at the IP, $(\qOverPt, d_0, z_0,\phi,\theta)$, in bins of truth $\pT$ for \ours and the baseline conformal-tracking reconstruction, on the reconstructable tracks that both reconstruct. The $\qOverPt$ residual is relative to the truth value.
    }
    \label{fig:ttbar_tracking_params}
\end{figure}

Figures~\ref{fig:ttbar_tracking_eff} and~\ref{fig:ttbar_tracking_params} compare track finding and parameter resolution with conformal tracking. Table~\ref{tab:ttbar_tracking_residuals} summarises the integrated residuals for the 162\,000 tracks reconstructed by both methods.

\begin{table}[!htbp]
    \centering
    \small
    \caption{Median and IQR of track-parameter residuals for tracks reconstructed by both methods in the $\ttbar{}$ benchmark at $\sqrts=365$~GeV. Residuals are reconstructed minus true values; the $\qOverPt$ residual is relative to its true value and expressed as a percentage.}
    \label{tab:ttbar_tracking_residuals}
        \renewcommand{\null}{\hbox{}}
        \begin{tabular}{lrrrr}
            \toprule
            & \multicolumn{2}{c}{\ours{}} & \multicolumn{2}{c}{Conformal tracking} \\
            Parameter & Median & IQR & Median & IQR \\
            \midrule
            $\qOverPt$ (relative, \%) & 0.014 & 0.68 & 0.014 & 0.77 \\
            $d_0$ [$\mu$m] & $-0.26$ & 21 & $-0.0068$ & 30 \\
            $z_0$ [$\mu$m] & 0.0 & 25 & $-0.042$ & 37 \\
            $\phi$ [mrad] & 0.0075 & 1.3 & 0.00018 & 1.2 \\
            $\theta$ [mrad] & 0.00048 & 0.88 & 0.0011 & 0.90 \\
            \bottomrule
        \end{tabular}
    \end{table}
    
The silicon-hit assignments are more complete for \ours{}: the mean hit-on-track \IoU{} is 0.947, compared with 0.798 for conformal tracking. The fraction of tracks carrying exactly the silicon hits of their matched truth particle is 68.1\% and 7.6\%, respectively.

\subsubsection{Additional Charged-Particle Reconstruction Diagnostics}
\label{app:ttbar_tracking}

\begin{figure}[!htbp]
    \centering
    \includegraphics[width=\textwidth]{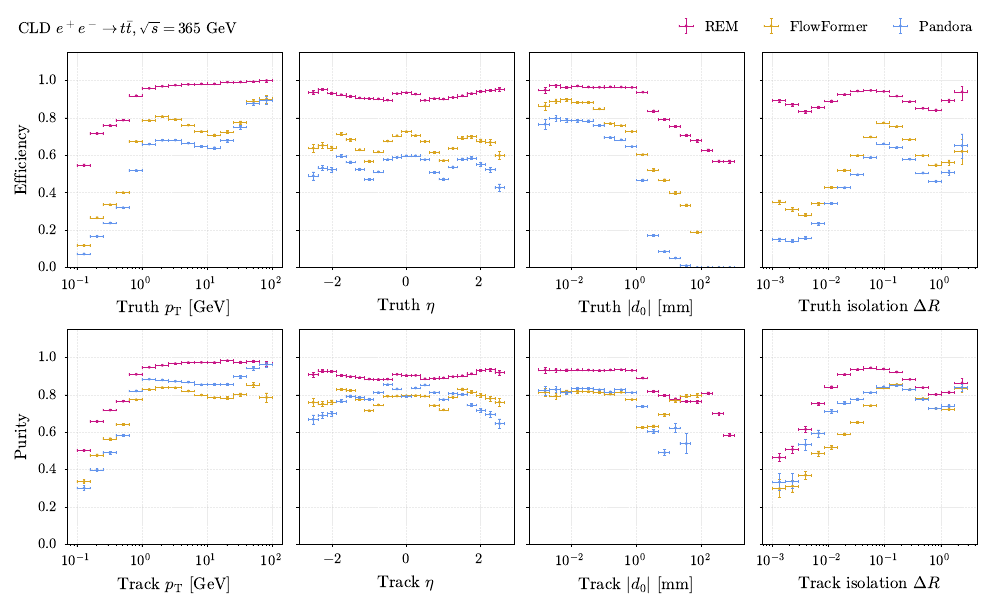}
    \caption{
    Reconstruction performance for charged particles with a reconstructable track in simulated
    $e^+e^- \to \ttbar{}$ events at $\sqrts=365~\mathrm{GeV}$, comparing \ours, \flowf and \ppfa.
    The top row shows the reconstruction efficiency as a function of truth
    $\pT$, $\eta$, transverse impact parameter $|d_0|$, and isolation; the bottom row shows the purity as a function of the same quantities of the reconstructed track. Isolation is the smallest $\Delta R$ to another charged particle (truth) or to another reconstructed particle with a track (reconstructed). Error bars are 68\% Clopper--Pearson intervals.
    }
    \label{fig:app_all_charged_eff_pur}
\end{figure}

Additional charged-particle reconstruction diagnostics in simulated
$e^+e^- \to \ttbar{}$ events at $\sqrts=365~\mathrm{GeV}$ are shown for charged particles with a reconstructable track: charged hadrons, electrons and muons that do not decay in the tracker, with $\pT > 100~\mathrm{MeV}$, $|\cos\theta|<0.99$ and at least four silicon hits. Reconstructed particles must carry a track satisfying the same $\pT$, $\cos\theta$ and silicon-hit requirements; for \ours, a track is a particle predicted as charged with at least four silicon hits. Matching and purity treatment follow Sections~\ref{sec:metrics} and~\ref{sec:targets}. For these diagnostics, the double-majority criterion is applied separately to silicon and calorimeter hits, and the second matching uses silicon hits.
Figure~\ref{fig:app_all_charged_eff_pur} shows the efficiency and purity as a function of transverse momentum, pseudorapidity, transverse impact parameter and isolation.

\begin{figure}[!htbp]
    \centering
    \includegraphics[width=\textwidth]{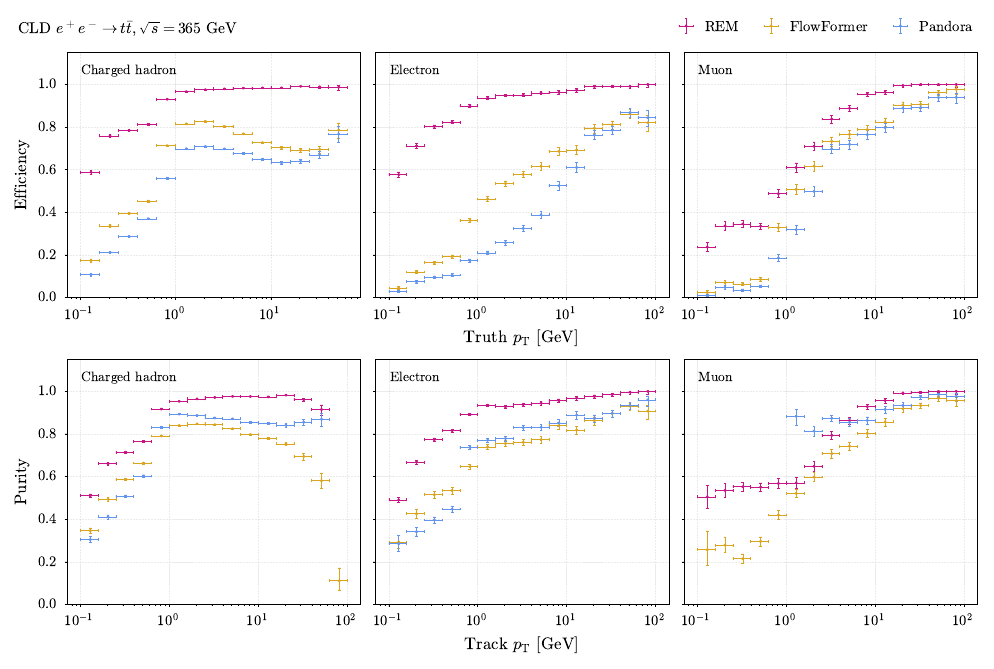}
    \caption{
    Charged-particle reconstruction performance by particle species in simulated
    $e^+e^- \to \ttbar{}$ events at $\sqrts=365~\mathrm{GeV}$, for charged particles with a reconstructable track as in \cref{fig:app_all_charged_eff_pur}.
    The top row shows reconstruction efficiency as a function of truth
    $\pT$ for charged hadrons, electrons, and muons, by truth species. The bottom row
    shows the corresponding reconstruction purity as a function of the reconstructed track $\pT$, by predicted species. \ours{} is compared with
    \flowf{} and \ppfa{} using the same truth-matching and object definitions.
    }
    \label{fig:app_charged_species_pt}
\end{figure}

Figure~\ref{fig:app_charged_species_pt} shows the species-level charged-particle performance as a function of
transverse momentum.

\FloatBarrier

\subsection{Calorimeter Hit Assignment}
\label{app:ttbar_calo}

\begin{figure}[hb]
    \centering
    \includegraphics[width=\textwidth]{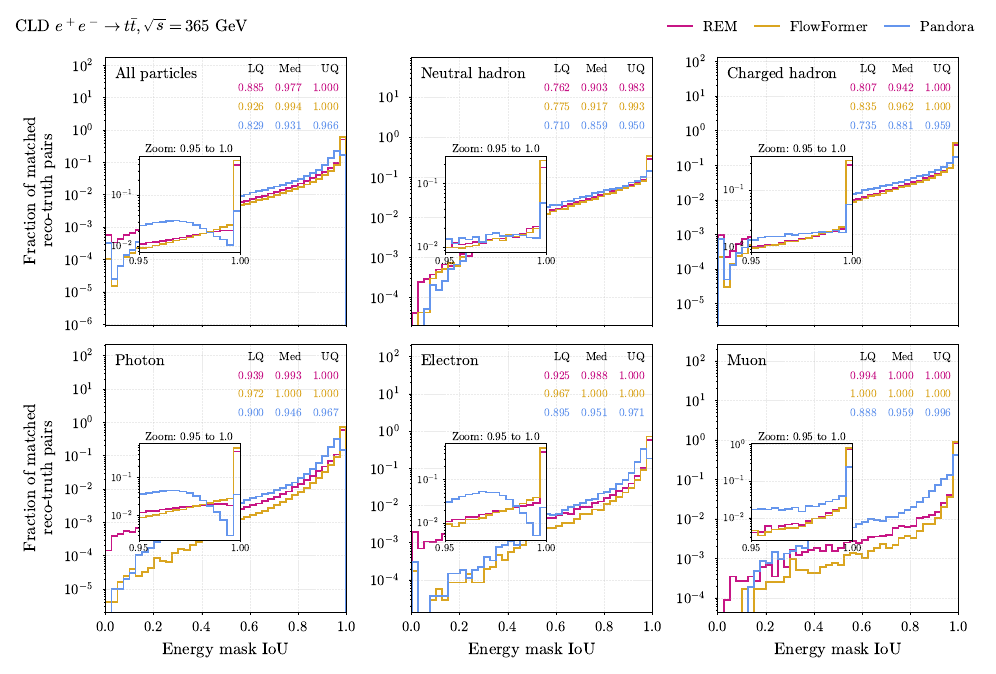}
    \caption{
    Energy mask \IoU{} between reconstructed particles and matched truth particles in $e^+e^- \to \ttbar{}$ events at $\sqrts=365~\mathrm{GeV}$.
    The energy mask \IoU{} is shown from left to right for all particles, neutral hadrons, and charged hadrons on the top,
    and for photons, electrons and muons on the bottom.
    }
    \label{fig:ttbar_energy_maskIOU}
\end{figure}

Figure~\ref{fig:ttbar_energy_maskIOU} compares the energy-weighted calorimeter mask \IoU{} defined in Section~\ref{sec:results_240_calo}. Both learned methods capture the truth-particle deposits more completely and with less contamination than \ppfa{}.

\FloatBarrier

\subsection{Particle Classification}
\label{app:ttbar_classification}

\begin{figure}[h]
    \centering
    \includegraphics[width=\textwidth]{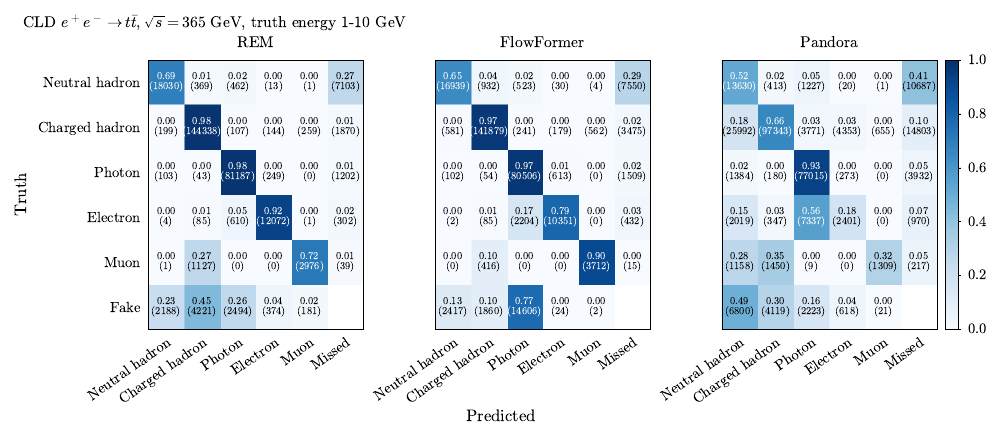}
    \caption{Confusion matrices for particles with $1<E<10$~GeV for \ppfa{}, \flowf, and \ours.
    Rows are normalised to unity.}
    \label{fig:confusion_ttbar}
\end{figure}

Figure~\ref{fig:confusion_ttbar} compares particle identification for truth energies of 1--10~GeV. The integrated charged-hadron comparison helps explain the largest gains over \ppfa{}.

Charged hadrons, the \addedtext{second} most abundant class \addedtext{after photons} (37\% of the target particles), account for the largest share of the improvement: 45\% of the additional correctly identified particles of \ours{} relative to \ppfa{}. Most of this gain comes from charged hadrons without a reconstructed track.

In this sample, 30\% of the charged hadrons have no matched silicon track, either because they leave fewer than four silicon hits or because no track satisfies the double-majority match. Both learned methods correctly identify a much larger fraction of these particles than \ppfa{} (Table~\ref{tab:ttbar_trackless_charged_hadrons}). These particles account for 61\% of the charged-hadron gain of \ours{} over \ppfa{}.

\begin{table}[!htbp]
    \centering
    \caption{Reconstruction efficiency after particle identification for truth charged hadrons without a matched silicon track in the $\ttbar{}$ benchmark at $\sqrts=365$~GeV. The highest efficiency is shown in bold.}
    \label{tab:ttbar_trackless_charged_hadrons}
    \begin{tabular}{lc}
        \toprule
        Algorithm & Efficiency [\%] \\
        \midrule
        \ours{} & \textbf{91.9} \\
        \flowf{} & 84.5 \\
        \ppfa{} & 19.1 \\
        \bottomrule
    \end{tabular}
\end{table}

\ppfa{} builds charged particles from tracks associated with calorimeter clusters, and a cluster without an associated track becomes a neutral particle~\cite{Thomson:2009rp}. Accordingly, every charged hadron \ppfa{} reconstructs carries a track, and the trackless charged hadrons it does reconstruct are labelled neutral hadrons (61\%) or photons (10\%). This structurally caps \ppfa{}'s charged-hadron efficiency after identification and, at the same time, depresses its neutral-hadron purity: 43\% of the objects \ppfa{} labels as neutral hadrons are matched to truth charged hadrons, which explains its neutral-hadron purity of 19.3\%.

\FloatBarrier

\subsection{Particle Reconstruction}
\label{app:ttbar_particles}

\begin{figure}[htb]
    \centering
    \includegraphics[width=\textwidth]{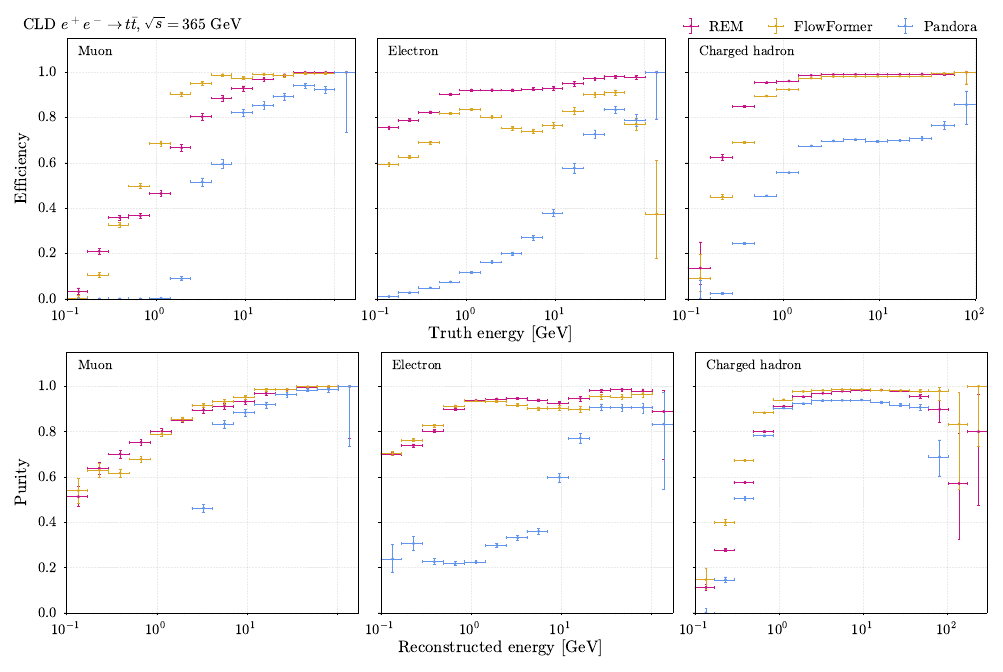}
    \caption{
    Particle-level reconstruction performance for muons, electrons, and charged hadrons (left to right) in simulated $e^+e^- \to \ttbar{}$ events at $\sqrts=365~\mathrm{GeV}$.
    The top row shows reconstruction efficiency as a function of truth energy; the bottom row shows purity as a function of reconstructed energy.
    The full particle-identification requirement is applied. \ours is compared with \flowf and \ppfa using the same truth-matching and object definitions.
    }
    \label{fig:ttbar_charged_particle_eff_purity}
\end{figure}

\begin{figure}[t]
    \centering
    \includegraphics[width=0.68\textwidth]{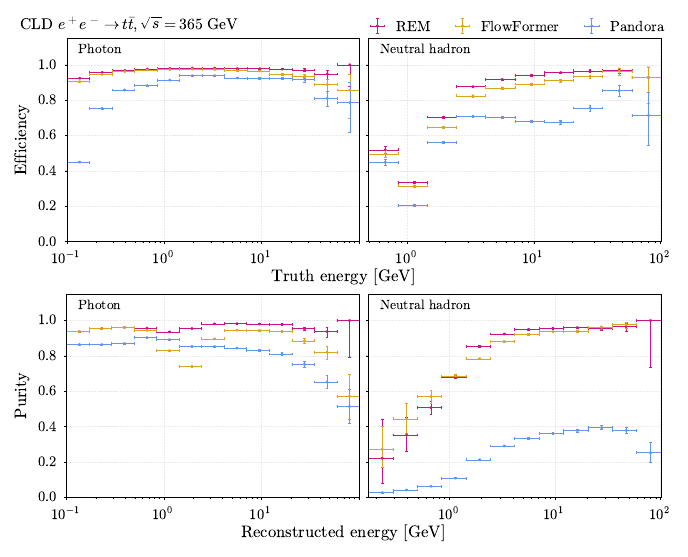}
    \caption{
    Particle-level reconstruction performance for photons and neutral hadrons (left to right) in simulated $e^+e^- \to \ttbar{}$ events at $\sqrts=365~\mathrm{GeV}$.
    The top row shows reconstruction efficiency as a function of truth energy; the bottom row shows purity as a function of reconstructed energy.
    The full particle-identification requirement is applied. \ours is compared with \flowf and \ppfa using the same truth-matching and object definitions.
    }
    \label{fig:ttbar_neutral_particle_eff_purity}
\end{figure}

The particle reconstruction efficiency and purity for charged particles after full particle identification are shown in \cref{fig:ttbar_charged_particle_eff_purity} and for neutral particles in \cref{fig:ttbar_neutral_particle_eff_purity}.
\ours and \flowf generally achieve higher efficiency and purity than \ppfa. Their differences show how reconstruction changes when tracking is included in the learned model, although the training targets and objectives also differ. \cref{tab:particle_eff_pur_by_class} gives the integrated efficiency and purity for each particle type.

\begin{table}[htb]
    \centering
    \small
    \setlength{\tabcolsep}{3pt}
    \newcommand{\metricpair}[2]{%
        \makebox[3em][c]{#1}\makebox[3em][c]{#2}}
    \newcommand{\metrictriple}[3]{%
        \makebox[3.5em][c]{#1}%
        \makebox[3.5em][c]{#2}%
        \makebox[4em][c]{#3}}

    \caption{Integrated reconstruction efficiency and purity for all target particles and for each particle class in $e^+e^- \to \ttbar{}$ events at $\sqrts=365~\mathrm{GeV}$. Matching and metric definitions follow Section~\ref{sec:metrics}. The best value in each column is shown in bold.}
    \label{tab:particle_eff_pur_by_class}

    \begin{tabular}{lc|c|c|c|c|c}
        \toprule
        \multirow{2}{*}{Algorithm}
        & All & CH & NH & $\gamma$ & $e$ & $\mu$ \\
        & \metrictriple{eff}{pur}{fakes/ev}
        & \metricpair{eff}{pur}
        & \metricpair{eff}{pur}
        & \metricpair{eff}{pur}
        & \metricpair{eff}{pur}
        & \metricpair{eff}{pur} \\
        \midrule
        \ours{}
        & \metrictriple{\textbf{0.9121}}{\textbf{0.9166}}{7.01}
        & \metricpair{\textbf{0.971}}{0.900}
        & \metricpair{\textbf{0.714}}{\textbf{0.877}}
        & \metricpair{\textbf{0.917}}{\textbf{0.949}}
        & \metricpair{\textbf{0.841}}{0.838}
        & \metricpair{0.564}{\textbf{0.847}} \\
        \flowf{}
        & \metrictriple{0.8876}{0.9050}{\textbf{6.70}}
        & \metricpair{0.942}{\textbf{0.944}}
        & \metricpair{0.673}{0.847}
        & \metricpair{0.910}{0.890}
        & \metricpair{0.699}{\textbf{0.845}}
        & \metricpair{\textbf{0.643}}{0.835} \\
        \ppfa{}
        & \metrictriple{0.6167}{0.7349}{6.96}
        & \metricpair{0.615}{0.895}
        & \metricpair{0.539}{0.193}
        & \metricpair{0.718}{0.869}
        & \metricpair{0.125}{0.357}
        & \metricpair{0.251}{0.804} \\
        \bottomrule
    \end{tabular}
\end{table}

Both learned methods reconstruct substantially more target particles than \ppfa{}, with higher integrated efficiency and purity after identification (Table~\ref{tab:particle_eff_pur_by_class}). Before requiring the correct particle class, the integrated reconstruction efficiencies are 93.3\% for \ours{}, 92.2\% for \flowf{} and 77.8\% for \ppfa{}. All three algorithms produce a similar number of fake and duplicate particles.

\FloatBarrier

\subsection{Particle Kinematic Regression}
\label{app:ttbar_regression}

\begin{figure}[tbh]
    \centering
    \includegraphics[width=\textwidth]{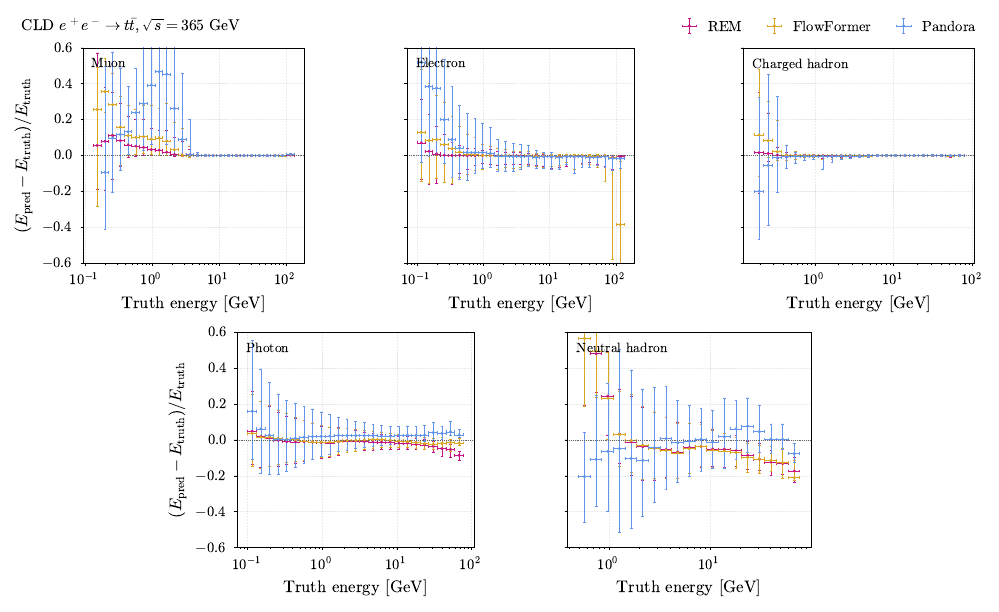}
    \caption{
    Particle-level energy performance in simulated $e^+e^- \to \ttbar{}$ events at $\sqrts=365~\mathrm{GeV}$.
    The fractional energy response (Section~\ref{sec:metrics}) is shown from left to right for muons, electrons, and charged hadrons on the top,
    and for photons and neutral hadrons on the bottom.
    \ours is compared with \flowf and \ppfa using the same truth-matching and object definitions.
    }
    \label{fig:ttbar_particle_energy_response}
\end{figure}

\cref{fig:ttbar_particle_energy_response} shows the fractional energy response. Markers indicate the median in each truth-energy bin, and vertical bars span the IQR. Both \ours and \flowf generally have smaller biases and narrower responses than \ppfa.

\begin{figure}[!htbp]
    \centering
    \includegraphics[width=\textwidth]{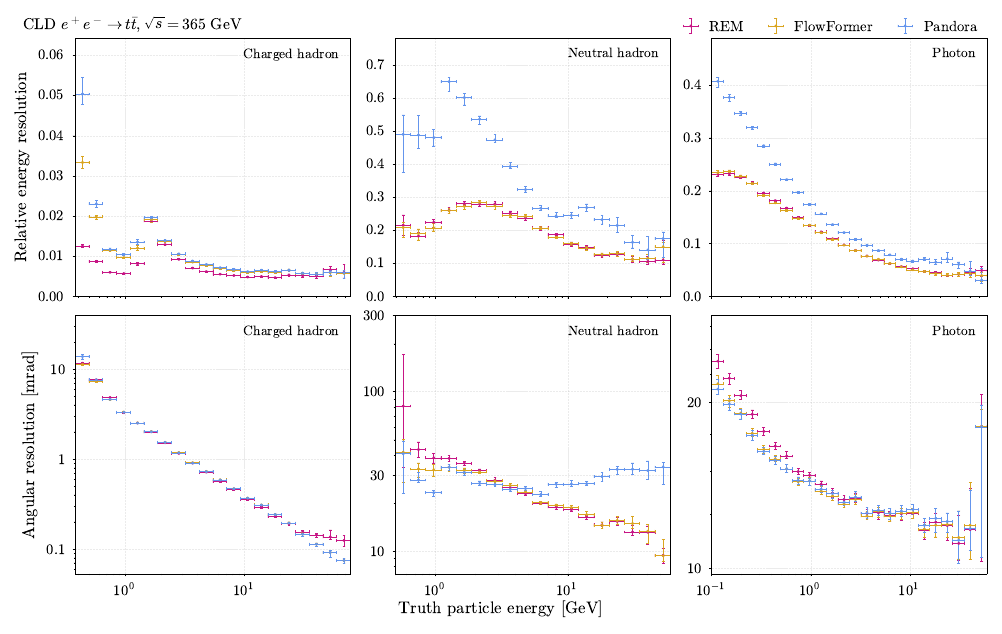}
    \caption{Energy (top) and angular (bottom) resolution of charged hadrons (left), neutral hadrons (middle) and photons (right) as a function of truth energy in simulated $e^+e^- \to \ttbar{}$ events at $\sqrts=365~\mathrm{GeV}$, comparing \ours, \flowf and \ppfa. Only truth particles matched by every reconstruction with an object of their own particle class are used, classified by their truth class. Resolution definitions follow Section~\ref{sec:metrics}. Error bars are 16th to 84th percentile intervals from bootstrap resampling of events, and bins with fewer than 20 particles are not shown. The charged-hadron energy resolution is shown above 0.38~GeV. The energy of a charged hadron with a track is computed from the track momentum assuming the charged-pion mass (Appendix~\ref{app:zqq_ch_momentum}).}
    \label{fig:ttbar_particle_resolution}
\end{figure}

Figure~\ref{fig:ttbar_particle_resolution} shows the energy and angular resolution of charged hadrons, neutral hadrons and photons for the truth particles that every reconstruction matches with an object of their own particle class, so that all reconstructions are compared on the same particles. The rise of the charged-hadron energy resolution between about 1 and 2~GeV is discussed in Appendix~\ref{app:zqq_ch_momentum}.

\FloatBarrier

\subsection{Event-Level Observables}

\begin{figure}[htb]
    \centering
    \includegraphics[width=0.49\textwidth]{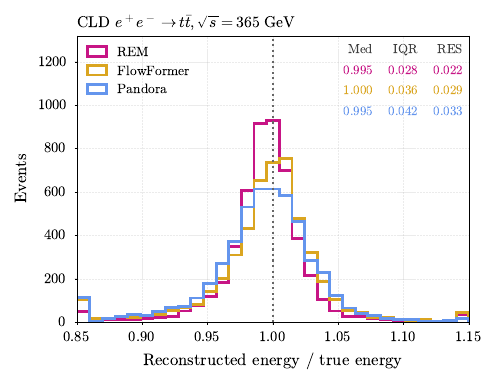}
    \includegraphics[width=0.49\textwidth]{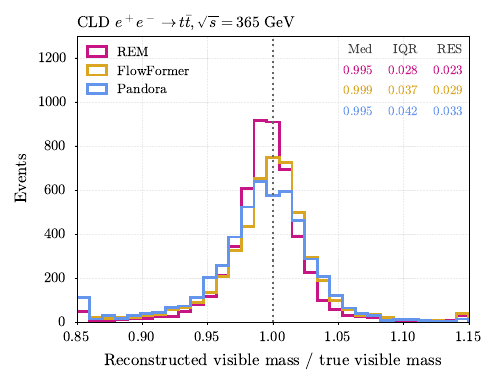}
    \includegraphics[width=0.49\textwidth]{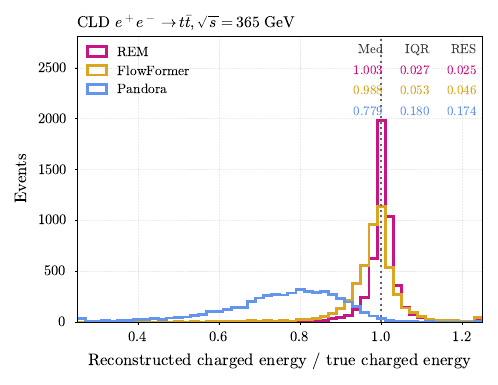}
    \includegraphics[width=0.49\textwidth]{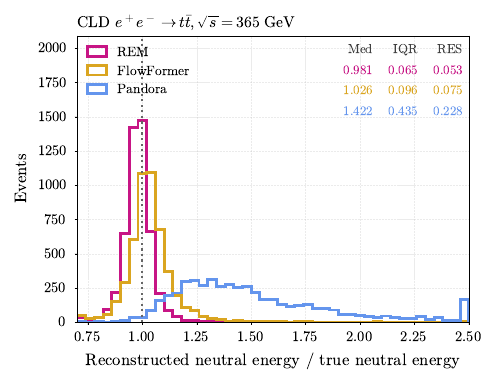}
    \caption{Event-level reconstructed-to-truth ratios in simulated $\ttbar{}$ events at $\sqrts=365$~GeV, comparing \ours{}, \flowf{} and \ppfa{}. Top: total visible energy (left) and visible invariant mass (right). Bottom: charged (left) and neutral (right) energy. Object selections follow Section~\ref{sec:targets}.}
    \label{fig:ttbar_event_level}
\end{figure}

The event-level observables follow the definitions in Section~\ref{sec:metrics}.
Figure~\ref{fig:ttbar_event_level} compares the total visible-energy and mass responses and their charged and neutral components. All three methods give similar total-energy closure, while the learned methods recover a more accurate charged--neutral partition. \ours{} gives the narrowest visible-energy response, measured by the IQR as in Section~\ref{sec:metrics}; the values are summarised in Table~\ref{tab:ttbar_headline}.

\begin{figure*}[p]
    \centering

    \includegraphics[width=0.4\textwidth]{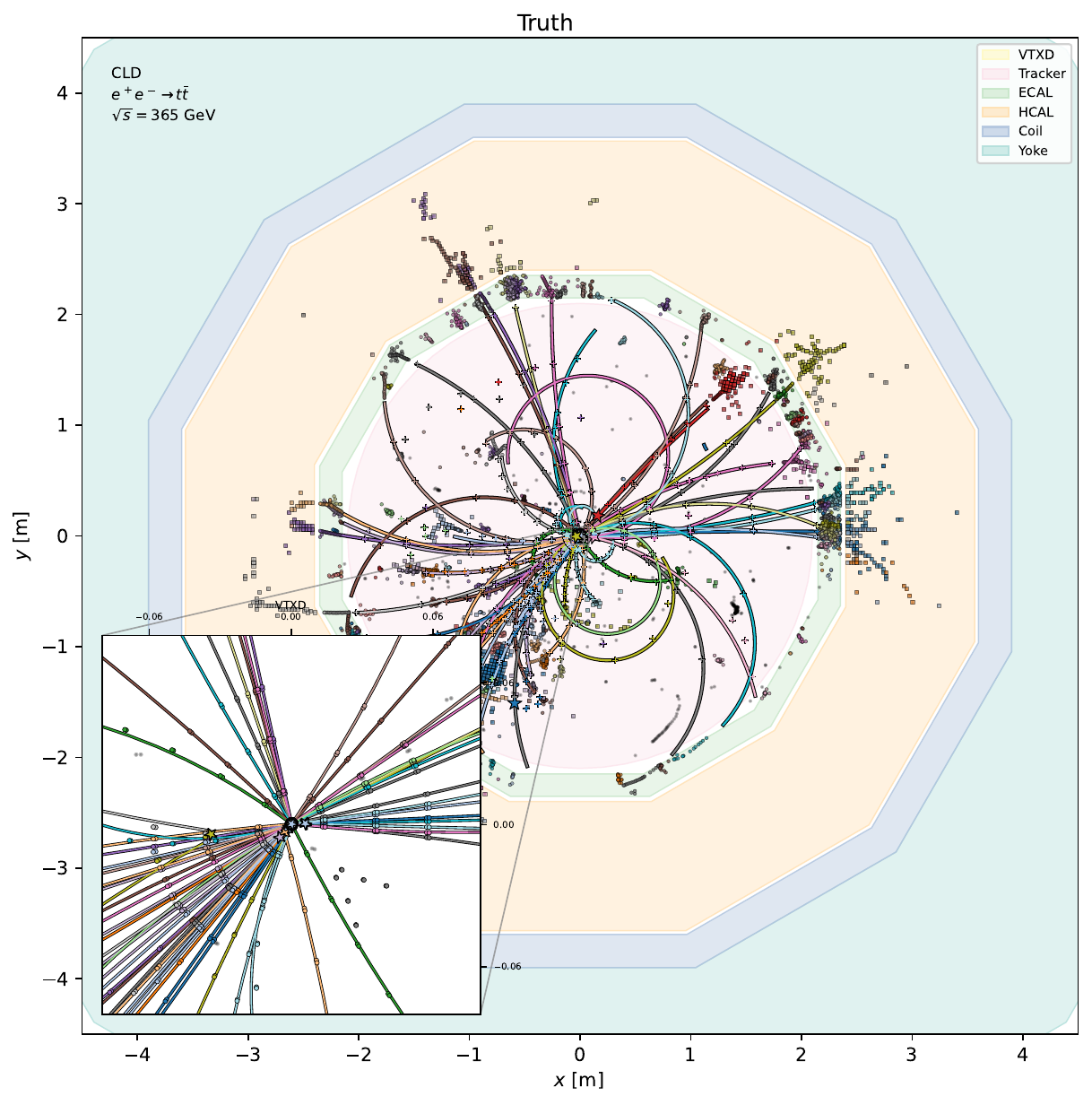}
    \includegraphics[width=0.4\textwidth]{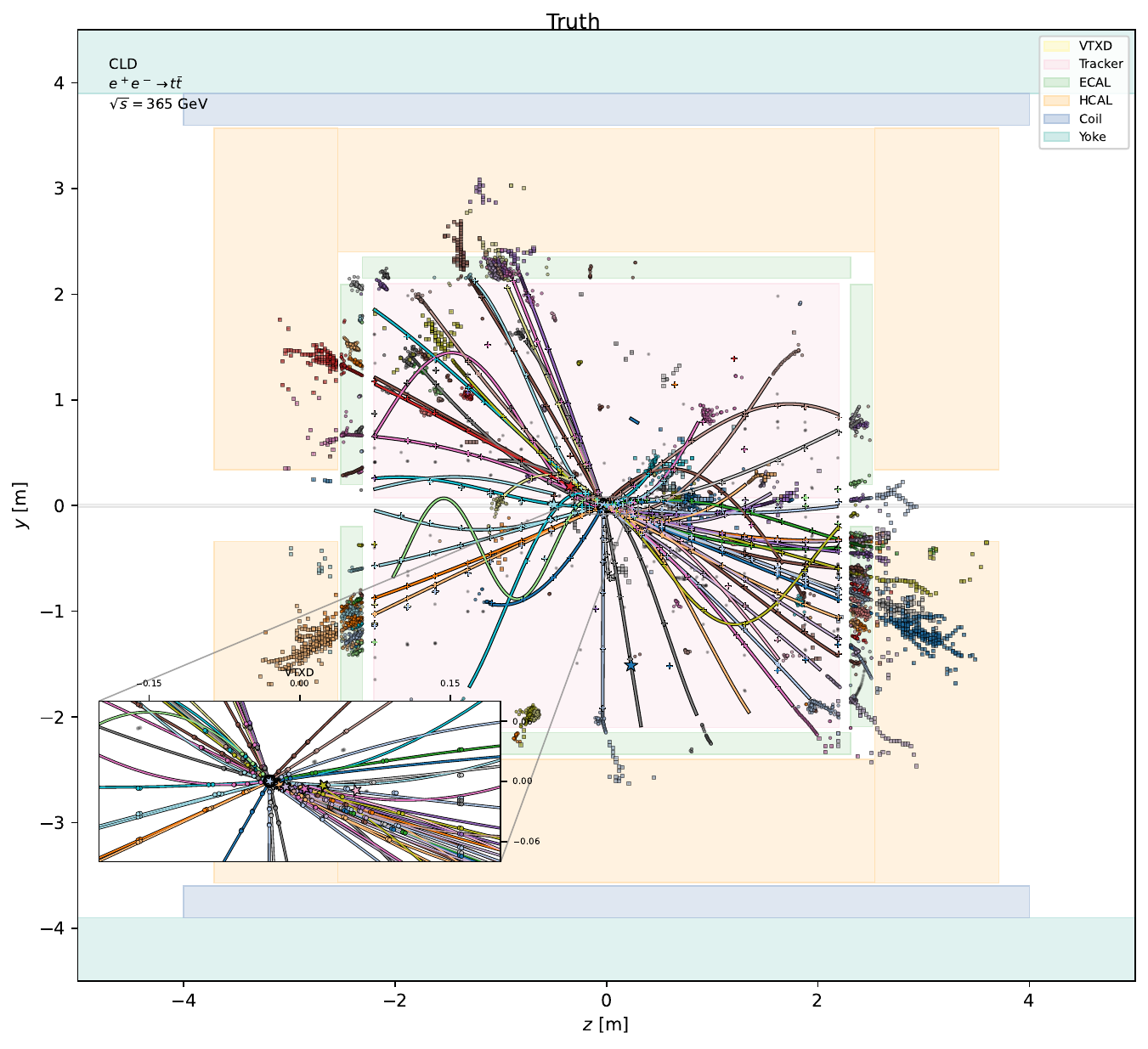}\\[-0.05cm]

    \includegraphics[width=0.4\textwidth]{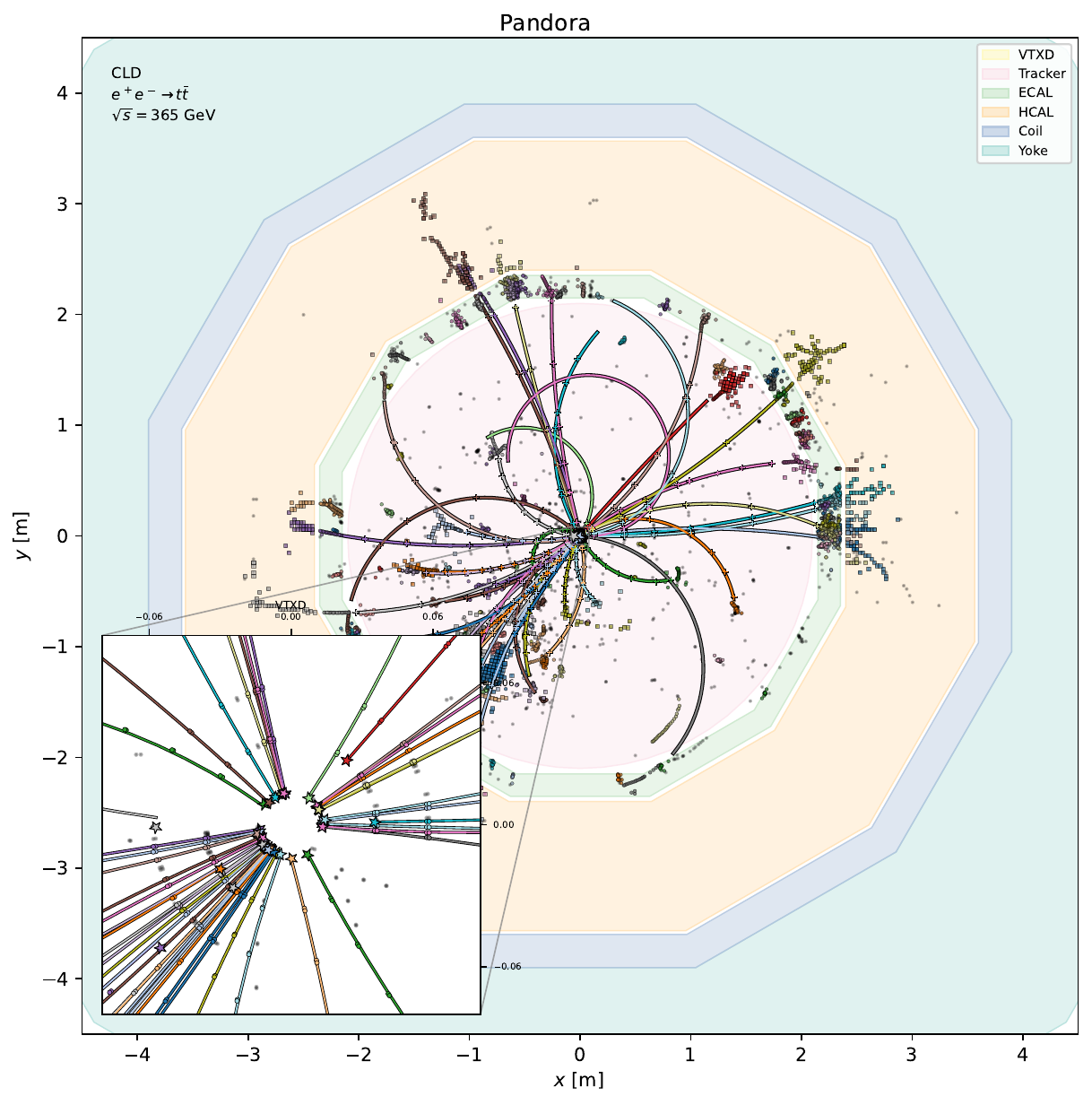}
    \includegraphics[width=0.4\textwidth]{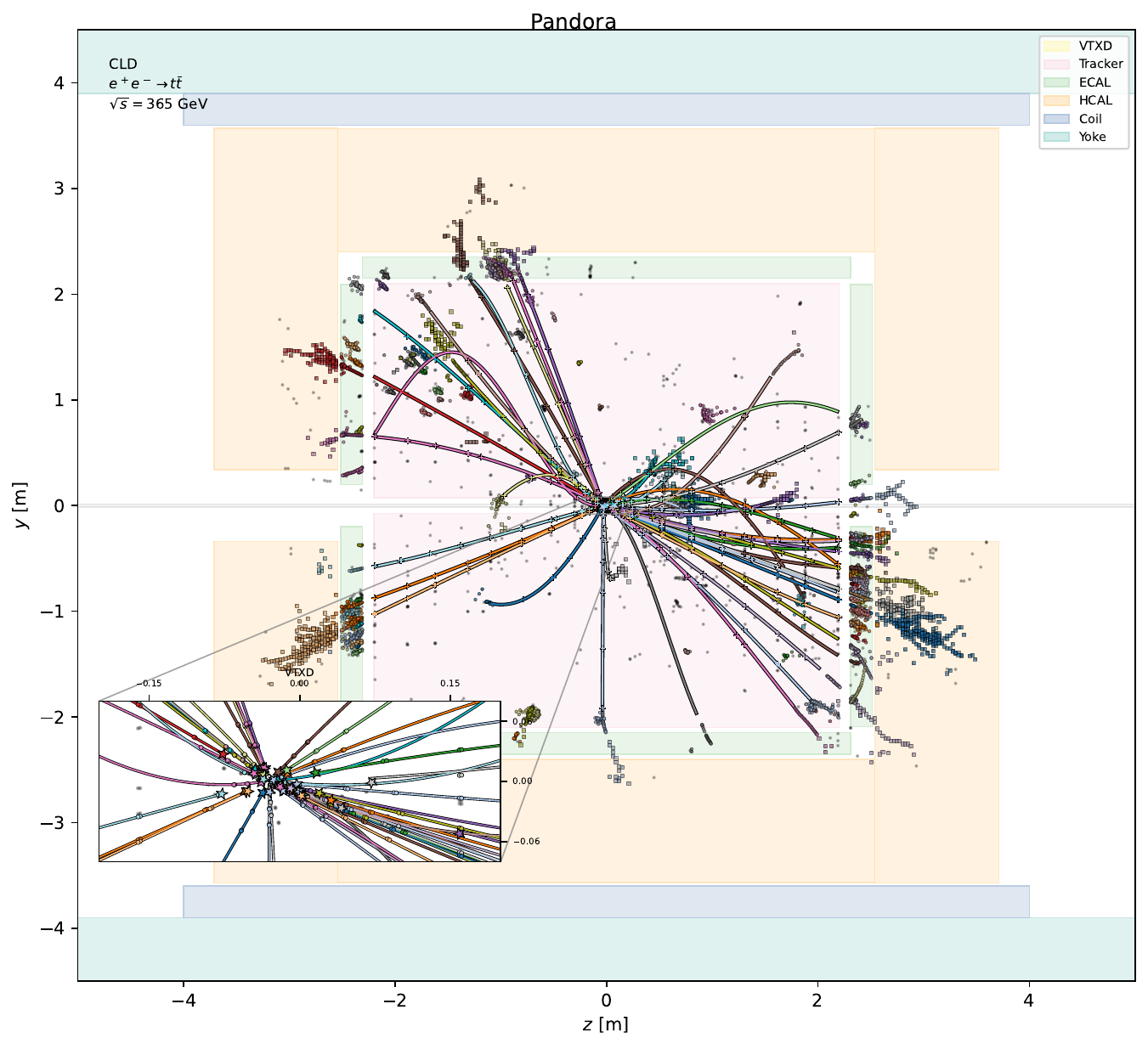}\\[-0.05cm]
    \includegraphics[width=0.4\textwidth]{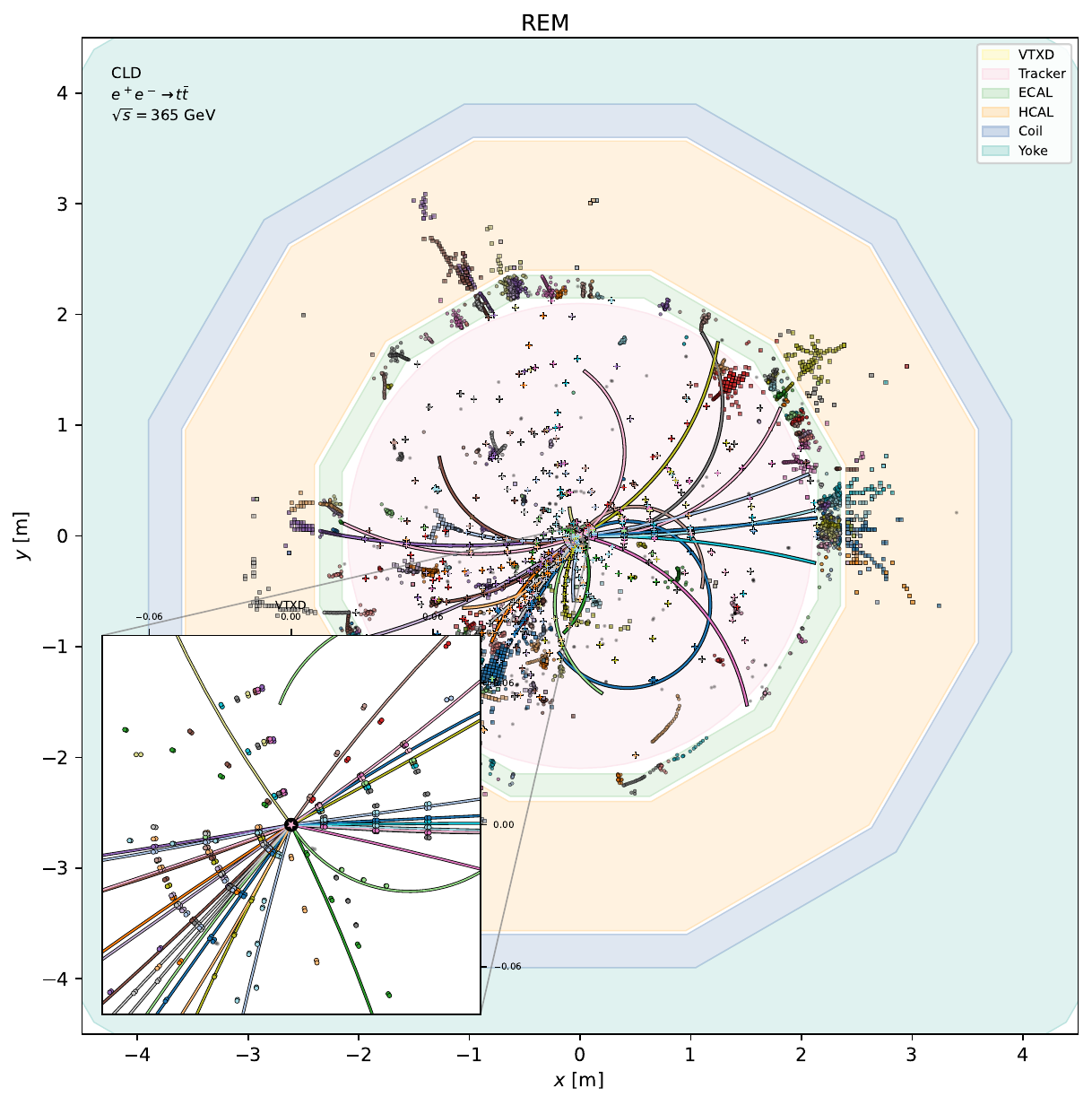}
    \includegraphics[width=0.4\textwidth]{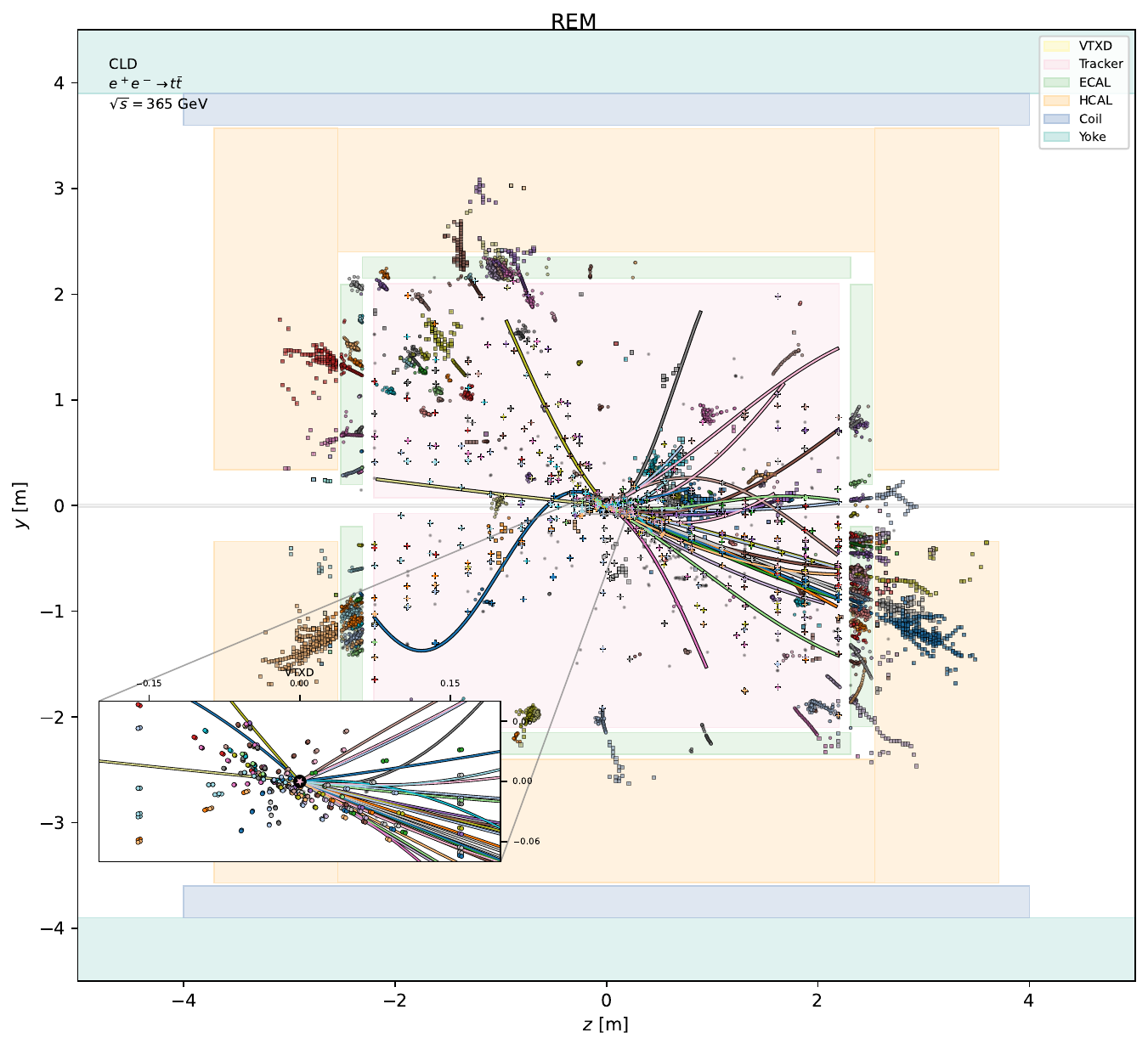}\\[-0.05cm]

    \caption{Event display for a simulated $\ttbar{}$ event at $\sqrts=365$~GeV in CLD. Columns show transverse $x$--$y$ (left) and longitudinal $z$--$y$ (right) views. Rows show truth hit assignments, \ppfa{} and \ours{} (top to bottom). Tracker, ECAL, HCAL and muon-system hits are shown with charged-particle trajectories overlaid.}
    \label{fig:event_display_ttbar}
\end{figure*}

A comparison of event displays is found in \cref{fig:event_display_ttbar}.

\FloatBarrier

\FloatBarrier
\clearpage

\clearpage

\printbibliography
\end{document}